\documentclass[11pt]{article}
\usepackage[colorlinks=true,linkcolor=darkblue,citecolor=darkblue]{hyperref}
\usepackage{fullpage,times,fourier,charter,graphicx,color}
\usepackage{amsmath,amssymb,amsthm,enumerate,tocloft}
\usepackage{cleveref}
\crefformat{equation}{(#2#1#3)}
\crefmultiformat{equation}{(#2#1#3)}{ and~(#2#1#3)}{, (#2#1#3)}{ and~(#2#1#3)}
\usepackage{booktabs}

\def\maketitle{\par\noindent{\color{darkblue}\LARGE\bf\sffamily\thetitle}\\[1.4ex]
{\large\theauthor}\\[0.6ex]
\textit{\thefirstaddress}\\[0.1ex]
\textit{\thesecondaddress}\\[0.4ex]
{\small\today}\par\vglue1.4\bigskipamount}
\def\title#1{\def\thetitle{#1}}
\def\author#1{\def\theauthor{#1}}
\def\firstaddress#1{\def\thefirstaddress{#1}}
\def\secondaddress#1{\def\thesecondaddress{#1}}
\allowdisplaybreaks
\definecolor{darkred}{rgb}{0.9,0,0}
\definecolor{darkgray}{rgb}{0.4,0.4,0.4}
\definecolor{darkblue}{rgb}{0,0,0.7}
\definecolor{brown}{rgb}{0.6,0.2,0.2}
\def\F{\mathop{\mathcal F}\nolimits}
\def\cn{\mathop{\rm cn}\nolimits}
\def\sn{\mathop{\rm sn}\nolimits}
\def\dn{\mathop{\rm dn}\nolimits}
\def\am{\mathop{\rm am}\nolimits}
\def\phase{\cos\alpha x + \sin \alpha y}
\def\wavelength{L} 
\newcommand\backvec[1]{\reflectbox{$\vec{\reflectbox{\!$#1$}}$}}

\let\epsilon=\varepsilon

\makeatletter
\renewcommand\section{\@startsection {section}{1}{\z@}%
  {-2.2ex \@plus -1ex \@minus -.2ex}%
  {1ex \@plus.1ex}%
  {\normalfont\bf\sffamily\color{darkblue}}}
\renewcommand\subsection{\@startsection{subsection}{2}{\z@}%
  {-1.6ex\@plus -0.4ex \@minus -.2ex}%
  {0.6ex \@plus .1ex}%
  {\normalfont\small\bf\sffamily}}
\renewcommand\subsubsection{\@startsection{subsubsection}{3}{\z@}%
  {-0.6ex\@plus -0.2ex \@minus -.2ex}%
  {0.4ex \@plus .1ex}%
  {\normalfont\normalsize\it}}
\renewcommand\paragraph{\@startsection{paragraph}{4}{\z@}%
  {0.2ex \@plus0.2ex \@minus0.1ex}{-0.5em}%
  {\normalfont\normalsize\bfseries}}
\makeatother

\def\fbf#1{\setbox0=\hbox{$#1$}\kern-0.10\wd0
  \lower0.02em\copy0\kern-\wd0 \lower0.02em\hbox{\kern+0.05em\copy0}\kern-\wd0
  \raise0.02em\copy0\kern-\wd0 \raise0.02em\hbox{\kern-0.05em\box0}}
\let\truenabla=\nabla
\def\nabla{{\fbf\truenabla}}

\makeatletter
\let\tru@int=\int
\def\int{\mathop{\textstyle\tru@int}\limits}
\def\overl@ss#1#2{\vcenter{\offinterlineskip
        \ialign{$\m@th#1\hfil##\hfil$\crcr#2\crcr<\crcr } }}
\def\overgr@at#1#2{\vcenter{\offinterlineskip
        \ialign{$\m@th#1\hfil##\hfil$\crcr#2\crcr>\crcr } }}
\def\gl{\mathrel{\mathpalette\overl@ss>}}
\def\lg{\mathrel{\mathpalette\overgr@at<}}

\def\Integer{\mathbb{Z}}
\def\Real{\mathbb{R}}
\def\Complex{\mathbb{C}}

\def\Im{\mathop{\rm Im}\nolimits}
\def\sn{\mathop{\rm sn}\nolimits}
\def\cn{\mathop{\rm cn}\nolimits}
\def\dn{\mathop{\rm dn}\nolimits}

\def\pvint{\int\kern-0.94em-\kern0.2em}
\def\sech{\mathop{\rm sech}\nolimits}

\def\dn{\mathrm{dn}}
\let\^=\hat
\let\==\bar
\let\~=\tilde
\let\@=\mathbf
\let\_=\mathsf

\let\le=\leqslant
\let\ge=\geqslant

\def\K{K_m}
\def\R{\mathcal{R}}

\def\d{\mathrm{d}}
\def\e{\mathrm{e}}
\def\rhobar{\overline{\rho}}
\def\slowgradient{\mathop{\bf D}\nolimits}
\def\backnabla{\backvec{~\nabla}}
\def\backslowgradient{\backvec{~\slowgradient}}
\def\div{\mathop{\rm div}\nolimits}

\def\t{{\mathrm{t}}}
\def\txtfrac#1#2{{\textstyle\frac{#1}{#2}}}

\def\partialderiv#1#2{{\frac{\partial #1}{\partial #2}}}
\def\diag{\mathop{\rm diag}\nolimits}
\def\bflambda{\boldsymbol\lambda}

\def\half{{\textstyle\frac12}}
\def\O#1{^{(#1)}}

\makeatother

\advance\parskip 4pt
\advance\textwidth -2em
\advance\hoffset 1em
\advance\textheight 8mm
\advance\voffset -2mm

\def\be{\begin{equation}}
\def\ee{\end{equation}}
\def\bse{\begin{subequations}}
\def\ese{\end{subequations}}

\numberwithin{equation}{section}
\theoremstyle{plain}

\theoremstyle{definition}

2

\begin{document}
\title{Whitham modulation theory for the Davey-Stewartson\\[3pt]
system and stability analysis of its periodic traveling\\[3pt]wave solutions}
\author{Gino Biondini$^1$, Alexander Chernyavsky$^{1,2}$, Haodong Lin$^1$ and John Ringland$^1$}
\firstaddress{$^1$ Department of Mathematics, State University of New York at Buffalo, Buffalo, NY - USA}
\secondaddress{$^2$ Akian College of Science and Engineering, American University of Armenia, Yerevan - Armenia}
\maketitle

\begin{quotation}
\noindent\small\textbf{Abstract.}
The Davey-Stewartson (DS) system is a canonical model for a weakly nonlinear wave packet in two spatial dimensions that is resonantly coupled to the slowly varying mean flow it generates, and as such it governs modulated wave trains in settings as different as finite-depth water waves, quadratic optical media and magnetic films.
Here we develop the Whitham modulation theory for the DS system and use it to study the stability of its periodic traveling wave solutions.
Specifically:
1.~We perform a detailed study of the parametric dependence of the periodic solutions of all four variants of the DS system, including their harmonic and soliton limits. 
2.~We present the formulation of Whitham modulation theory for the DS system and write down the resulting Whitham modulation equations for all four variants of the DS system.
3.~We use the resulting modulation equations to study the stability of the periodic traveling wave solutions of all four variants of the DS system.  
4.~We validate the theoretical predictions by comparing them with a direct linearization of the DS system as well as with the direct numerical simulations of the DS system, showing excellent agreement.
The results indicate that the periodic solutions of the defocusing DSII system are linearly stable,
whereas those of all other variants of the DS system are unstable.
The calculations also evidence many structural features (non-evolutionary modulation equations, constraints on the admissible initial data, and auxiliary mean-field constants absent from the traveling wave solutions yet indispensable to the modulation system) are expected to recur in any two-dimensional envelope equation with mean-field coupling.
\end{quotation}


\section{Introduction}

Equations of nonlinear Schr\"odinger (NLS) type are universal models
for the evolution of weakly nonlinear dispersive wave trains
(e.g., see \cite{BenneyNewell1967,CalogeroEckhaus1987})
and are ubiquitous in nature, arising in such disparate fields as water waves, optics, acoustics, Bose-Einstein condensates and beyond
\cite{AC1991,AS1981,NMPZ1984,HasegawaKodama,IR2000}.
As such, they have been studied extensively \cite{AC1991,AS1981,Newell1985,NMPZ1984,SulemSulem}. 
In many other cases, when more than one spatial dimension is relevant,
the governing equations are often systems of nonlinear Schr\"odinger type with 
coupling to mean terms; e.g., see \cite{PRE63p046605,AS1979,BenneyRoskes,Leblond,Zakharov1972}.
The physical origin of the coupling in these systems is easy to understand.
Whenever a quasi-monochromatic wave packet propagates in a medium in which the envelope can drive a non-oscillatory, long-scale response (a mean flow, a static rectified field, a mean magnetization), that response feeds back on the packet through the nonlinearity, and in two spatial dimensions the feedback is generically nonlocal, because the mean field is slaved to the envelope through an elliptic or hyperbolic PDE with no time derivative of its own.
For example, in surface gravity waves on fluids of finite depth, which is the context in which the DS system was originally derived~\cite{DaveyStewartson,AS1979,BenneyRoskes}, the envelope of the free-surface displacement is tied to the wave-induced mean flow
by the depth-averaged mass balance that the latter must satisfy.
Systems with similar structure describe the interaction of an optical envelope with the static field it rectifies in a quadratic nonlinear medium~\cite{PRE63p046605}, magnetization waves in ferromagnetic films~\cite{Leblond}, and Langmuir wave packets coupled to the ion-acoustic response of a plasma~\cite{Zakharov1972}.
In every case, the presence of the mean field sets apart the resulting dynamics from that in the NLS equation.

A distinguished limit of such kinds of equations is the so-called Davey-Stewartson (DS) system~\cite{DaveyStewartson};
see~\cite{KleinSaut} for a comprehensive account of the DS system and related nonlinear dispersive 
equations from the combined perspectives of asymptotic modeling, PDE theory and integrable systems.
Like the NLS equation in one spatial dimension, the DS system is completely integrable,
and as such it possesses a remarkable mathematical structure, including the existence of a Lax pair (e.g., see \cite{AC1991}), 
the solvability of the initial value problem by a suitable inverse scattering transform
\cite{AC1991,FokasAblowitz1983,FokasAblowitz1984}, 
and the existence of a large variety of exact solutions \cite{BLMP:PLA1988,FokasPelinovskySulem,FS1990,ManasSantini,TajiriArai}.
The DS system continues to receive considerable interest in recent years, including the study of well-posedness \cite{Perry2016}, boundary value problems \cite{Fokas2003}, the development of efficient high-precision numerical methods and numerical studies of blow-up \cite{KleinStoilovI,KleinStoilovII,KleinMcLaughlinStoilovPRSA2019}, the study of 
periodic solutions \cite{EAJAM2022}, multi-soliton solutions \cite{JPA55p305701},
generalizations thereof \cite{GuoMichalache2022,RaoHeCheng2022,Xu2017},
and the stability of solutions \cite{BrownPerry,Pelinovsky1994,SenChowdhury,TajiriTakeuchiArai}.
In particular, a problem of recent theoretical and practical interest has been the study of the semiclassical limit of the DS system \cite{Assainova,KleinMcLaughlinStoilovPhysD2019,KleinRoidot2014,WangWang2019}, 
i.e., the scenario in which dispersive effects are small compared to nonlinear ones,
as well as the singular limit of the dispersionless DS system \cite{GungorOzemir,GeYi2020,GeYi2022,YiLiaoTianWang2022}.

An effective tool to study small dispersion problems is Whitham modulation theory \cite{Whitham1974}, which has been applied with great success to study 
a variety of problems (e.g., see the review article \cite{ElHoefer2016}).
Roughly speaking, Whitham modulation theory describes slow modulations of a rapidly oscillating periodic wave train via a small set of slowly varying parameters (such as amplitude, wavenumber and mean level) which satisfy a closed system of first-order PDEs governing their evolution, thereby reducing a dispersive problem to a quasi-linear, hydrodynamic-type one.
Until recently, the use of Whitham modulation theory was mostly limited to systems in one spatial dimension.
In~\cite{RSPA2017}, a general methodology was formulated to extend the formulation of Whitham modulation theory to systems with two or more spatial dimensions, starting with the 
Kadomtsev-Petviashvili (KP) equation.
A similar approach was then successfully applied to several other nonlinear equations 
such as the two-dimensional Benjamin-Ono equation, modified KP equation, NLS equation in  
two and three spatial dimensions, the Zakharov-Kuznetsov equation and the Benney-Luke equation in \cite{ABH_JPA2023,ABW_2DBO,ABR2018,BC2024,ACR_2DNLS,HuLuoWang2024}.
Moreover, the resulting modulation equations were used in \cite{HuLuoWang2024,JFM2026,JFM2021,NLTY2021,RSPA2022} to effectively study a variety of dynamical problems, including in particular
a study of the stability of the underlying periodic traveling wave solutions of the original system. 

The above works evidenced structural features of multi-dimensional modulation theory not present in one dimension, such as the appearance of non-evolutionary equations in the modulation system and of compatibility constraints on the admissible initial data.
The DS system shares these features and adds two more.
First, the mean field is governed by an equation without a time derivative that is elliptic for DSII and hyperbolic for DSI.  Consequently the modulation system inherits \textit{two} non-evolutionary equations, and, as we show in section~\ref{s:whitham}, the additive constants for \textit{both} of the corresponding fields must be retained as dependent variables even though one of them does not appear in the traveling wave solutions.  
Second, the character of the periodic solutions themselves depends on the direction of propagation: the sign of a key quantity [defined in~\eqref{e:kappadef}], which selects between two structurally different families of elliptic solutions, is fixed once and for all for DSII, but for DSI it changes across the main diagonals of the $xy$-plane.
We are unaware of another system in which the qualitative form of the wave train depends on its inclination in this way.
The DS system therefore represents a significant generalization of the methodology of~\cite{RSPA2017}, and the lessons drawn here are expected to apply to any two-dimensional envelope equation with mean-field coupling, integrable or not.

More specifically, the purpose of this work is severalfold. 
1.~We perform a detailed study of the parametric dependence of the periodic solutions of all variants of the DS system, including their harmonic and soliton limits. 
2.~We present the formulation of Whitham modulation theory for the DS system and we obtain explicitly the resulting Whitham modulation equations for all variants of the DS system. 
3.~We use the resulting modulation equations to study the stability of the periodic traveling wave solutions of all variants of the DS system. 
4.~We validate the theoretical predictions by comparing them with a direct linearization of the DS system as well as with the direct numerical simulations of the DS system, showing excellent agreement.
5.~Finally, we identify a number of structural features of the resulting modulation system that we expect to be generic, and we formulate several open problems to which the availability of the DS-Whitham system gives access: the integrability of the modulation system itself, in analogy with the situation for the KP equation~\cite{NLTY2020,NLTY2024}; the rigorous justification of the modulation predictions in the framework of \cite{ClarkeMarangellPerkins,JohnsonOregero,JohnsonOregeroPerkins,JohnsonPerkins}; the well-posedness of the DSI system with periodic boundary conditions, which is a delicate issue because in that case \eqref{e:Ppde} is a forced linear wave equation; and the use of the modulation equations to study genuinely two-dimensional initial value problems, such as oblique soliton-mean flow interactions and dispersive shock waves, along the lines of what was recently done for the KP equation in \cite{JFM2021,NLTY2021,RSPA2022,JFM2026}.
We return to each of these in section~\ref{s:discussion}.

We consider the Davey-Stewartson (DS) system in a semiclassical scaling, which is the system of coupled nonlinear partial differential equations
\bse
\label{e:DS}
\begin{gather}
i\epsilon\psi_t + \half\epsilon^2 (\psi_{xx}-\sigma \psi_{yy}) 
	+ \sigma \Psi \psi = 0, 
\label{DS-qqq}\\
\Psi_{xx} + \sigma \Psi_{yy} = -\nu(|\psi|^2)_{xx} + \sigma \nu (|\psi|^2)_{yy}. 
\label{DS-psi}
\end{gather}
\ese
Here, $\psi$ is a complex-valued function of $x,y,t$ describing a slowly-varying envelope, while $\Psi$ is a real-valued function of $x,y,t$ describing a mean field,
and as usual subscripts $x,y,t$ denote partial differentiation.
The choice of parameters $\sigma=\pm 1$ and $\nu=\pm 1$ determines four possible variants of the DS system. 
Specifically, $\sigma = -1$ and $\sigma = 1$ identify respectively the DSI and DSII equations, 
whereas $\nu= -1$ and $\nu=1$ identify respectively the focusing and defocusing variants.
Since we will repeatedly need to deal with each variant of the DS system individually, we find it convenient to list them explicitly here:
\\[1ex]\hglue1em
1. Defocusing DSII equation: $\sigma=1$ and $\nu=1$;
\\[0.3ex]\hglue1em
2. Focusing DSII equation: $\sigma=1$ and $\nu=-1$;
\\[0.3ex]\hglue1em
3. Defocusing DSI equation: $\sigma = -1$ and $\nu=1$;
\\[0.3ex]\hglue1em
4. Focusing DSI equation: $\sigma = -1$ and $\nu = -1$.

It is important to observe, however, that, while the labels ``focusing'' and ``defocusing'' are unambiguous for the DSII systems, for the DSI systems the behavior of the reductions to solutions that are independent of~$y$
differs from that of the reductions to solutions independent of~$x$, 
e.g., see \cite{EAJAM2022}.
Indeed, the defocusing and focusing DSI systems can easily be transformed into each other via the map 
\vspace*{-1ex}
\be
(\psi(x,y,t),\Psi(x,y,t),\nu) \leftrightarrow (\psi(y,x,-t),-\Psi(y,x,-t),-\nu)\,.
\label{e:DSIsymmetry}
\ee
Thus, for the DSI variants, the labels ``focusing'' and ``defocusing'' are somewhat arbitrary.
For this reason, in this work we will often deal with a single DSI system
and use the above symmetry to obtain the corresponding results for the other variant.

\smallskip
The DS systems are also often written using a different auxiliary variable via the dependent variable 
transformation 
\be
P(x,y,t) = \Psi(x,y,t) - \nu |\psi(x,y,t)|^2\,.
\label{e:Pdef}
\ee
Then \eqref{e:DS} becomes
\bse
\label{e:DS1}
\begin{gather}
i\epsilon \psi_t + \half\epsilon^2 (\psi_{xx}-\sigma \psi_{yy}) + \sigma (P+\nu |\psi|^2) \psi = 0, 
\label{e:psipde}
\\
P_{xx} + \sigma P_{yy} + 2\nu(|\psi|^2)_{xx} = 0\,.
\label{e:Ppde}
\end{gather}
\ese
The system~\eqref{e:DS} is also completely integrable and can be expressed as a compatibility condition of a suitable Lax pair (e.g., see \cite{AC1991}), but integrability will not play a role in our analysis.
In all the calculations of the following sections, we will assume that $\psi(x,y,t)$ and $P(x,y,t)$, as well as all the related dependent variables, are sufficiently smooth that all differential expressions are meaningful.

Since Whitham modulation equations are first-order systems, it is also convenient for our purposes to convert the resulting PDE~\eqref{e:Ppde} for $P(x,y,t)$ into a pair of first-order PDEs by introducing a second auxiliary field $Q(x,y,t)$, obtaining
\bse
\label{e:DS2}
\begin{gather}
i\epsilon \psi_t + \half \epsilon^2 (\psi_{xx}-\sigma \psi_{yy}) 
	+ \sigma (P+\nu |\psi|^2)\psi = 0, 
\label{e:DS-q}
\\
P_x = \sigma Q_y - 2\nu (|\psi|^2)_x,
\qquad 
Q_x = -P_y \,.
\label{e:DS-p}
\end{gather}
\ese
When $\sigma = 1$, \eqref{e:DS-p} is equivalent to inhomogeneous Cauchy-Riemann conditions,
whereas when $\sigma = -1$, \eqref{e:DS-p} is the first-order decomposition of an inhomogeneous one-dimensional wave equation.

Finally, it is important to realize that \eqref{e:Ppde} only determines $P(x,y,t)$ up to an arbitrary solution of the associated homogeneous PDE
(i.e., Laplace's equation for the DSII equation and the wave equation for the DSI equation).
Therefore, in order to formulate a well-posed Cauchy problem for the DS system~\eqref{e:DS1}, 
one needs to specify such a homogeneous solution for all $t\in\Real^+$ in addition to the initial conditions for $\psi(x,y,t)$ and $P(x,y,t)$. 
Throughout this work, we will always take such homogeneous solutions to be constant.

The rest of this work is organized as follows.
In section~\ref{s:madelung} we apply the Madelung transformation to the DS system, derive the associated conservation laws, and introduce the multiple scales ansatz that underlies the modulation theory.
In section~\ref{s:periodic} we construct the periodic traveling-wave solutions of the DS systems and study their soliton limit.
In section~\ref{s:whitham} we carry out the Whitham averaging procedure and derive the modulation equations governing the slow evolution of the wave parameters.
In section~\ref{s:reductions} we examine the harmonic and soliton reductions of the Whitham modulation equations.
In section~\ref{s:stability} we analyze the stability of the periodic solutions, comparing the predictions of Whitham modulation theory with those obtained from direct linearization of the DS system.
Finally, in section~\ref{s:discussion} we summarize our results and outline directions for future work.
A few technical results are confined to various appendices.

\section{Madelung transformation, conservation laws, multiple scales ansatz}
\label{s:madelung}

\subsection{Madelung transformation and conservation laws}

As is common in studies of Schr\"odinger-type evolution equations, for the purposes of the present analysis it is convenient to introduce the transformation to the so-called ``Madelung fluid'' ansatz as
\be
\psi(\@x,t) = \sqrt{\rho(\@x,t)}\,\e^{i\Phi(\@x,t)}\,,
\label{Madelung}
\ee
with $\@x = (x,y)^\t$ (the superscript ``$\mathrm t$'' denoting matrix transpose), 
where $\rho(\@x,t)$ and $\Phi(\@x,t)$ are real-valued functions, and with the requirement that $\rho(\@x,t)$, which represents a fluid density in this analogy, be non-negative.
No equivalent transformation is required for the mean fields $P(\@x,t)$ and $Q(\@x,t)$ since they are already real-valued.

Plugging~\eqref{Madelung} into the system~\eqref{e:DS2} and separating real and imaginary parts in \eqref{e:DS-q}, one obtains
\vspace*{-1ex}
\bse
\begin{gather}
-\epsilon \Phi_t + \frac1{2\sqrt\rho}\epsilon^2 \left( 
	(\sqrt{\rho})_{xx} - \sigma(\sqrt{\rho})_{yy} \right)
	- \frac12\epsilon^2 (\Phi_x^2 - \sigma\Phi_y^2)
	+ \sigma(P+\nu\rho) = 0, 
\label{DS-real}\\
\rho_t + \epsilon \left( \rho_x \Phi_x - \sigma\rho_y \Phi_y 
	+ \rho \Phi_{xx} - \sigma\rho \Phi_{yy}\right) = 0, 
\label{DS-imag}\\
\noalign{\noindent whereas \eqref{e:DS-p} yields}
P_x - \sigma Q_y + 2\nu\rho_x = 0, \qquad
Q_x + P_y = 0. 
\label{DS-prho}
\end{gather}
\ese
Note that $\Phi$ never appears directly in the above equations, only its derivatives do.
It is therefore convenient to introduce the new vector-valued dependent variable
\be
\@ u(\@x,t) = (u_1,u_2)^\t = \epsilon\,\nabla\Phi(\@x,t)
\label{e:udef}
\ee
[where as usual $\nabla = (\partial_x,\partial_y)^\t$], 
which plays the role of fluid velocity for the Madelung fluid.
Note that~\eqref{e:udef} immediately implies the irrotationality constraint
\begin{equation}
\nabla\times\@ u = 0\,
\label{e:uirrotational}
\end{equation}
(or, explicitly, $u_{1,y} = u_{2,x}$).
Using $u_1$ and $u_2$ and simplifying the resulting equations, we then have the DS system in Madelung variables as
\bse
\begin{gather}
    -\epsilon \Phi_t + \tfrac12 (\sigma u_2^2-u_1^2) + \sigma(P + \nu\rho)
	+ \tfrac1{8} \epsilon^2 \left(\frac{1}{\rho^2}(\sigma\rho_y^2 - \rho_x^2) 
	+ \frac{2}{\rho}(\rho_{xx}-\sigma\rho_{yy})\right) = 0, \label{phiS}\\
    \rho_t + (\rho u_1)_x - \sigma(\rho u_2)_y = 0, 
    \\
    P_x - \sigma Q_y + 2\nu\rho_x = 0,\qquad
    Q_x + P_y = 0\,. 
\label{e:Madelungsystem1c}
\end{gather}
\ese
Differentiating the first equation with respect to $x$ and $y$ and using the definition of $\@u$, as well as the irrotationality constraint~\eqref{e:uirrotational}, we can write the resulting system in vector form as
\bse
\label{e:Madelungsystem}
\begin{gather}
    \@u_t + \tfrac12 \nabla (\@u^\t\Sigma_3\@u)
	- \nabla \left(\sigma(P+\nu\rho)
        +\frac{\epsilon^2}{8}\left( \Box_\sigma\log\rho 
	+ \frac{1}{\rho}\Box_\sigma\rho\right)\right) = \@0, 
\label{e:Madelungsystem1a}\\
    \rho_t + \nabla\cdot (\rho\Sigma_3 \@u) = 0,
\label{e:Madelungsystem1b}
\end{gather}
\ese
together with~\eqref{e:Madelungsystem1c},
where 
the Pauli matrices $\sigma_1,\sigma_2,\sigma_3$ are defined as usual so that  $\sigma_3 = \diag(1,-1)$, the modified d'Alembertian operator is defined by  $\Box_\sigma u = u_{xx}-\sigma u_{yy}$, and where we introduced the matrix
$\Sigma_3 = \diag(1, -\sigma)$.
Equivalently, \eqref{e:Madelungsystem1a} can also be rewritten as
\begin{gather}
    (\rho\@u)_t + (\rho \@u\otimes \Sigma_3\@u)\backnabla
		- \rho\nabla\left(\sigma(P+\nu\rho)+\frac{\epsilon^2}{8}
	\left(\Box_\sigma\log\rho + \frac{1}{\rho}\Box_\sigma\rho\right)\right) = 0, 
\label{e:Madelungsystem2a}
\end{gather}
where for any two vectors
$\@a = [a_1,a_2]^\t$ and $\@b = [b_1,b_2]^\t$ the dyadic operator $\otimes$ is defined as
\be
\@a\otimes \@b = \@a \@b^\t = \begin{pmatrix} a_1 \\ a_2\end{pmatrix} \big(b_1\,\,\,b_2\big) = 
\begin{pmatrix} a_1 b_1 & a_1 b_2 \\ a_2 b_1 & a_2 b_2 \end{pmatrix}\,,
\ee
and where for any $2\times2$ matrix~$A$ we have defined $(A)\backnabla = (\nabla^\t A^\t)^\t$.
Note however that \eqref{e:Madelungsystem1a} and~\eqref{e:Madelungsystem2a} are still not completely in hydrodynamic form, however,
because of the last term in the first of~\eqref{e:Madelungsystem1a} and~\eqref{e:Madelungsystem2a}.
To fix this issue, it is useful to use the following identities, which 
which can be easily verified by direct calculation:
\bse
\label{e:vectordifferentialidentities}
\begin{gather}
\sigma \rho\,\nabla(P+\nu\rho) = \sigma \nabla\left(\frac12\nu\rho^2 + P\rho\right) - \sigma P\,\nabla\rho, 
\label{e:vectordiffiden1}
\\
\rho \nabla\left(\Box_\sigma\log\rho + \frac{1}{\rho}\Box_\sigma\rho\right) = 
   2 \nabla \Box_\sigma\rho - 2 \left( \frac1\rho \nabla\rho\otimes\nabla\rho \right) \Sigma_3 \backnabla \,.
\end{gather}
\ese
Moreover, to rewrite the last term in the right-hand side of~\eqref{e:vectordiffiden1} in hydrodynamic form,
it is straightforward to verify that 
\bse
\label{e:divrelation}
\be
\sigma P\nabla\rho = \div T\,,
\ee
where we have introduced the matrix
\be
T = \begin{pmatrix} 
    \dfrac{Q^2-\sigma P^2}{4\nu} & \dfrac{PQ}{2\nu} \\
			\sigma\rho Q+\dfrac{\sigma PQ}{2\nu}&
			\sigma P\rho+\dfrac{\sigma P^2-Q^2}{4\nu} 
    \end{pmatrix}
\ee
\ese
and where, for any $T = (T_{ij})$ we have defined the row-wise divergence as $\div T = \big( d_1(T), d_2(T) \big)^\t$ with
$d_j(T) = \partial_x T_{j1} + \partial_y T_{j2}$ for $j=1,2$.
Then, using the identities~\eqref{e:vectordifferentialidentities} as well as~\eqref{e:divrelation}, we can finally write \eqref{e:Madelungsystem2a} in hydrodynamic form,
obtaining the full system of conservation laws as
\bse
\label{e:conservationlaws}
\begin{gather}
    (\rho\@u)_t + (\rho\@u\otimes\Sigma_3\@u) \backnabla
	- \sigma\nabla \left(\frac12\nu\rho^2 + P\rho\right) + 
      \div T 
		- \frac{\epsilon^2}{4}\left(\nabla\Box_\sigma\rho - \left(\frac1\rho \nabla\rho\otimes \nabla\rho\right)\Sigma_3\backnabla \right) = \@0,
        \\
    \rho_t + \nabla\cdot (\rho\Sigma_3 \@u) = 0\,, 
\end{gather}
\ese
again supplemented with~\eqref{e:Madelungsystem1c}.

\subsection{Two-phase multiple scales ansatz}

We now begin studying the slow modulations of the periodic solutions 
of the DS system.
Owing to the complex-valuedness of the field $\psi(x,y,t)$, and 
similarly to the corresponding calculations for the NLS equation \cite{ABH_JPA2023,ACR_2DNLS},
we introduce a two-phase ansatz for the solutions of the DS system.
Specifically, we define two fast variables $Z$ and~$S$ as well as slow spatial and temporal variables $X = x$, $Y = y$ and $T=t$, and look for solutions in the form
\bse
\label{e:ansatz}
\begin{gather}
\rho(x,y,t) = \rho(Z,X,Y,T)\,,\qquad
\Phi(x,y,t) = \phi(Z,X,Y,T) + S\,,\\
\noalign{\noindent together with}
P(x,y,t) = P(Z,X,Y,T)\,,\quad 
Q(x,y,t) = Q(Z,X,Y,T)\,,
\end{gather}
\ese
and where $\rho$, $\Phi$, $P$ and~$Q$ are all are assumed to be periodic in $Z$ with period~1.
We also define
\bse
\label{e:wavevectors}
\begin{gather}
\nabla Z = \frac{\@k(X,Y,T)}\epsilon\,,\qquad 
Z_t = - \frac{\omega(X,Y,T)}\epsilon\,,
\\
\nabla S = \frac{\@v(X,Y,T)}\epsilon\,,\qquad 
S_t = - \frac{\mu(X,Y,T)}\epsilon\,,
\end{gather}
\ese
where the local wavevectors $\@k = (k_1,k_2)^\t$ and $\@v = (v_1,v_2)^\t$ and the local
frequencies $\omega$ and $\mu$ 
are functions of $X,Y,T$, i.e., the slow variables only. 
The above definitions imply
\begin{equation}
\partialderiv{ }x \mapsto 
  \frac{k_1}\epsilon \partialderiv{ }Z + \frac{v_1}\epsilon \partialderiv{ }S
  + \partialderiv{ }X\,,\quad
\partialderiv{ }y \mapsto 
  \frac{k_2}\epsilon \partialderiv{ }Z + \frac{v_2}\epsilon \partialderiv{ }S
  + \partialderiv{ }Y\,,\quad
\partialderiv{ }t \mapsto - \frac\omega\epsilon \partialderiv{ }Z 
  - \frac\mu\epsilon \partialderiv{ }S + \partialderiv{ }T\,.
\label{der-xyt}
\end{equation}
Hereafter, we will use primes to denote differentiation with respect to~$Z$. 
With some abuse of notation, we will also use $x,y,t$ to denote the slow variables.

Notably, the equality of the mixed partial derivatives of $Z$ and $S$ immediately leads to the 
``conservation of waves'' evolution equations
\begin{equation}
\@k_T + \slowgradient\omega = 0\,,\qquad
\@v_T + \slowgradient\mu = 0\,,
\label{e:dkdt}
\end{equation}
where from now on $\slowgradient=(\partial_X,\partial_Y)^\t$ will denote differentiation with respect to the slow spatial variables,
together with the scalar constraints
\begin{equation}
\slowgradient\times\@k = 0\,,\qquad \slowgradient\times\@v = 0\,,
\label{e:constraints}
\end{equation}
Equations~\eqref{e:dkdt} will constitute the first Whitham modulation equations,
whereas \eqref{e:constraints} will be useful.

In the following sections we will substitute the above ansatz into the DS system and collect terms at increasing orders in $\epsilon$.
With the above definitions, \eqref{e:udef} now yields%
\be
\label{e:uvdef}
\@u(Z,X,Y,T) = \@k(X,Y,T)\,\phi'(Z) + \@v(X,Y,T)\,.
\ee
Note also that \eqref{e:uvdef} implies $\@k\times\@u' = 0$. 
Moreover, at leading order,
the irrotationality constraint~\eqref{e:uirrotational} yields
\be
k_2u_1' - k_1 u_2' = 0\,.
\label{e:kluprime}
\ee

\section{Periodic solutions of the Davey-Stewartson systems}
\label{s:periodic}

The periodic solutions of the DS systems were obtained in \cite{EAJAM2022}
by first obtaining stationary solutions and then using the Galilean symmetry of the DS system.
Here, we recover these solutions at the leading order in the above multiple scales expansion.
We also explicitly discuss the harmonic and soliton limits of all four variants of the DS system, as well as certain features of these solutions that to the best of our knowledge are not well known.
We point out that here we will use a parametrization for the periodic solutions different from 
the one in \cite{EAJAM2022}, as this will be more convenient for the formulation of modulation theory.
Hereafter, $\overline{f}$ will denote the average of a periodic function $f$ of $Z$ over the period $[0,1]$, namely, 
\vspace*{-1ex}
\begin{equation}
\overline{f} = \int_0^1\,f(Z)\,\d Z\,.
\end{equation}

\subsection{Periodic solution for the fluid velocity and auxiliary fields}
\label{s:periodicauxiliary}

Substituting~\eqref{e:ansatz}, \eqref{e:wavevectors} and~\eqref{der-xyt}
into the system~\eqref{e:Madelungsystem}, yields,
to leading order, 
\bse
\label{e:hydroleading}
\begin{gather}
\big( - \omega  + k_1 u_1 - \sigma k_2 u_2 \big)\, u_j' 
  - \sigma k_j (P+\nu\rho)'
  - \frac18 k_j( k_1^2 - \sigma k_2^2)
      \left( (\log\rho)'' + \frac{\rho''}{\rho} \right)' = 0\,,
  \qquad j = 1,2\,,
  \label{e:uprime}
\\
-\omega \rho' + k_1(\rho u_1)' - \sigma k_2(\rho u_2)' = 0,  \label{rhop} \\
k_1P' - \sigma k_2 Q' + 2k_1\nu\rho' = 0, 
\qquad
k_1Q' + k_2 P' = 0, 
\label{Ppp}
\end{gather}
\ese
where again primes denote derivatives with respect to the fast variable~$Z$.
Integrating the last two equations in~\eqref{e:hydroleading} in $Z$ yields, 
as long as $k_1^2 + \sigma k_2^2\ne0$, 
\vspace*{-0.6ex}
\begin{equation}
P(Z) = -\frac{2\nu k_1^2}{k_1^2+\sigma k_2^2}\rho + P_o, \quad 
Q(Z) = \frac{2\nu k_1k_2}{k_1^2+\sigma k_2^2}\rho + Q_o,
\label{e:PQperiodic}
\end{equation}
where  $P_o$ and $Q_o$ are arbitrary integration constants. 
Note however that even though $P_o$ and $Q_o$ are independent of the fast variable~$Z$, they are still functions of slow variables $x,y,t$.
(The degenerate case $k_1^2 + \sigma k_2^2=0$ will be discussed separately in section~\ref{s:phase}.) 
Namely, $P_o = P_o(x,y,t)$ and $Q_o=Q_o(x,y,t)$.
Next, integrating equation~\eqref{rhop} yields
\vspace*{-1ex}
\begin{equation}
k_1u_1 - \sigma k_2u_2 = \frac{J(k_1^2-\sigma k_2^2)}{\rho} + \omega\,,
\label{ku-lv}
\end{equation}
where $J(k_1^2-\sigma k_2^2)$ is an arbitrary constant of integration.
Then, recalling \cref{e:uvdef}, \eqref{ku-lv} uniquely determines the derivative of the phase $\phi(Z)$:
\vspace*{-1ex}
\be
\phi'(Z) = \frac{J}{\rho(Z)} - \phi_{o,1}\,,
\label{e:phiprime}
\ee
where
$\phi_{o,1}= (k_1v_1 - \sigma k_2v_2 - \omega)/(k_1^2-\sigma k_2^2)$
(and the reason for this particular name choice will become apparent in section~\ref{s:phase}). 
Since we assumed that $\phi(Z)$ be periodic with period~1, however, the average of~\eqref{e:phiprime} over one period must vanish.
Enforcing this condition, we therefore obtain the constraint
\begin{equation}
    \omega = k_1v_1 - \sigma k_2v_2 - (k_1^2-\sigma k_2^2)J\,\overline{\rho^{-1}}\,,
\label{e:omega}
\end{equation}
which implies  $\phi_{o,1}= J\,\overline{\rho^{-1}}$.

Equation~\cref{e:omega} is a nonlinear dispersion relation that determines the velocity of the traveling wave solution
(as determined by $\omega$) in terms of the other solution parameters.
Using \eqref{e:omega}, we can also equivalently rewrite \cref{e:phiprime} as
\begin{equation}
\phi'(Z) = J/\rho(Z)-J\,\overline{\rho^{-1}}\,.
\label{e:phiprimenew}
\end{equation}
We emphasize that the quantity $\overline{\rho^{-1}}$ (which should not be confused with $1/\rhobar$) diverges in some limiting cases.
However, throughout this work, this quantity will never appear by itself, 
but will always be multiplied by $J$ (or another equivalent term),
and the product $J\,\overline{\rho^{-1}}$ will always be finite.

\subsection{Periodic solution for the fluid density}
\label{s:leadingorderdensity}

\paragraph{General setting.}
Next, to determine the fluid density~$\rho$, we substitute~\cref{ku-lv} into~\cref{e:uprime}, obtaining
\vspace*{-1ex}
\begin{gather}
\frac{Ju_j'}{\rho} + \frac{\sigma\nu k_j}{k_1^2+\sigma k_2^2}\rho' 
	- \frac1{8} k_j \left( (\log\rho)'' + \frac{\rho''}{\rho}\right)' = 0\,,
\qquad j = 1,2\,.
\label{e:uvprime}
\end{gather}
The two ODEs for $j=1$ and $j=2$ are equivalent [as can be verified directly by multiplying the
first equation by $k_2$, the second equation by $k_1$, subtracting and using 
the constraint~\cref{e:kluprime}].
Conversely, subtracting $\sigma k_2$ times the second equation from $k_1$ times the first and using~\cref{ku-lv} yields, after simple algebraic manipulations,
\vspace*{-1ex}
\begin{equation}
-J^2\rho' + \frac{1}{\kappa}\rho'\rho^3 
	- \frac{1}{8}
	\left( (\log\rho)'' + \frac{\rho''}{\rho}\right)'\rho^3 = 0\,,
\label{J2rho}
\end{equation}
where we introduced the quantity 
\begin{equation}
\kappa= \sigma\nu (k_1^2+\sigma k_2^2)\,,
\label{e:kappadef}
\end{equation}
which will be used extensively throughout this work, and whose sign will discriminate among different kinds of solutions for the DSI equation. 
Next, using the identity~\eqref{a:lnrho''}, 
we can integrate~\cref{J2rho} to obtain (see Appendix~\ref{a:periodicdensity} for details) 
\begin{equation}
(\rho')^2 = 4 \left( \frac1{\kappa} \rho^3 - c_2 \rho^2 + c_1 \rho - J^2 \right)\,,
\label{e:rho'sqr}
\end{equation}
where $c_1$ and $c_2$ are arbitrary integration constants. 
Or, equivalently, 
\begin{equation}
(\rho')^2 = P_3(\rho)\,,\qquad 
P_3(\rho) = \frac{4}{\kappa}(\rho-\lambda_1)(\rho-\lambda_2)(\rho-\lambda_3),
\label{rhopSq}
\end{equation}
where $\lambda_1,\lambda_2,\lambda_3$ are the roots of the cubic polynomial $P_3(\rho)$.
The solution of~\cref{rhopSq} can be written in terms of Jacobi elliptic functions by letting
\begin{equation}
\rho(Z) = A+By^2(Z)\,.
\label{e:rho2ymap}
\end{equation}
Substituting~\eqref{e:rho2ymap} into~\cref{e:rho'sqr} yields (again, see Appendix~\ref{a:periodicdensity} for details) 
\begin{gather}
\kappa c_1 = s_2(\bflambda)\,,\qquad 
\kappa c_2 = s_1(\bflambda)\,,\qquad 
\kappa J^2 = s_3(\bflambda)\,.  
\label{ccJJ}
\end{gather}
where $s_1(\bflambda),s_2(\bflambda),s_3(\bflambda)$ are the symmetric polynomials of the roots $\lambda_1,\dots,\lambda_3$:
\begin{equation}
s_1(\bflambda) = \lambda_1 + \lambda_2 + \lambda_3\,,\qquad 
s_2(\bflambda) = \lambda_1\lambda_2 + \lambda_2\lambda_3 + \lambda_3\lambda_1\,,\qquad 
s_3(\bflambda) = \lambda_1\lambda_2\lambda_3\,.
\end{equation}
The last step is to recall the nonlinear first-order ODEs \eqref{e:cnODE} and~\eqref{e:snODE}
satisfied by the Jacobian elliptic functions 
$y(z) = \sn(cz,m)$ and $y(z) = \cn(cz,m)$, respectively. 
Matching \eqref{yprime} to \eqref{e:cnODE} [i.e., taking $y(z) = \cn(cz,m)$]
then yields
\bse
\label{e:rhocoeffs}
\begin{align}
&\kappa c_1 = 3A^2 + 8(1-2m)A\kappa\K^2 - 16{\kappa}^2(1-m)m \K^4\,, \qquad
\kappa c_2 = 3A + 4\kappa(1-2m) \K^2\,, \\
&\kappa J^2 = A(A-4\kappa m \K^2)(A+4\kappa(1-m) \K^2)\,, \qquad
B = -4\kappa m \K^2\,,
\label{e:Bdef}
\end{align}
\ese
where $\K = K(m)$ is the complete elliptic integral of the first kind \cite{NIST}, and where we chose $c=2K_m$ in order to normalize period of $y$ to 1. 
This normalization will be important later, when we write the Whitham modulation equations and average the conservation laws.
Note also that, since $\rho(Z)$ only depends on $y^2(Z)$, \eqref{e:cnsqr+snsqr=1} implies that matching \eqref{yprime} to~\eqref{e:snODE} instead of~\eqref{e:cnODE} would simply amount to a reparametrization of the solution.

Until this point the calculations are completely general, in the sense that they apply to all four variants of the DS system.
From this point on, however, the resulting expressions for the periodic solutions will differ among the various cases.
As it will become clear in the following paragraphs,
the key feature that discriminates among the various variants is the sign of the quantity $\kappa$ in~\eqref{e:kappadef},
as well as the requirement that $\rho(Z)$ be non-negative for all $Z\in\Real$.
Without loss of generality, in what follows we will always take the roots $\lambda_1,\lambda_2,\lambda_3$
to be well-ordered, so that
$\lambda_1\le \lambda_2\le \lambda_3$.

\paragraph{Defocusing DSII system ($\sigma = \nu=1$).}
In this case \eqref{e:kappadef} implies that $\kappa\ge0$.
From \eqref{e:rhocoeffs} and~\cref{ccJJ} it then immediately follows that
\begin{equation}
\lambda_1 = A - 4\kappa m \K^2, \quad
\lambda_2 = A, \quad 
\lambda_3 = A + 4\kappa(1-m) \K^2, 
\label{lambda-Am}
\end{equation}
and also
\begin{equation}
A = \lambda_2, \quad m=\frac{\lambda_2-\lambda_1}{\lambda_3-\lambda_1}, 
\quad \kappa=\frac{\lambda_3-\lambda_1}{4 \K^2}, \quad 
B = \lambda_1-\lambda_2\,.
\label{e:AaB}
\end{equation}
We then obtain the expression for the fluid density up to an arbitrary translation constant
(which is inessential because of the space translation invariance of the DS system) 
as 
\begin{equation}
\rho(Z) = \lambda_2 - (\lambda_2-\lambda_1)\cn^2\left(2K_m Z, m \right)
= \lambda_1 + (\lambda_2-\lambda_1)\sn^2
	\left(2K_m Z, m \right)\,,
\label{e:rhoDSIIdefocusing}
\end{equation}
with $m$ as in~\eqref{e:AaB}.
Note from~\eqref{ccJJ} that $\kappa J^2=\lambda_1\lambda_2\lambda_3\ge 0$ in this case.  
Also, \eqref{e:rhoDSIIdefocusing} implies that
$\rho(Z)$ oscillates between $\lambda_1$ and $\lambda_2$.
Since we need to have $\rho(Z)\ge0$, we therefore need $0\le \lambda_1\le \lambda_2\le \lambda_3$ in this case
to ensure self-consistency.
Moreover, since the period averages of $\cn^2(z,m)$ and $\sn^2(z,m)$ are respectively less than and greater than 1/2 for all $m>0$ [cf. \eqref{a:cnperiodaverages}],
\eqref{e:rhoDSIIdefocusing} implies that on average the solution represents a localized intensity dip from $\lambda_2$ to $\lambda_1$.
In particular, if $\lambda_1 = 0$, the minimum of the solution is exactly zero.

\paragraph{Focusing DSII system ($\sigma = 1$, $\nu=-1$).}
In this case \eqref{e:kappadef} implies that $\kappa\le 0$.  
In order to preserve the ordering between the roots,  
we therefore swap the definitions of $\lambda_1$ and $\lambda_3$ from~\eqref{lambda-Am},
obtaining:
\begin{equation}
\lambda_1 = A + 4\kappa(1-m) \K^2, \quad
\lambda_2 = A, \quad 
\lambda_3 = A-4\kappa m \K^2, 
\label{lambda-Am:variant}
\end{equation}
where $\lambda_1\le \lambda_2\le \lambda_3$, and also
\begin{equation}
A = \lambda_2, \quad m=\frac{\lambda_3-\lambda_2}{\lambda_3-\lambda_1}, 
\quad \kappa=\frac{\lambda_1-\lambda_3}{4 \K^2}, \quad 
B = \lambda_3-\lambda_2\,,
\label{e:AaB:variant}
\end{equation}
and we can thus write the solution as
\begin{equation}
\rho(Z) = \lambda_2 + (\lambda_3-\lambda_2)\cn^2\left(2K_m Z, m \right)
= \lambda_3 - (\lambda_3-\lambda_2)\sn^2\left(2K_m Z, m \right)\,,
\label{e:rhoDSIIfocusing}
\end{equation}
with $m$ as in \eqref{e:AaB:variant}.
Note that $\kappa J^2=\lambda_1\lambda_2\lambda_3\le 0$ in this case, and $\rho(Z)$ oscillates between $\lambda_2$ and $\lambda_3$ now.
Hence, to ensure self-consistency, in this case we need $\lambda_1\le 0 \le \lambda_2\le \lambda_3$.
Moreover, since the period averages of $\cn^2(z,m)$ and $\sn^2(z,m)$,
are respectively less than and greater than 1/2 for all $m>0$ [cf. \eqref{a:cnperiodaverages}],
\eqref{e:rhoDSIIfocusing} implies that on average the solution represents a localized intensity increase from $\lambda_2$ to $\lambda_3$.
In particular, if $\lambda_2=0$, the background value of the solution is exactly zero.

\paragraph{DSI system ($\sigma = -1$).} 
In this case the situation is more complicated, because \eqref{e:kappadef} implies that 
$\kappa$ could be positive or negative depending on the relative magnitude of $k_1$ and $k_2$.
What this means is that the periodic solutions of both variants of the DSI equation differ depending on whether their inclination angle is less than or greater than $\pm\pi/4$.
This should not be surprising, since, when $\sigma = -1$, \eqref{e:Ppde} is a forced wave equation
with unit propagation speed.

In particular, for the defocusing DSI equation ($\sigma = -1$ and $\nu = 1$) we have that, 
when $|k_1|>|k_2|$, the sign of $\kappa$ is the same as for the focusing DSII equation, i.e., $\kappa<0$. 
Therefore, equations \eqref{lambda-Am:variant}, \eqref{e:AaB:variant} and~\eqref{e:rhoDSIIfocusing} in the previous paragraph apply.  
On the other hand, when $|k_1|<|k_2|$, we have $\kappa>0$ as for the defocusing DSII equation, and therefore equations 
\eqref{lambda-Am}, \eqref{e:AaB} and~\eqref{e:rhoDSIIdefocusing} apply.
The situation is reversed for the focusing DSI equation ($\sigma = -1$ and $\nu = -1$),
by virtue of the symmetry~\eqref{e:DSIsymmetry}.
The singular case $|k_1| = |k_2|$, for which $\kappa=0$, is discussed separately in section~\ref{s:phase}.

\subsection{Frequency constant, phase dependence, harmonic limit and flat-phase solutions}
\label{s:phase}

\paragraph{Frequency constant.}
The periodic solutions derived in the previous section are not completely determined until the value of $\mu$ is obtained.
To find $\mu$, we need to plug in our solution into the undifferentiated equation~\cref{phiS}.
We insert the multiple scale ansatz~\eqref{e:ansatz} into~\eqref{phiS} and 
collect terms at the leading order in $\epsilon$
in the resulting expression.
Recalling~\eqref{e:ansatz} and~\eqref{e:wavevectors}, the time derivative yields
\begin{equation}
-\epsilon\Phi_t = \omega\phi' + \mu + O(\epsilon)\,.
\end{equation}
Then, using \eqref{e:phiprime}, \eqref{e:omega}, \eqref{e:kappadef}, \eqref{rhopSq} and~\eqref{ccJJ},
and after tedious but straightforward simplifications which are omitted for brevity, one can verify that the equation obtained from \eqref{phiS} is indeed identically satisfied for all $Z$ at leading order in $O(\epsilon)$ provided that 
\begin{equation}
\mu  = 
  \frac12( v_1^2 - \sigma v_2^2 ) - \frac12 ( k_1^2-\sigma k_2^2) (J\overline{\rho^{-1}})^2
  + \frac12( k_1^2 - \sigma k_2^2 ) \,\frac{s_1}{\kappa}
  - \sigma P_o\,,
\label{e:mu}
\end{equation}
which explicitly determines~$\mu$.
[Recall that $s_1 = \lambda_1 + \lambda_2 + \lambda_3$ and $\kappa$ is expressed in terms of $k_1$ and $k_2$ by \eqref{e:kappadef}.]

Summarizing,
up to arbitrary translation and phase rotation constants, 
the exact, periodic traveling wave solutions of the Davey-Stewartson system are determined by the seven independent constant parameters
$A$, $m$, $\@k$, $\@v$ and $P_o$.
(Note that the value of $Q_o$ does not affect the solution.) 
Or, equivalently, one can express the solutions in terms of the parameters
$\lambda_1,\lambda_2,\lambda_3$, $q = k_2/k_1$ 
(together with either of the possible sign choices for $k_1$ and $k_2$),
$\@v$ and $P_o$.
(For comparison purposes, recall that the periodic traveling wave solutions of the two-dimensional NLS equation are completely determined by six real parameters \cite{ABH_JPA2023,ACR_2DNLS}.)

\paragraph{Phase dependence.}
The last ingredients needed to make these solutions fully explicit are the complete determination of 
$\phi(Z)$ and the calculation of $J\overline{\rho^{-1}}$.  
We turn to these tasks next.
(We point out that the explicit form of $\phi(Z)$ will not be needed in the derivation of the Whitham modulation equations, similarly to \cite{ABH_JPA2023}.  On the other hand, $\phi(Z)$ is needed for direct numerical simulations of the DS system.)

Integrating \eqref{e:phiprimenew}, we can write the phase function $\phi(Z)$ up to an arbitrary additive constant (which is inessential because of the phase invariance of the DS system) as
\bse
\begin{gather}
\phi(Z) = \phi_o(Z) - \phi_{o,1}Z\,,
\\
\noalign{\noindent where}
\phi_o(Z) = J \,\int_0^Z \rho^{-1}(\zeta)\,\d\zeta\,,
\end{gather}
\ese
and where, as before, $\phi_{o,1}= J\,\overline{\rho^{-1}}$.
Therefore, the overall phase function $\Phi(Z)$ in~\eqref{e:ansatz} becomes
\bse
\be
\Phi(Z) = \phi_o(Z) + S_o\,,
\label{e:Phinew}
\ee
where $S_o = S - \phi_{o,1}Z$.
(Note however that $\phi_o(Z)$ is not a periodic function of~$Z$.
Indeed, we equivalently have $\phi_{o,1} = \phi_o(1)$, which explains the name choice for $\phi_{o,1}$ in section~\ref{s:periodicauxiliary}.)
In particular, for the exact periodic solutions, for which all solution parameters are constant,
\be
S_o = [(v_1 - \phi_{o,1}k_1)x + (v_2 - \phi_{o,1}k_2)y - (\mu - \phi_{o,1}\omega)t ] \epsilon\,.
\ee
\ese
We therefore see that the effect of the term $J\,\overline{\rho^{-1}}$ simply amounts to a shift for 
the quantities $v_1$ and $v_2$ (as well as the associated constants $\omega$ and $\mu$).

We next turn to the explicit calculation of $\phi_o(Z)$.
Recall that $\rho(Z)$ is given by \eqref{e:rho2ymap}, where $y(Z)= \cn(2K_mZ,m)$ and with suitable 
values for $A$ and $B$.
Using~\eqref{a:dnidentity}, we then have
\bse
\be
\phi_o(Z) = \frac{J}{2(A+B)\K}\Pi\bigg(\frac{B}{A+B};\am(2K_mZ,m),m\bigg)\,,
\label{e:phiO}
\ee
where 
$\am(z,m)$ is the Jacobi elliptic amplitude, 
$\Pi(z;\phi,m)$ is the incomplete elliptic integral of the third kind \cite{NIST}
and 
the specific values of $A$ and $B$ depend of course on which of the four variants of the DS system is being considered.
In turn, \eqref{e:phiO} implies
\be
J\,\overline{\rho^{-1}} = \phi_o(1) = \frac{J}{(A+B)\K} \Pi\bigg(\frac{B}{A+B},m\bigg)\,,
\label{e:phiavg}
\ee
\ese
where $\Pi(z,m) = \Pi(z;\pi/2,m)$ is the complete elliptic integral of the third kind,
and where we used~\eqref{a:Jacobiamplitude}.

For the defocusing DSII system ($\sigma=\nu=1$) and for the DSI focusing/defocusing systems ($\sigma=-1$ and $\nu=\pm 1$) when $\kappa>0$ we have~\eqref{e:AaB}.
Recalling that $J$ is determined via~\eqref{ccJJ},
\eqref{e:phiO} and~\eqref{e:phiavg} then become, respectively,
\bse
\label{e:DSIIdefocusing_explicitphase}
\begin{gather}
\phi_o(Z) =\sqrt{\frac{\lambda_2\lambda_3}{\lambda_1(\lambda_3-\lambda_1)}}\,
  \Pi\bigg(1- \frac{\lambda_2}{\lambda_1};\am(2K_mZ,m),m\bigg)\,,
\\
J\,\overline{\rho^{-1}} = 2\sqrt{\frac{\lambda_2\lambda_3}{\lambda_1(\lambda_3-\lambda_1)}}\,\Pi\bigg(1- \frac{\lambda_2}{\lambda_1},m\bigg)\,.
\end{gather}
\ese
Conversely, for the focusing DSII system ($\sigma=1$, $\nu=-1$) and the focusing/defocusing  DSI equations ($\sigma=-1$ and $\nu=\pm1$) 
when $\kappa<0$ we have
\bse
\label{e:rhoDSIIfocusing_explicitphases}
\begin{gather}
\label{e:rhoDSIIfocusing_explicitphase}
\phi_o(Z) =\sqrt{\frac{-\lambda_1\lambda_2}{\lambda_3(\lambda_3-\lambda_1)}}\,
  \Pi\bigg(1- \frac{\lambda_2}{\lambda_3};\am(2K_mZ,m),m\bigg)\,,
\\
J\,\overline{\rho^{-1}} = 2\sqrt{\frac{-\lambda_1\lambda_2}{\lambda_3(\lambda_3-\lambda_1)}}\,\Pi\bigg(1- \frac{\lambda_2}{\lambda_3},m\bigg)\,.
\end{gather}
\ese
Note that
in some of the explicit cases that will be discussed below, we will take $\lambda_1=0$, but 
all the above quantities remain finite in the limit $\lambda_1\to0$ by virtue of \eqref{a:ellipticpiidentity}.

\paragraph{Harmonic limit of the periodic solutions.} 
The periodic solutions derived in the previous section admit distinguished limits when $m\to0$ and $m\to1$. 
In this section and the following one we discuss these limiting cases, which are referred to respectively as the harmonic limit (since the Jacobi elliptic functions reduce to trigonometric functions in this case) and the soliton limit (since this limit will yield the soliton solutions of the DS system).

As we show next, when $m\to 0$, the above periodic solutions reduce to plane waves. 
Explicitly, \eqref{e:Bdef} yields $B = 0$, which in turn implies $\rho(Z) \equiv A$.
Then \eqref{e:phiprimenew} implies $\phi'(Z) \equiv 0$.
Therefore, the solution is completely independent of $Z$ and therefore of~$\@k$.
Moreover, without loss of generality, we can take $\phi(Z) \equiv 0$ using the phase invariance of the DS system.
Note that the harmonic limit corresponds to $\lambda_2 \to\lambda_1$ when $\kappa\ge0$ (i.e., for the defocusing DSII system) and to
$\lambda_2 \to\lambda_3$ when $\kappa\le0$ (i.e., for the focusing DSII system), respectively.
In either case, however, since $\rho(Z)$ becomes constant in the limit $m\to0$, \eqref{e:phiprimenew} implies that 
$\phi(Z)$ is also a constant, which without loss of generality we can take to be zero.
Employing a suitable Galilean transformation, one obtains
\be
\psi(\@x,t) = A\,\e^{i(\@v\cdot\@x - \mu t)/\epsilon}\,,\qquad
P(\@x,t) = P_o\,,
\label{e:uniformbackground}
\ee
with a redefinition of the constants $A$ and $P_o$, and 
with 
$\mu = \half(v_1^2-\sigma v_2^2) - \sigma\nu A^2 - \sigma P_o$.
(In particular, $A = \sqrt{\lambda_1}$ when $\kappa>0$ and $A = \sqrt{\lambda_3}$ when $\kappa<0$.)
Thus, the solution in this limit is completely determined by the four independent parameters $A$, $\@v$ and $P_o$,
while $k_1$ and $k_2$ disappear completely from the solution.
The situation should be compared to the case of the two-dimensional NLS equation, for which the plane wave solutions only have three independent parameters: $A$ and $\@v$ (e.g., see \cite{ABH_JPA2023}).

\paragraph{Flat-phase periodic solutions.}
Another simple class of solutions is obtained when $\lambda_1 = 0$, 
in which case one can take $\phi(Z) = 0$ without loss of generality
(thanks to the phase invariance of the DS system).
We call corresponding solutions ``flat-phase'' periodic solutions by analogy with the corresponding class of solutions of the NLS equation
\cite{DeconinckBottman,DeconinckSegal}.
Taking $v_1 = v_2 = 0$ without loss of generality (thanks to the Galilean invariance of the DS system), we have:

For the defocusing DSII equation ($\sigma=\nu=1$),
\bse
\begin{gather}
    \psi(\@x,t) = \sqrt{m}\sn( \phase ,m)\,\e^{-\frac12i(1+m) \cos (2\alpha )t}\,,\\
    P(\@x,t) = -2m\, \cos^2 \alpha\,\sn^2(\phase,m)\,,
\end{gather}
\ese
where we used the parametrization $\@k = ( \cos\alpha, \sin\alpha)^\t/(2K_m)$. 
In particular, when $\alpha=0$,
one obtains the simplest flat-phase solution of the defocusing DSII equation as
\be
\psi(\@x,t) = \sqrt{m}\sn(x,m)\,\e^{-\frac12i(1+m) t}\,,\qquad
P(\@x,t) = -2m\,\sn^2(x,m)\,.
\label{e:DSIIdef_sn}
\ee
[Note that, as $m\to0$, \eqref{e:DSIIdef_sn} reduces to a trivial solution, not to the uniform background~\eqref{e:uniformbackground}.
In order to obtain~\eqref{e:uniformbackground}, one must take the limit $m\to0$ with $\lambda_1\ne0$.]
For the focusing DSII equation ($\sigma=1$ and $\nu=-1$), 
\bse
\begin{gather}
    \psi(\@x,t) = \dn(\phase,m)\,\e^{i(1-\frac 12m)\cos (2\alpha ) t}\,,\\
    P(\@x,t) = 2\, \cos^2 \alpha \,\dn^2(\phase,m)\,,
\end{gather}
\ese
again with $\@k = ( \cos\alpha, \sin\alpha)^\t/(2K_m)$. 
In particular, when $\alpha=0$,
one obtains the simplest flat-phase solution of the focusing DSII equation as
\be
\psi(\@x,t) = \dn(x,m)\,\e^{i(1-\frac 12m) t}\,,\qquad
P(\@x,t) = 2\,\dn^2(x,m)\,.
\label{e:DSIIfoc_dn}
\ee
The above solutions are essentially reversed for the DSI equation.  
Namely, for the focusing DSI equation ($\sigma=\nu=-1$) with $|k_2|<|k_1|$, 
\bse
\begin{gather}
    \psi(\@x,t) = \sqrt{m \cos(2\alpha)}\sn(\phase,m)\,\e^{-\frac12i(1+m)  t}\,,\\
    P(\@x,t) = 2m\, \cos^2 \alpha \,\sn^2(\phase,m)\,.
\end{gather}
\ese
In particular, when $\alpha=0$,
one obtains the simplest flat-phase solution of the focusing DSI equation as
\be
\psi(\@x,t) = \sqrt{m}\sn(x,m)\,\e^{-\frac12i(1+m) t}\,,\qquad
P(\@x,t) = 2m\sn^2(x,m)\,.
\label{e:DSIfoc_sn}
\ee
Note that $\psi(\@x,t)$ is exactly the same as in the solution~\eqref{e:DSIIdef_sn} of the defocusing DSII equation, while $P(\@x,t)$ has the opposite sign.
Finally, for the defocusing DSI equation ($\sigma=-1$ and $\nu=1$)
with $|k_2|>|k_1|$, 
\bse
\begin{gather}
    \psi(\@x,t) = \sqrt{\cos(2\alpha)} \,\dn(\phase,m)\,\e^{i(1-\frac 12m) t}\,,\\
    P(\@x,t) = - 2\,\cos^2 \alpha\,\dn^2(\phase,m)\,.
\end{gather}
\ese
In particular, when $\alpha=0$,
one obtains the simplest flat-phase solution of the defocusing DSI equation as
\be
\psi(\@x,t) = \dn(x,m)\,\e^{i(1-\frac 12m) t}\,,\qquad
P(\@x,t) = - 2\,\dn^2(x,m)\,.
\label{e:DSIdef_dn}
\ee
Again, note that $\psi(\@x,t)$ is exactly the same as in the solution~\eqref{e:DSIIfoc_dn}
of the focusing DSII equation, while $P(\@x,t)$ has the opposite sign.
In section~\ref{s:stability} we will study the stability of all these solutions with respect to transverse perturbations.

\paragraph{Solutions of the DSI system at 45 degrees.}
For the defocusing DSI system ($\sigma=-1$), a degenerate situation is obtained when $|k_1|=|k_2|$, since in this case $\kappa= 0$ by~\eqref{e:kappadef},
which invalidates several steps in the calculations of sections~\ref{s:periodicauxiliary} and~\ref{s:leadingorderdensity}.
On the other hand, it is straigthforward to see that in this case the only self-consistent solution is the trivial one, 
obtained when $\lambda_1 = \lambda_2 = \lambda_3$, implying $\rho(Z) = \lambda_2 = \mathrm{const}$.
Therefore, the corresponding solutions in this scenario coincide with those in the harmonic limit discussed earlier.

\subsection{Soliton limit of the periodic solutions} 
\label{sec:soliton-limit}

As mentioned earlier, the soliton limit corresponts to $m\to1$.
While for the harmonic limit we were able to treat all four variants of the DS system simultaneously, the soliton limit is singular, however, because $\K\to\infty$ as $m\to1$ [cf.\ \eqref{a:Kinfty}], and thus requires more care.  
We therefore discuss the various cases separately.

\paragraph{Soliton limit, defocusing DSII equation ($\sigma = \nu=1$).}
The limit $m\to1$ corresponds to $\lambda_2\to\lambda_3$ in this case.
Since $\K\to\infty$, 
\eqref{e:AaB} implies that $\kappa\to0$ in this limit, and therefore $k_1$ and $k_2$ also tend to zero.
On the other hand, the quantity $\~\kappa= 4\kappa K_m^2 = \lambda_3 -\lambda_1$ remains finite in the limit.
It is therefore convenient to name
the rescaled fast variable appearing in \eqref{e:rhoDSIIdefocusing} as
$z = 2K_m Z = \sqrt{\lambda_3-\lambda_1}(\@k_o\cdot\@x - \omega_o t)/\epsilon$,
with ${\@k}_o = (k_{o,1},k_{o,2})^\t$,
where $k_{o,j} =k_j/\sqrt{\kappa}$ for $j=1,2$ and $\omega_o = \omega/\sqrt{\kappa}$, which also remain finite as $m\to1$.
In fact, it is trivial to see that $\|\@k_o\| = 1$, so that one can take $\@k_o = (\cos\alpha,\sin\alpha)^\t$ with $\alpha\in\Real$.
Thus, the divergence of $\K$ is simply reflected in the fact that the unit period $(-\half,\half)$ for the variable $Z$ is mapped to the range $(-\infty,\infty)$ for the variable $z$.
We can then write the fluid density $\rho$ in terms of an expression that remains finite as $m\to1$ as
\vspace*{-1ex}
\begin{gather}
\rho(z) =  \lambda_1 + (\lambda_3-\lambda_1)\tanh^2z\,.
\label{e:rhodefocusDSIIsoliton}
\end{gather}
As for the phase $\phi(Z)$, using the explicit expressions~\eqref{e:DSIIdefocusing_explicitphase},
and noting that the ratio $J/(2K_m) = \sqrt{\lambda_1\lambda_2\lambda_3/(\lambda_3-\lambda_1)}$ remains finite as $m\to1$, by virtue of~\eqref{ccJJ} and~\eqref{e:AaB},
one obtains
\vspace*{-1ex}
\be
\phi(z) = \arctan\left(\sqrt{\frac{\lambda_3}{\lambda_1}-1}\,\tanh z\right) \,.
\label{e:phisolitonlimit}
\ee
\noindent 
It is then straightforward to see that \eqref{e:phisolitonlimit} implies
\vspace*{-0.2ex}
\begin{equation}
e^{i\phi(z)} = \big(1/\sqrt{\rho(z)}\big)\,\big(\sqrt{\lambda_1} +i\sqrt{\lambda_3-\lambda_1}\tanh z\big)\,.
\end{equation}
We then obtain the soliton solutions of the defocusing DSII equation as
\bse
\begin{gather}
\psi(\@x,t) =  A\,\e^{i(\@v\cdot\@x - \mu_o t)/\epsilon}
    \big(\cos\theta +i\sin\theta\tanh z\big)\,,
\\
P(\@x,t) = - 2A^2\cos^2\alpha \,\big(\cos^2\theta + \sin^2\theta\tanh^2z \big) + P_o\,,
\end{gather}
\ese
with 
\be
z(\@x,t) = A\sin\theta\,(\@k_o\cdot\@x - \omega_o t)/\epsilon\,,
\ee
and $\@k_o = (\cos\alpha,\sin\alpha)^\t$ as before,
and where we defined $A = \sqrt{\lambda_3}$,
together with $\cos\theta = \sqrt{\lambda_1/\lambda_3}$ and $\sin\theta = \sqrt{1 - \lambda_1/\lambda_3}$, 
and with
\vspace*{-1ex}
\begin{gather}
\label{e:omegamudefousingDSII}
\omega_o = v_1\cos\alpha - v_2\sin\alpha - A\,\cos\theta\,\cos(2\alpha)\,,\qquad
\mu_o = \half(v_1^2 - v_2^2) + A^2\cos(2\alpha) - P_o\,,
\end{gather}
where we used~\eqref{e:omega},~\eqref{e:mu} and the fact that 
$\sqrt{\kappa}J\overline{\rho^{-1}}=\sqrt
{\lambda_1}=A\cos\theta$, see~\eqref{a:Pim1asymptotic}. 
In summary, the soliton solutions of defocusing DSII are determined by
six parameters: $A$, $\theta$, $\alpha$, $\@v$ and $P_o$.

These solutions are dark line solitons.  They are line solitons since for each $z_o\in\Real$, the equation $z= z_o$ defines a line in the $xy$-plane. 
They are also dark solitons, representing an amplitude dip from $A$ (as $z\to\pm\infty$) to $A\cos\theta$ (when $z=0$).  
In particular, when $\theta=\pi/2$ (i.e., $\lambda_1=0$), one obtains the so-called black solitons, whose intensity goes all the way to zero.  
As a special case, when $\alpha=0$ or $\alpha=\pi/2$ we obtain stationary solutions aligned respectively with the $y$-axis or the $x$-axis.
In particular, when $A=\epsilon=1$ and $\@v=\@0$, we have
\bse
\begin{gather}
\psi(\@x,t) = \e^{-it}\tanh x, \quad P(\@x,t) = -2\tanh^2 x\,,
\label{e:DSIIdefblacksolitonx}\\
\psi(\@x,t) = \e^{it}\tanh y, \quad P(\@x,t) = 0\,,
\label{e:DSIIdefblacksolitony}
\end{gather}
\ese
respectively for $\alpha=0$ and $\alpha=\pi/2$.
Note that, by adding a constant term to $P(\@x,t)$ in~\eqref{e:DSIIdefblacksolitonx},
one obtains the equivalent solution
$\psi(\@x,t) = \e^{it}\tanh x$ and $P(\@x,t) = 2\sech^2 x$.
Note also that, at $t=0$, the density $\rho(\@x,t)$ of the solution obtained by taking any value of $\alpha$ is identical to the one obtained by taking the value $\alpha+\pi$, 
but these two solutions have opposite phases and travel in opposite directions in time.  

\paragraph{Soliton limit, focusing DSII equation ($\sigma=1$ and $\nu=-1$).}
In this case, the limit $m\to 1$ corresponds to $\lambda_2\to \lambda_1^+$.
As before, the periodic solutions reduce to the soliton solutions in this limit. 
Since $\lambda_1\le 0\le \lambda_2$, however both $\lambda_1$ and $\lambda_2$ must vanish in this limit. 
As before, the limit is singular, because $\K\to\infty$ as $m\to1$.
Note from \eqref{e:AaB:variant} that $\kappa\to0$ in the same limit, while $\~\kappa= 4|\kappa|\K^2 = \lambda_3 - \lambda_1$ remains finite as $m\to1$.
As before, we therefore introduce the quantity
$z = 2K_m Z = 2\sqrt{|\kappa|} \K(\@k_o\cdot\@x - \omega_o t)/\epsilon$, 
with $\@k_o=\@k/\sqrt{|\kappa|}$ and $\omega_o =\omega/\sqrt{|\kappa|}$, 
so that the divergence of $\K$ is simply reflected in the fact that the range $(-\half,\half)$ for the variable $Z$ is mapped to the range $(-\infty,\infty)$ for the variable $Z$.
We can then write the fluid density $\rho$ in terms of an expression that remains finite as $m\to1$ as
\begin{gather}
\rho(z) =  \lambda_3\sech^2z\,.
\end{gather}
As for the solution phase, using the explicit expressions~\eqref{e:rhoDSIIfocusing_explicitphase},
and noting that the ratio $J/(2K_m) = \sqrt{\lambda_1\lambda_2\lambda_3/(\lambda_1-\lambda_3)}\to 0$  as $m\to1$, by virtue of~\eqref{ccJJ} and~\eqref{e:AaB:variant}, we have $\phi(Z)=0$.
We then obtain the soliton solutions of the focusing DSII equation as
\bse
\begin{gather}
\psi(\@x,t) =  A\,\e^{i(\@v\cdot\@x - \mu_o t)/\epsilon} \sech z\,,
\\
P(\@x,t) = 2A^2\cos^2\alpha \sech^2z + P_o\,,
\end{gather}
\ese
with 
\be
z(\@x,t) = A\,(\@k_o\cdot\@x - \omega_o t)/\epsilon\,,
\ee
and with $\@k_o = (\cos\alpha,\sin\alpha)^\t$ as before,
where we defined $A = \sqrt{\lambda_3}$, and, finally,
\vspace*{-1ex}
\begin{gather}
\omega_o = v_1\cos\alpha - v_2\sin\alpha\,,\qquad
\mu_o = \half(v_1^2 - v_2^2) -\half A^2\cos(2\alpha) - P_o\,.
\end{gather}
In summary, the soliton solutions of focusing DSII are determined by
five parameters: $A$, $\alpha$, $\@v$ and $P_o$.
(Note that there is one less parameter than in the defocusing DSII equation because in this case the soliton limit requires both $\lambda_1$ and $\lambda_2$ to be zero.)

The above solutions are bright line solitons, since $\rho(\@x,t)=\lambda_3$ along the line $z=0$ and 
$\rho(\@x,t)\to0$ as $\|\@x\|\to\infty$ in any direction not parallel to the line $z=0$.
As a special case, when $\alpha=0$ or $\pi/2$, we obtain solutions parallel to either the $x$-axis or the $y$-axis.
In particular, when $A=\epsilon=1$, $P_o=0$ and $\@v=\@0$, one obtains
\vspace*{-0.6ex}
\bse
\begin{gather}
\psi(\@x,t) = \e^{it/2} \sech x, \qquad 
P(\@x,t) = 2\sech^2 x\,,
\label{e:DSIIfocblacksolitonx}\\ 
\psi(\@x,t) = \e^{-it/2} \sech y\,,\qquad 
P(\@x,t) = 0\,,
\label{e:DSIIfocblacksolitony}
\end{gather}
\ese
respectively for $\alpha=0$ and $\alpha=\pi/2$.

\paragraph{Soliton limit, defocusing DSI equation ($\sigma=-1$ and $\nu=1$).}
Like with the periodic solutions, the soliton limit for the DSI systems differs depending on whether $|k_1|\gtrless|k_2|$. 
As before, 
$\~\kappa= 4\kappa K_m^2$ remains finite in the limit $m\to 1$. 
Recalling~\eqref{e:kappadef}, it is convenient to consider separately the two cases
$\kappa>0$ and $\kappa<0$, which arise depending on the relative magnitude of $k_1$ and~$k_2$.
Explicitly, 
since $\kappa= k_2^2 - k_1^2$ here, and 
defining $\tan\alpha = k_2/k_1$ as before, we have:
(i)~$\kappa<0$ when $|k_1|>|k_2|$,
implying $\alpha\in(-\pi/4,\pi/4)\cup(3\pi/4,5\pi/4)$, 
$|\tan\alpha|<1$ and $\cos(2\alpha)>0$,  
and 
(ii)~$\kappa>0$ when $|k_1|<|k_2|$,
implying $\alpha\in(\pi/4,3\pi/4)\cup(5\pi/4,7\pi/4)$,
$|\tan\alpha|>1$ and $\cos(2\alpha)<0$.
Note here we chose the range for $\alpha$ to be $(-\pi/4,7\pi/4)$ without loss of generality. 

When $\kappa<0$ (i.e., when $|k_1|>|k_2|$), similarly to the focusing DSII equation, we obtain
the bright soliton solutions of the defocusing DSI equation:
\vspace*{-0.4ex}
\bse
\begin{gather}
\psi(\@x,t) =  A\,\e^{i(\@v\cdot\@x - \mu_o t)/\epsilon} \sech z\,,
\\
P(\@x,t) = -2A^2\cos^2\alpha\sec 2\alpha\, \sech^2z + P_o\,,
\end{gather}
with 
\ese
\be
z(\@x,t) = A\,\sqrt{\sec(2\alpha)}\,(\@k_o\cdot\@x - \omega_o t)/\epsilon\,,
\ee
and $\@k_o = (\cos\alpha,\sin\alpha)^\t$ as before,
where we defined $A = \sqrt{\lambda_3}$, and, finally,
\vspace*{-0.4ex}
\begin{gather}
\omega_o = v_1\cos\alpha + v_2\sin\alpha\,,\qquad
\mu_o = \half(v_1^2 + v_2^2) -\half A^2\sec2\alpha - P_o\,.
\end{gather}
In summary, the soliton solutions of the defocusing DSI equation for $|k_1|>|k_2|$ are bright solitons, determined by five parameters: $A$, $\alpha$, $\@v$ and $P_o$.
As a special case, when $A = \epsilon = 1$, $\alpha=P_o = 0$ and $\@v=\@0$  
we obtain solitons parallel to the $y$ axis:
\vspace*{-0.4ex}
\be
\psi(\@x,t) = \e^{it/2}\sech x, \quad P(\@x,t) = -2\sech^2 x\,.
\label{e:DSIdefblacksolitonx}
\ee

Conversely, when $\kappa>0$ (i.e., when $|k_1|<|k_2|$), similarly to the defocusing DSII equation, we obtain the following dark soliton solutions of the defocusing DSI equation:
\bse
\begin{gather}
\psi(\@x,t) =  A\,\e^{i(\@v\cdot\@x - \mu_o t)/\epsilon}
    \big(\cos\theta +i\sin\theta\tanh z\big)\,,
\\
P(\@x,t) = - 2 A^2\cos^2\alpha\sec2\alpha \,\big(\cos^2\theta + \sin^2\theta\tanh^2z \big) + P_o\,,
\end{gather}
\ese
with 
\be
z(\@x,t) = A\sin\theta\,(\@k_o\cdot\@x - \omega_o t)/\big(\epsilon\sqrt{|\cos2\alpha|}\big)\,,
\ee
and with $\@k_o = (\cos\alpha,\sin\alpha)^\t$ as before,
where we defined $A = \sqrt{\lambda_3}$,
together with $\cos\theta = \sqrt{\lambda_1/\lambda_3}$ and $\sin\theta = \sqrt{1 - \lambda_1/\lambda_3}$, 
and with
\vspace*{-0.6ex}
\begin{gather}
\omega_o = v_1\cos\alpha + v_2\sin\alpha - A\,\cos\theta/\sqrt{|\cos2\alpha|}\,,\qquad
\mu_o = \half(v_1^2 + v_2^2) - A^2\sec2\alpha - P_o\,,
\end{gather}
where we used~\eqref{e:omega},~\eqref{e:mu} and the fact that 
$\sqrt{\kappa}J\overline{\rho^{-1}}=\sqrt
{\lambda_1}=A\cos\theta$, see~\eqref{a:Pim1asymptotic}. 
(Recall that $\cos2\alpha<0$ in this case.)
In summary, the soliton solutions of the defocusing DSI equation for $|k_1|<|k_2|$ are dark solitons, determined by six parameters: $A$, $\theta$, $\alpha$, $\@v$ and $P_o$.
As a special case, when $A=\epsilon=1$, $\theta = \alpha = \pi/2$, $P_o = 0$ and $\@v=\@0$ we obtain
\be
\psi(\@x,t) = \e^{- it}\tanh y\,,
\qquad
P(\@x,t) = 0\,.
\ee

\section{Whitham averaging}
\label{s:whitham}

\subsection{Averaged conservation laws; Whitham equations in physical variables}

\paragraph{General set-up.}
Recall that the periodic solutions derived in the previous section are given by \eqref{Madelung}, with $\phi(Z)$ and $\Phi(Z)$ given by
\eqref{e:ansatz}, $\phi(Z)$ given by~\eqref{e:phiO} and $\rho(Z)$ given alternatively by~\eqref{e:rhoDSIIdefocusing} or~\eqref{e:rhoDSIIfocusing},
and with $Z(X,Y,T)$ and $S(X,Y,T)$ given by \eqref{e:wavevectors}, 
with $\omega$ and $\mu$ given respectively by \eqref{e:omega} and~\eqref{e:mu} for each variant.
The solution is thus completely determined 
(up to inessential translation constants $S_o$ and $Z_o$, which are not captured by Whitham modulation theory anyway) in terms of $\lambda_1,\lambda_2,\lambda_3$, $q = k_2/k_1$, $\@v$ and $P_o$.
(Recall that $\kappa$ is given by~\eqref{e:kappadef} as well as \eqref{e:AaB} or~\eqref{e:AaB:variant}.)
When all these parameters are constant,
one recovers the exact periodic solutions of the DS system.
We now consider modulations of the periodic solutions obtained when the solution parameters are slowly varying with $X,Y,T$ and derive the corresponding Whitham modulation equations.

Note that, even though the auxiliary constant parameter $Q_o$ does not explicitly appear in the solution, it is necessary to include it when studying slow modulations of the solutions in order to obtain first-order modulation equations, similarly to what happens in the Kadomtsev-Petviashvili equation \cite{RSPA2017}.
Thus, overall, we have eight independent parameters
($\lambda_1,\lambda_2,\lambda_3$, $q = k_2/k_1$, $\@v$, $P_o$ and~$Q_o$).
Accordingly, we will need eight Whitham modulation equations.
(For comparison purposes, recall that the Whitham modulation system the two-dimensional NLS equation has six modulation equations \cite{ABH_JPA2023,ACR_2DNLS},
those for the Kadomtsev-Petviashvili and two-dimensional Benjamin-Ono equations have five equations \cite{RSPA2017,ABW_2DBO}
and the one for the Zakharov-Kuznetsov has four equations \cite{BC2024}.)

The first four modulation equations are simply the ``conservation of waves'' equations
\eqref{e:dkdt}.
Additionally, four further modulation equations are obtained by averaging the DS system in hydrodynamic form over the unit period, 
obtaining 
\vspace*{-0.2ex}
\bse
\label{e:averagedconservationlaws}
\begin{gather}
    \overline{\rho_t} + \overline{\nabla\cdot (\rho\Sigma_3 \@u)} = 0\,, 
\label{e:Wmass}
\\
    \overline{(\rho\@u)_t} + \overline{\nabla \cdot (\rho\@u\otimes\Sigma_3\@u)} 
       - \sigma\overline{\nabla \left(\half\nu\rho^2 + P\rho\right)}
       + \overline{\div T} 
      - \txtfrac14\epsilon^2 \overline{\nabla\,\Box_\sigma\rho}
      + \txtfrac14\epsilon^2 \overline{\left(\frac1\rho \nabla\rho\otimes \nabla\rho\right)\Sigma_3\backnabla } = 0\,,
\label{e:averagedmomentum}
\\
    \overline{P_x} - \sigma \overline{Q_y} + 2\nu\overline{\rho_x} = 0,\qquad
    \overline{Q_x} + \overline{P_y} = 0\,,
\label{e:averagedPQequations}
\end{gather}
\ese
where, as before, the overbar denotes average over the unit period.

The next task is the calculation of the period averages in \eqref{e:averagedconservationlaws},
which we turn to below.\break
We insert the multiple scales expansion in each of the equations in~\eqref{e:averagedconservationlaws} and recall~\cref{der-xyt}.
Note that if $f(Z)$ is any periodic function of $Z$ with unit period, 
$\overline{f'} = \int\nolimits_0^1 f'(Z)\,\d Z = 0$.
We will use this property repeatedly below.
We will also repeatedly use the fact that, from \eqref{e:udef}, we have
\be
\@u = \@k\,\phi'(Z) + \@v + O(\epsilon)\,.
\label{e:u=kphi'+v}
\ee
Note that all terms at $O(1/\epsilon)$ in~\eqref{e:Wmass2} cancel out, since the leading-order expression for $\rho(Z)$ and $\phi(Z)$ comprise an exact periodic solution of the DS system.

\paragraph{Averaged conservation laws.}
We start from the averaged mass equation~\eqref{e:Wmass}.  
Inserting the multiple scales ansatz yields
\vspace*{-0.5ex}
\be
\int_0^1 \bigg[ -\frac\omega\epsilon\rho' + \rho_T
  + \frac{k_1}{\epsilon}(\rho u_1)'  - \sigma \frac{k_2}{\epsilon}(\rho u_2)' 
  + (\rho u_1)_X - \sigma(\rho u_2)_Y \bigg]\,\d Z = 0\,.
\label{e:Wmass2}
\ee
Then, since derivatives with respect to $X,Y,T$ commute with the integration with respect to $Z$, 
using \eqref{e:u=kphi'+v} and~\eqref{e:phiprime}, at $O(1)$ we obtain
\be
\rhobar_T + \slowgradient\cdot\Sigma_3\big((J - \phi_{o,1}\,\rhobar)\,\@k + \rhobar\,\@v \big) = 0\,,
\label{e:drhobardT}
\ee
where $\phi_{o,1}= J\,\overline{\rho^{-1}}$ and 
$\slowgradient = (\partial_X,\partial_Y)^\t$ as before.

Next, in order to rewrite averaged momentum equation~\eqref{e:averagedmomentum}, it is convenient to consider it term by term.
Since all terms at $O(1/\epsilon)$ will eventually cancel out, here we only write the $O(1)$ terms.
The first term yields, similarly to \eqref{e:drhobardT},
\vspace*{-0.5ex}
\be
\overline{(\rho\@u)_t} = \int_0^1 \bigg( - \frac\omega\epsilon \partialderiv{ }Z + \partialderiv{ }T \bigg) \big((J - \phi_{o,1}\rho)\,\@k + \rho\,\@v \big) \d Z = 
  \big( \rhobar\,\@v + ( J - \phi_{o,1}\rhobar)\,\@k \big)_T \,.
\ee
For the second term, after inserting the multiple scales ansatz again, using~\eqref{e:u=kphi'+v} and~\eqref{e:phiprime} and expanding the products, at $O(1)$ we have
\be
\overline{(\rho\@u\otimes\Sigma_3\@u)\backnabla} = 
  \big[ - \phi_{o,1}(J - \phi_{o,1}\rhobar)\,\@k\otimes\Sigma_3\@k  
    + (J - \phi_{o,1}\rhobar)\,\big( \@k\otimes\Sigma_3\@v + \@v\otimes\Sigma_3\@k \big)
    + \rhobar\,\@v\otimes\Sigma_3\@v  \big] \backslowgradient\,,
\ee
where $\backslowgradient$ is defined similarly to $\backnabla$,
namely $(A)\backslowgradient = (\slowgradient^\t(A^\t))^\t$,
and where 
we used the fact that $\overline{J^2/\rho} = J \phi_{o,1}$.
Next, using~\eqref{e:PQperiodic}, we have
\be
 \sigma\overline{\nabla \left(\half\nu\rho^2 + P\rho\right)} =
  \half\,\slowgradient\bigg( \frac{\sigma k_2^2 - 3 k_1^2}{\kappa}\,\overline{\rho^2} \bigg) 
 + \sigma\slowgradient(P_o\rhobar)\,. 
\label{e:NLterm}
\ee
Similarly,
\be
\overline{\div T} = \overline{T} \backslowgradient = 
\begin{pmatrix}
\tfrac14\nu\big( \overline{Q^2}-\sigma\overline{P^2}\big) &
\half\nu\overline{PQ} \\
\sigma\overline{\rho Q} + \half\sigma\nu\overline{PQ} &
\sigma\overline{P\rho} + \tfrac14\nu\big(\sigma\overline{P^2}-\overline{Q^2}\big)
\end{pmatrix}\backslowgradient\,,
\ee
where 
\bse
\begin{gather}
\overline{P\rho} = -\frac{2k_1^2}{\kappa\sigma}\overline{\rho^2} + P_o\rhobar \,, \qquad
\overline{Q\rho} = \frac{2k_1k_2}{\kappa\sigma}\overline{\rho^2} + Q_o\rhobar\,,
\\
\overline{PQ} = -\frac{4k_1^3k_2}{{\kappa}^2}\overline{\rho^2} + \frac{2(k_1k_2P_o-k_1^2Q_o)}{\kappa\sigma}\rhobar + P_oQ_o\,,
\\
\overline{Q^2} - \sigma\overline{P^2} = \frac{4k_1^2(k_2^2-\sigma k_1^2)}{{\kappa}^2}\overline{\rho^2} + \frac{4(k_1k_2Q_o+\sigma k_1^2P_o)}{\kappa\sigma}\rhobar + Q_o^2 - \sigma P_o^2\,.
\end{gather}
\ese

Lastly, we look at the dispersive terms.  At $O(1)$ we have
$\epsilon^2 \overline{\nabla\,\Box_\sigma\rho} = 0$
since $\overline{\rho'} = \overline{\rho''} = 0$ thanks to the periodicity of~$\rho$, we have
\be
  \tfrac14\epsilon^2 \overline{\left(\frac1\rho \nabla\rho\otimes \nabla\rho\right) \Sigma_3\backnabla }
  = \tfrac14 \overline{(\rho')^2/\rho} \,(\@k\otimes\@k)\Sigma_3\backslowgradient\,.
\ee
Combining all terms, we then finally obtain the averaged momentum equation as follows:
\begin{multline}
\big( \rhobar\,\@v + ( J - \phi_{o,1}\rhobar)\,\@k \big)_T +   \big[ - \phi_{o,1}(J - \phi_{o,1}\rhobar)\,\@k\otimes\Sigma_3\@k  
    + (J - \phi_{o,1}\rhobar)\,\big( \@k\otimes\Sigma_3\@v + \@v\otimes\Sigma_3\@k \big)
    + \rhobar\,\@v\otimes\Sigma_3\@v  \big] \backslowgradient
\\
+\half\,\slowgradient\bigg( \frac{ 3 k_1^2 - \sigma k_2^2 }{\kappa}\,\overline{\rho^2} \bigg) - \sigma \slowgradient(P_o\rhobar) + \overline{T}\backslowgradient 
+ \tfrac14 \overline{{(\rho')^2}/\rho} \, (\@k\otimes\@k)\Sigma_3\backslowgradient = 0 \,.
\end{multline}
Finally, following similar steps as above, for the averaged mean field equations~\eqref{e:averagedPQequations} we obtain 
\begin{gather}
\frac{2\sigma}{\kappa} \begin{pmatrix} \sigma k_2^2 & - \sigma k_1k_2 \\ k_1k_2 &  - k_1^2 \end{pmatrix} \slowgradient \rhobar
  + \slowgradient P_o + \Sigma_1 \slowgradient Q_o = 0\,.
\end{gather}

\paragraph{Whitham modulation equations in physical variables.}
Now that all dependence on the fast spatial and temporal coordinates has been eliminated, 
hereafter we will, with some abuse of notation, use $x,y,t$ instead of $X,Y,T$ to denote the slow coordinates,
since there is no further possibility of confusion.
Summarizing the previous paragraphs, the Whitham modulation equations ``in physical variables'' are:
\bse
\label{Whitham-vector}
\begin{gather}
\@k_t + \nabla \omega = 0\,, 
\label{e:whithamphysical_dkdt}
\\
\@v_t + \nabla \mu = 0\,, 
\label{e:whithamphysical_dvdt}
\\
\rhobar_t + \nabla\cdot\Sigma_3\big((J - \phi_{o,1}\,\rhobar)\,\@k + \rhobar\,\@v \big) = 0\,,
\label{e:whithamphysical_drhodt}
\\
\kern-6em \big( \rhobar\,\@v + ( J - \phi_{o,1}\rhobar)\,\@k \big)_t +   \big[ - \phi_{o,1}(J - \phi_{o,1}\rhobar)\,\@k\otimes\Sigma_3\@k  
    + (J - \phi_{o,1}\rhobar)\,\big( \@k\otimes\Sigma_3\@v + \@v\otimes\Sigma_3\@k \big)
    + \rhobar\,\@v\otimes\Sigma_3\@v  \big] \backnabla \nonumber
\\
\kern6em
+ \half\,\nabla\bigg( \frac{ 3 k_1^2 - \sigma k_2^2 }{\kappa}\,\overline{\rho^2} \bigg) - \sigma \nabla (P_o\rhobar) + \overline{T} \backnabla
+\tfrac14 \overline{{(\rho')^2}/\rho} \, (\@k\otimes\@k)\Sigma_3\backnabla = 0\,,
\label{e:averagedmomentum2}
\\
\frac{2\sigma}{\kappa} \begin{pmatrix} \sigma k_2^2 & - \sigma k_1k_2 \\ k_1k_2 &  - k_1^2 \end{pmatrix} \nabla \rhobar
  + \nabla P_o + \Sigma_1 \nabla Q_o = 0\,,
\label{e:whithamphysical_dPQdt}
\end{gather}
\ese
where the dispersion relations take the following vector form:
\bse
\begin{align}
\omega +(\mathbf{k}\cdot(\Sigma_3\mathbf{k}))\overline{J\rho^{-1}} 
	-\mathbf{k}\cdot(\Sigma_3\mathbf{v}) &=0, \label{e:omega1}\\
\mu + \sigma P_o - \frac12 \left(
	(\mathbf{k}\cdot (\Sigma_3\mathbf{k}))
	(c_2 - (\overline{J\rho^{-1}})^2) 
	+ (\mathbf{v}\cdot(\Sigma_3\mathbf{v}))\right) &= 0\,.
\end{align}
\ese
with $c_2$ given by equation~\eqref{ccJJ}.
Note that, like in the KP equation~\cite{RSPA2017}, not all equations are evolutionary.
Moreover, the above equations are supplemented by the conditions~\eqref{e:constraints},
which, like with the KP equation, impose constraints on the admissible set of initial conditions in order for solutions of the system~\eqref{Whitham-vector} to actually represent modulations of one-phase solutions of the DS system.

\paragraph{Period averages.}
The period averages in~\cref{Whitham-vector} can be evaluated using the expression~\eqref{e:rho2ymap} for $\rho(Z)$ with $y(Z) = \cn(2K_m Z,m)$, obtaining%
\bse
\begin{align}
\overline{\rho} &= \int_0^1 \rho(z)dz = A + B C_2(m)\,,\\
\overline{\rho^2} &= \int_0^1 \rho^2(z)dz = A^2 + 2AB C_2(m) + B^2 C_4(m)\,,
\end{align}
\ese
together with~\eqref{e:phiavg},
where, as before, $\K=K(m)$ and $E_m=E(m)$ are the complete elliptic integrals of the 
first and second kind, respectively, whose  derivatives are given by \eqref{a:KEPiderivs},
and where $C_2(m)$ and $C_4(m)$ are given by \eqref{a:cnperiodaverages}.
Equivalently, $\overline{\rho^2}$ 
can be obtained from~\eqref{e:rho''} as
\be
\overline{\rho^2} = \frac{2}{3}{\kappa}c_2\overline{\rho} - \frac{1}{3}{\kappa}c_1
\ee
Moreover, using~\eqref{e:rho'sqr}, we can express the quantity $\overline{(\rho')^2/\rho}$ that appears in~\eqref{e:averagedmomentum2} as
\be
\overline{\frac{(\rho')^2}{\rho}} =  4 \left( \frac1{\kappa} \overline{\rho^2} - c_2 \overline{\rho} + c_1 - J \phi_{o,1}\right)\,.
\ee
We will use these expressions in order to write down the modulation equations explicitly.

\subsection{Dependent variables for transverse variations; convective derivatives}

Before we write down the modulation equations explicitly, 
it is convenient to perform various changes of dependent variables to rewrite them in a more suitable way, adopting similar transformations to those that were used in \cite{ABH_JPA2023,ACR_2DNLS}.

\paragraph{Spatial convective derivatives.}
We introduce the dependent variable $q = q(x,y,t)$ as
\be
q = k_2/k_1\,
\label{e:qdef}.
\ee
In light of~\eqref{e:qdef}, 
the compatibility condition~\eqref{e:kluprime} and the definition~\eqref{e:uvdef} then imply
\be
u_2 = q u_1 + p\,,\qquad 
v_2 = q v_1 + p\,,
\ee
where $p = p(x,y,t)$ is an arbitrary function of the slow spatial and temporal coordinates that will be one of the dependent variables of the 
Whitham modulation equations.
It is also convenient to introduce the vector $\@U = \@U(x,y,t) = (U_1,U_2)^\t$ as
\begin{equation}
\@U = \@v - J\overline{\rho^{-1}}\,\@k\,.
\label{e:Udef}
\end{equation}
Then,
noting that $U_2 = qU_1 + p $, 
using equation~\eqref{e:omega} we can express $\omega$ as
\begin{equation}
\omega = k_1 v_1 - \sigma k_2 v_2 - J(k_1^2-\sigma k_2^2)\overline{\rho^{-1}} = 
k_1 (g_-  U_1 - \sigma p q) \,,
\label{e:omeganew}
\end{equation}
where we have introduced the quantities
\be
g_\pm = 1 \pm \sigma q^2\,,
\ee
which we will use extensively below.
Similarly, recalling~\cref{e:mu}, we have:
\be
\mu =  \half \left(
k_1^2 g_- c_2 - g_- U_1^2 - \sigma p^2\right) + v_1 (g_- U_1 - \sigma p q) - \sigma P_o \,,
\label{e:munew}
\ee
where $c_2 = s_1/\kappa$ as per \eqref{ccJJ}.
Moreover, \eqref{e:dkdt} yields the following evolution equation for~$q$:
\begin{equation}
q_t + \frac{D_y \omega}{k_1} = 0\,,
\label{e:qt}
\end{equation}
were we introduced the convective derivatives
\begin{equation}
D_x = \partial_x - \sigma q \partial_y, \qquad 
D_y = \partial_y - q \partial_x\,,
\end{equation}
which we will use throughout the rest of this work.
Similarly, using~\eqref{e:constraints} one has
\begin{equation}
\label{e:qxqy}
q_x = \frac{D_y k_1}{k_1}, \qquad q_y = \frac{D_y k_2}{k_1} = D_y q + q q_x.
\end{equation}
Then, using equations~\eqref{e:omeganew} and~\eqref{e:qxqy}, equation~\eqref{e:qt} takes the form
\begin{equation}
q_t + D_y (g_-  U_1 - \sigma p q) + (g_-  U_1 - \sigma p q)q_x = 0\,.
\label{e:dqdt}
\end{equation}
Equation~\eqref{e:dqdt} is the first modulation equation in the system.

The second modulation equation is an evolution equation for~$p$.
In order to obtain it, we recall the second compatibility condition in~\eqref{e:constraints}, namely
$v_{1,y} = v_{2,x}$, which we can write as
\begin{equation}
\label{e:Dyv1}
p_x = D_y v_1 - q_x v_1\,.
\end{equation}
Then, recalling the system of equations~\eqref{e:dkdt} for $\@v$, we have 
$p_t + D_y \mu + v_1 q_t = 0$,
and, using equations~\eqref{e:dqdt} and~\eqref{e:Dyv1}, we get
\be
p_t + D_y \mu - D_y (v_1(g_- U_1 - \sigma p q)) + p_x (g_- U_1 - \sigma p q) = 0\,.
\label{e:dpdt0}
\ee
Then, using~\eqref{e:munew}, \eqref{e:dpdt0} becomes
\be
p_t + \half D_y\left( -g_- U_1^2 -\sigma p^2 - 2\sigma P_o + k_1^2 g_-  c_2 \right) 
+ p_x (g_- U_1 - \sigma p q) = 0\,.
\label{e:dpdt}
\ee
Equation~\eqref{e:dpdt} is the second modulation equation in the system.

Four additional modulation equations govern the parameters of the cnoidal wave solutions, and are as follows.
From equation~\eqref{e:whithamphysical_dkdt}, we have
\be
k_{1,t} + (k_1(g_-  U_1 - \sigma p q))_x = 0\,.
\label{e:dk1dt}
\ee
From equation~\eqref{e:whithamphysical_dvdt}, we have
\be
v_{1,t} - \half \left(
g_-  U_1^2 + \sigma p^2 + 2\sigma P_o-k_1^2 g_-  c_2  - 2v_1 (g_- U_1 - \sigma p q)\right)_x = 0\,.
\label{e:dv1dt}
\ee
Equivalently, recalling~\eqref{e:Udef}, we also have
\begin{equation}
U_{1,t} = v_{1,t} -  \overline{\rho^{-1}} (k_1 J)_t \,.
\label{e:dU1dttodk1Jdt}    
\end{equation}
Similarly, rewriting~\eqref{e:whithamphysical_drhodt} in terms of the variables~$p$, $q$ and $U_1$, we get
\be
\overline{\rho}_t + D_x (k_1 J+U_1\overline{\rho}) - \sigma (k_1 J+U_1 \overline{\rho})q_y - \sigma (p\overline{\rho})_y=0\,.
\label{e:drhobardt}
\ee
Moreover, from equation~\eqref{e:averagedmomentum2}, we get
\begin{multline}
\label{e:drhobarudt}
(k_1J + U_1\overline{\rho})_t  
+ D_x \left( v_1 k_1 J + U_1 (k_1 J+U_1\overline{\rho}) \right)  
- \sigma \big( v_1 k_1 J + U_1(k_1 J +U_1 \rhobar ) \big) q_y
\\
  - \sigma \big( p(k_1J+U_1\overline{\rho}) \big)_y  
  + \Bigg( \frac{k_1^2 (2+g_-)}{2\kappa}\overline{\rho^2} \Bigg)_x
  - \sigma (P_o \rhobar)_x + \Bigg( \dfrac{\overline{Q^2}-\sigma\overline{P^2}}{4\nu} \Bigg)_x
  + \Bigg( \dfrac{\overline{PQ}}{2\nu} \Bigg)_y
\\
  + \Bigg( k_1^2 \bigg(\frac{1}{\kappa}\overline{\rho^2} + c_1 - c_2\overline{\rho}\bigg)
  - k_1J(v_1 - U_1) \Bigg)_x 
  - \sigma\Bigg( k_1^2 q\bigg(\frac{1}{\kappa}\overline{\rho^2} 
		+ c_1 - c_2\overline{\rho}\bigg) - qk_1J(v_1-U_1) \Bigg)_y = 0 \,.
\end{multline}

Finally, the last two modulation equations are non-evolutionary equations for the additive ``constants'' in the auxiliary fields $P$ and $Q$, namely, equations~\eqref{e:whithamphysical_dPQdt}, which are simply
\bse
\label{e:dPodxdQodx}
\begin{gather}
\Bigg( -\frac{2k_1^2}{\kappa\sigma}\overline{\rho} + P_o + 2\nu \overline{\rho} \Bigg)_x 
  - \sigma \Bigg( \frac{2k_1^2 q}{\kappa\sigma}\overline{\rho} + Q_o \Bigg)_y = 0\,, \\
\Bigg( \frac{2k_1^2 q}{\kappa\sigma}\overline{\rho} + Q_o \Bigg)_x 
  + \Bigg( -\frac{2k_1^2}{\kappa\sigma}\overline{\rho} + P_o \Bigg)_y = 0\,.
\end{gather}
\ese

Summarizing, \eqref{e:dqdt}, \eqref{e:dpdt}, \eqref{e:dk1dt}, \eqref{e:dv1dt}, 
\eqref{e:drhobardt}, \eqref{e:drhobarudt} and the two equations~\eqref{e:dPodxdQodx} 
provide eight modulation equations for the eight dependent variables 
$q$, $p$, $k_1$, $U_1$, $\overline\rho$, $k_1J$, $P_o$ and $Q_o$.
Next we further simplify these equations.

\paragraph{Temporal convective derivative.}
It is convenient to rewrite the above equations by introducing a third and final convective derivative $D_t$ as
\be
D_t = \partial_t + U_1\partial_x - \sigma (qU_1 + p)\partial_y\,.
\label{e:Ddef}
\ee
In appendix~\ref{a:convectivederivatives} we show that, using~\eqref{e:Ddef}, we can rewrite the modulation equations as:
\bse
\label{e:physicalDSWS}
\begin{align}
    \frac{1}{k_1} D_t k_1 + W_1 &= 0, 
\label{e:DkDt}\\
    D_t U_1 + D_t S + \dfrac{U_1 - S}{k_1}D_t k_1 + W_2 &= 0,
\label{e:DU1}\\
    D_t \overline{\rho} -\frac{\rhobar}{k_1} D_t k_1 + W_3 &= 0, 
\label{e:Drhobar}\\
    D_t(k_1 J) + \overline{\rho}D_t U_1 - 2J D_tk_1 + W_4 &= 0, 
\label{e:Dk1JDt}\\
    D_t q + W_q &= 0, 
\label{e:DqDt}\\
    D_t p + W_p &= 0, 
\label{e:DpDt}\\
    (P_o)_x - \sigma (Q_o)_y +
    2 \big( k_1^2q^2\rhobar/\kappa\big)_x - 2  \big( k_1^2 q \rhobar/\kappa\big)_y
    &= 0, 
\label{e:P0X}\\
    \sigma (P_o)_y + \sigma (Q_o)_x 
        - 2D_y \big( k_1^2\rhobar/\kappa\big) + 2\big( k_1^2\rhobar/\kappa\big)\,q_x &= 0\,,
\label{e:Q0X}
\end{align}
\ese
where we introduced the quantity
\begin{equation}
S = k_1 J \overline{\rho^{-1}}
\label{e:Sdef}
\end{equation}
(not to be confused with the fast variable with the same name used in section~\ref{s:madelung}),
and where
\bse
\label{e:Ws}
\begin{align}
W_1 &= \frac{1}{k_1} \sigma (qU_1 +p) D_y k_1  + (g_- U_1-\sigma p q )_x \,,
\\
W_2 &= \sigma (q U_1 + p )\big( (U_1+S)q_x+\tfrac{U_1-S}{k_1}D_yk_1+ p_x \big) + 2U_1(g_- U_1-\sigma p q )_x
\nonumber\\ & \kern10em
    - \half \left(g_- U_1^2 + \sigma p^2 + 2\sigma P_o-k_1^2 g_- c_2\right)_x
\\
W_3 &=\rhobar D_xU_1 + D_x (k_1 J) - \sigma((k_1 J+U_1 \overline{\rho})q_y +\overline{\rho} p_y)  - \rhobar W_1 \,, 
\\
W_4 & = 2 k_1 J D_xU_1 
      - \sigma U_1 k_1 J  q_y -\sigma k_1 J p_y 
      - 2\sigma J (qU_1 +p) D_y k_1 - 2k_1 J(g_- U_1-\sigma p q )_x 
\nonumber\\ & \kern10em
  + \Bigg( \frac{k_1^2 (2+g_-)}{2\kappa}\overline{\rho^2} \Bigg)_x
  - \sigma (P_o \rhobar)_x + \Bigg( \dfrac{\overline{Q^2}-\sigma\overline{P^2}}{4\nu} \Bigg)_x
  + \Bigg( \dfrac{\overline{PQ}}{2\nu} \Bigg)_y
\nonumber\\ & \kern10em
    + \bigg(k_1^2 \bigg(\frac{1}{\kappa}\overline{\rho^2} + c_1 -        c_2\overline{\rho}\bigg) \bigg)_x 
    - \sigma\left(k_1^2 q\bigg(\frac{1}{\kappa}\overline{\rho^2} 
		+ c_1 - c_2\overline{\rho}\bigg)\right)_y \,,
\\
W_q &= \sigma (qU_1 + p)D_y q + D_y(g_- U_1-\sigma p q)\,, 
\\
W_p &= \sigma (qU_1 + p)D_y p - \frac12 D_y (g_- U_1^2 + \sigma p^2
    + 2\sigma P_o - k_1^2 g_-  c_2)\,.
\end{align}
\ese
Importantly, even though the RHS of~\eqref{e:S_Pi} contains the elliptic integral of the third kind,
we will see below that all of the quantities in~\eqref{e:Ws} are free of it.

\subsection{Transformation to Riemann-type variables}

So far we have carried out the calculations simultaneously for all four variants of the DS system.
Next, we want to simplify the system by performing a change of dependent variables to the Riemann invariants of the 1D NLS equation.
This can be done by recalling that all dependent variables depend on the quantities
$\lambda_1,\lambda_2,\lambda_3$ that parametrize the periodic traveling wave solutions.
In particular, the second and fourth terms in~\eqref{e:DU1} can be rewritten as
\begin{equation}
D_tS  -S\dfrac{D_tk_1}{k_1} = 
  \sum_{j=1}^3 \bigg( \partialderiv S{\lambda_j}- \frac S{k_1} \partialderiv{k_1}{\lambda_j} \bigg) D_t\lambda_j\,.
\end{equation}
As we have already seen in the previous section, however, certain terms differ depending on the specific variant considered, as some of the quantities contain the parameters $\sigma$ and $\nu$.
Moreover, the explicit expression of $S$ changes depending on the particular variant of the DS system considered.
Therefore, from now on we will treat the different variants of the DS system separately.

\paragraph{Whitham equations for the defocusing DSII system.}
In this case $\sigma=\nu=1$, implying $\kappa\ge0$ and $g_\pm = 1 \pm q^2$. 
We have seen that, similarly to the one-dimensional case, several quantities are expressed in terms of the roots
$\lambda_1,\lambda_2,\lambda_3$.
Note first that we can write $S$ as a function of $\lambda_1,\lambda_2,\lambda_3$.
Specifically, we have $k_1 = \sqrt{\lambda_3-\lambda_1}/(2\sqrt{g_+}\K)$, and, using~\eqref{e:rhoDSIIfocusing_explicitphase} and~\eqref{e:Sdef},
\be
S = \frac{\sqrt{\lambda_2\lambda_3}}{\sqrt{g_+\lambda_1} \K} \Pi \left(\left. 1-\frac{\lambda_2}{\lambda_1}\right|m\right)\,.
\label{e:S_Pi}
\ee
Then, recalling the definitions of $m$, $k$ and $A$ in terms of $\lambda_1,\dots,\lambda_3$,  we find
\bse
\begin{gather}
\partialderiv S{\lambda_1} - \frac{S}{k_1} \partialderiv{k_1}{\lambda_1} = 
\dfrac{\sqrt{\lambda_2 \lambda_3} (E_m-\K)}{2 \sqrt{g_+} \K \sqrt{\lambda_1} (\lambda_1 - \lambda_2) }\,,
\\
\partialderiv S{\lambda_2} - \frac{S}{k_1} \partialderiv{k_1}{\lambda_2} =
\dfrac{ \sqrt{\lambda_1} \lambda_3 \big( (\lambda_3 - \lambda_2)\K + (\lambda_1 - \lambda_3) E_m \big) }{2 \sqrt{g_+} \K \sqrt{\lambda_2 \lambda_3} (\lambda_2 - \lambda_1)(\lambda_2 - \lambda_3)  }\,,
\\
\partialderiv S{\lambda_3} - \frac{S}{k_1} \partialderiv{k_1}{\lambda_3} = 
\dfrac{\sqrt{\lambda_1} \lambda_2 E_m}{2 \sqrt{g_+} \K \sqrt{\lambda_2 \lambda_3}  (\lambda_2 - \lambda_3) }\,.
\end{gather}
\ese
Crucially, note that none of the terms in the right-hand side contain the third elliptic integral. Therefore, none of the terms in~\eqref{e:Ws} will depend on $\Pi(\cdot,m)$.
We will see later that similar considerations apply for the other variants of the DS system.

Similarly to the case of the one-dimensional NLS equation, 
it is convenient to introduce the further change of dependent variables
from $\lambda_1,\dots,\lambda_3$ and $U_1$ to $\@r = (r_1,\dots,r_4)^\t$ with
\vspace*{-1ex}
\bse
\label{e:lambda_riemanninvariants_DSIIdefocusing}
\begin{gather}
U_1 = \half\,(r_1 + r_2 + r_3 + r_4), \qquad 
\lambda_1 = \tfrac14 g_+\,(r_1-r_2-r_3+r_4)^2\,, 
\\
\lambda_2 = \tfrac14 g_+\,(r_1-r_2+r_3-r_4)^2, \qquad 
\lambda_3 = \tfrac14 g_+\,(r_1+r_2-r_3-r_4)^2\,,
\end{gather}
\ese
where $r_1\le r_2 \le r_3 \le r_4$ are the real-valued Riemann invariants for the Whitham equations for the one-dimensional defocusing NLS equation.
We then have
\begin{equation}
k_1^2 = \frac{(r_3-r_1)(r_4-r_2)}{4K^2_m}, \qquad 
m = \frac{(r_2-r_1)(r_4-r_3)}{(r_3-r_1)(r_4-r_2)}.  
\end{equation}
It is also convenient to define 
\be
s_n(\@r) = r_1^n + r_2^n + r_3^n + r_4^n\,,\qquad n = 1,2,3\,.
\label{e:s_n(r)}
\ee
[Note the quantities $s_n(\@r)$ in~\eqref{e:s_n(r)} should not be confused with the quantities $s_1(\bflambda),\dots,s_3(\bflambda)$ used in section~\ref{s:leadingorderdensity}.]
We then have the following useful identities:
\bse
\label{sn-identities}
\begin{align}
U_1 &= v_1 - kJ\overline{\rho^{-1}} = \half s_1(\@r)\,, \label{e:U1s1} \\
\overline{\rho} &= g_+\bigg( \tfrac14 s_1^2(\@r) - (r_1 + r_2)(r_3+r_4) - (r_3-r_1)(r_4-r_2)\frac{E_m}{\K} \bigg)\,, 
\label{e:rhobarri}
\\
kJ &= \nu g_+ \big( \tfrac13 s_3(\@r) - \tfrac14 s_1(\@r) s_2(\@r) + \tfrac{1}{24}s_1^3(\@r) \big)
\label{e:kJs1s2s3}
\end{align}
\ese
(also recall~\eqref{ccJJ}),
where the parameter~$\nu$ was inserted in~\eqref{e:kJs1s2s3} so that the expression will remain valid for the defocusing DSII systems. 

We can express the convective derivatives of~$k$ in terms of Riemann invariants as
\vspace*{-0.4ex}
\be
D_t k = \sum_{i=1}^4 \frac{\partial k}{\partial r_i}D_t r_i\,,
\ee
and similarly for the other derivatives.
We can then rewrite \eqref{e:DkDt} as
\bse
\be
\frac{1}{k_1} \sum_{i=1}^4 \frac{\partial k_1}{\partial r_i}D_t r_i + W_1 = 0\,.
\label{RHSw1}
\ee
Next, for~\eqref{e:DU1}, we obtain
\begin{equation}
\frac{2}{k_1}\sum_{i=1}^4 r_i\frac{\partial k_1}{\partial r_i}D_t r_i + W_2 = 0\,.
\label{RHSw2}
\end{equation}
Next, in Appendix~\ref{a:convectivederivatives} we show that \eqref{e:Drhobar} and~\eqref{e:Dk1JDt} yield, respectively,
\begin{gather}
\frac{3}{k_1}\sum_{i=1}^4 r_i^2\frac{\partial k_1}{\partial r_i}D_t r_i + W_3' = 0,
\label{RHSw3}
\\
\frac{4}{k_1}\sum_{i=1}^4 r_i^3 \frac{\partial k_1}{\partial r_i} D_t r_i  + W_4' = 0,
\label{RHSw4}
\end{gather}
\ese
where
\bse
\begin{align}
W_3' &= \sigma \nu \frac{3}{2g_+}W_3 + \frac{3}{4}s_1 W_2 + \frac{3}{4}\left(s_2-\frac12 s_1^2\right)W_1 - \frac{3q\overline{\rho}}{g_+^2}W_q\,.
\label{e:W_3'}
\\
W_4' &= \sigma \nu\frac{1}{g_+}W_4 + \sigma\nu s_1 W_3' + \frac12 (s_2-s_1^2) W_2 + \left(s_3 - s_1s_2 + \frac{1}{4}s_1^3\right)W_1 
- 2\sigma q (k_1J/g_+^2)W_q 
\nonumber \\
    & = \sigma \nu\frac{1}{g_+}W_4 + \sigma \nu\frac{3s_1}{2g_+} W_3 + \frac14 (s_1^2+2s_2)W_2 + \left( s_3-\frac14 s_1 s_2 - \frac{1}{8}s_1^3\right)W_1 
- \frac{ q}{g_+^2} (3s_1 \overline{\rho} + 2\sigma k_1J)W_q\,.
\label{e:W_4'}
\end{align}
\ese
Importantly, note that all instances of the third elliptic integral $\Pi(\,\cdot\,,m)$ cancel out of $W_1$, $W_2$, $W_3'$ and~$W_4'$ when $q=p=0$, as it should be since in this case the DS system reduces to the focusing/defocusing NLS equation.
(Again, the parameters $\nu$ and $\sigma$ were inserted in $W_3'$ and $W_4'$ so that the expressions will remain valid for the defocusing DSII system as well for the DSI system.) 

The combination of the four equations~\eqref{RHSw1}, \eqref{RHSw2}, \eqref{RHSw3} and~\eqref{RHSw4} can be conveniently written in matrix form as
\be
\_V(\mathbf{r}) \nabla_{\@r}\big( \log k_1 \big)
\cdot D_t\mathbf{r} + \mathbf{W} = \mathbf{0},
\label{Mr-matrix}
\ee
where $\@r = (r_1,r_2,r_3,r_4)^\t$,
$\nabla_{\@r} = (\partial_{r_1},\partial{r_2},\partial{r_3},\partial{r_4})^\t$,
$\@W=(W_1,W_2,W_3',W_4')^\t$ and $\_V(\mathbf{r})$ is the Vandermonde matrix
\be
\_V(\mathbf{r}) = \left(\begin{array}{cccc}
1 & 1 & 1 & 1 \\
r_1 & r_2 & r_3 & r_4 \\
r_1^2 & r_2^2 & r_3^2 & r_4^2 \\
r_1^3 & r_2^3 & r_3^3 & r_4^3 \\
\end{array}\right)\,.
\ee
Inverting~\eqref{Mr-matrix} [see also Appendix~\ref{a:identities}] we then finally obtain
\begin{align}
D_t \@r + \@h(\@r) = \@0\,,
\label{e:rWhitham}
\end{align}
where $\@h(\@r) = (h_1,\dots,h_4)^\t$, with 
\be
h_j = \frac{(-1)^{j+1} \Delta_{ilm}}{|\Delta|(\partial k/\partial r_j)/k} 
\left[r_i r_l r_m W_1 - (r_i r_l + r_l r_m + r_m r_i)W_2 + (r_i + r_l + r_m)W_3' - W_4'\right], 
\quad j=1,\ldots,4\,,
\ee
where $j\ne i$, $j\ne l$, $j\ne m$, $i<l<m$, the summation of repeated indices is assumed, and 
\be
|\Delta|=\prod\limits_{j>l}^4 (r_j - r_l), \quad 
\Delta_{ilm} = (r_i - r_l)(r_l - r_m)(r_m - r_i).
\ee

Recall that the four-component equation~\eqref{e:rWhitham} is augmented by \eqref{e:DqDt}, \eqref{e:DpDt}, \eqref{e:P0X} and \eqref{e:Q0X}.
In particular, \eqref{e:P0X} and \eqref{e:Q0X}, written in terms of $r_1,\dots,r_4$, become:
\bse
\label{e:qpQ0P0}
\begin{align}
 (P_o)_x - \sigma (Q_o)_y + 2 (\sigma\nu q^2 \rhobar/g_+)_x - 2(\sigma\nu q \rhobar/g_+)_y = 0\,, 
\\
(P_o)_y + (Q_o)_x - 2 D_y( \nu \rhobar/g_+) + 2 \nu \rhobar q_x /g_+ = 0\,.
\end{align}
\ese

Together, equations~\eqref{e:rWhitham}, \eqref{e:DqDt}, \eqref{e:DpDt} and~\eqref{e:qpQ0P0} comprise the Whitham modulation equations for the defocusing DSII system in Riemann-type variables.
We can write the equations compactly in matrix form as
\be
\_A(\@R)\,\@R_t + \_B(\@R)\,\@R_x + \_C(\@R)\,\@R_y = 0,
\label{e:DSWS}
\ee
where 
\be
\@R = (r_1,r_2,r_3,r_4,q,p,P_o,Q_o)^\t,
\ee
and where $\_A(\@R)$, $\_B(\@R)$ and $\_C(\@R)$ are suitable $8\times 8$ matrices.
In particular, $\_A(\@R)$ is the $8\times8$ extension of the corresponding matrix for the Whitham modulation equations for the defocusing NLS equation, namely:
\bse
\begin{gather}
\label{e:defocusingDSIIAmatrix}
\_A(\@r) = \begin{pmatrix} \_A_{4\times4} & \_O_{4\times4} \\
  \_O_{4\times4} & \_I_{2,4} \end{pmatrix},\qquad
\_A_{4\times4} = \diag(a_1,\dots,a_4)\,\_V(\@r)\,,\\
\noalign{\noindent where}
\label{e:Inmdef}
\_I_{n,n+m} = \begin{pmatrix} \_I_n & \_O_{n,m} \\ \_O_{m,n} & \_O_{m,m} \end{pmatrix}\,,
\end{gather}
$\_I_n$ is the $n\times n$ identity matrix, 
$\_O_{m,n}$ is the $m\times n$ zero matrix, 
and for $j=1,\dots,4$ the coefficients $a_j = \partial {k_1}/(k_1 \partial r_j)$ are as follows:
\begin{gather}
a_1=\frac{\left(r_1-r_4\right) \K+ \left(r_4-r_2\right) E_m }{2 \left(r_1-r_2\right) \left(r_1-r_4\right) \K}\,,\qquad
a_2=\frac{\left(r_3-r_2\right) \K+ \left(r_1-r_3\right) E_m }{2 \left(r_1-r_2\right) \left(r_2-r_3\right)
\K}\,,\\
a_3=\frac{\left(r_2-r_3\right) \K+ \left(r_4-r_2\right) E_m }{2 \left(r_2-r_3\right)\left(r_3-r_4\right) \K}\,,\qquad
a_4=\frac{\left(r_1-r_4\right) \K+ \left(r_3-r_1\right) E_m }{2 \left(r_1-r_4\right) \left(r_4-r_3\right) \K}\,.
\end{gather}
\ese
Notably, $\_A(\@R)$ only depends on $r_1,\dots,r_4$, and not on $q$, $p$, $Q_o$ and~$P_o$.
Similarly, while $\_B(\@R)$ and $\_C(\@R)$ do depend on $q$ and~$p$, they are independent of $Q_o$ and $P_o$.
On the other hand, the entries of $\_B(\@R)$ and $\_C(\@R)$ are much more complicated than those of $\_A(\@R)$. Therefore, for brevity, they are not included here~\cite{supplement}.

Alternatively, one can multiply \eqref{e:DSWS} by a ``pseudo-inverse'' $\_A'_\mathrm{inv}$ of $\_A$ obtained by taking the inverse of $\_A_{4\times4}$,
so that $\_A'_\mathrm{inv}\_A = \_I_{6,8}$,
obtaining the equivalent system
\be
\_I_{6,8}\,\@R_t + \_B'(\@r)\,\@R_x + \_C'(\@r)\,\@R_y = 0,
\label{e:DSWS2}
\ee
where 
$\_B' = \_A'_\mathrm{inv}\_B$
and 
$\_C' = \_A'_\mathrm{inv}\_C$.
This transformation will be useful when considering the harmonic and soliton limits. 

The system~\eqref{e:DSWS}, or equivalently~\eqref{e:DSWS2}, together with their analogues for the other variants of the DS system, is one of the main results of this work.
Hereafter, we refer to~\eqref{e:DSWS} and its analogues for the other variants as the \textit{DS-Whitham system} (DSWS).

\paragraph{Whitham equations for the focusing DSII system.}
In this case $\sigma=1$ and $\nu = -1$, implying $\kappa\le 0$, while $g_\pm = 1 \pm q^2$ as before.
Now, however, equation~\eqref{e:S_Pi} is replaced by
\begin{equation}
S = \frac{\sqrt{-\lambda_1\lambda_2}}{\sqrt{g_+\lambda_3} \K} \Pi \left(\left. 1-\frac{\lambda_2}{\lambda_3}\right|m\right)\,,
\label{e:S_Pi_focDSII}
\end{equation}
which implies
\bse
\begin{gather}
\partialderiv S{\lambda_1} - \frac{S}{k_1} \partialderiv{k_1}{\lambda_1} = 
\dfrac{\sqrt{\lambda_2 \lambda_3} E_m}{2 \sqrt{g_+} \K \sqrt{-\lambda_1 }  }\,,
\\
\partialderiv S{\lambda_2} - \frac{S}{k_1} \partialderiv{k_1}{\lambda_2} =
\dfrac{ \sqrt{\lambda_3} \lambda_1 \big( (\lambda_1 - \lambda_2)\K + (\lambda_3 - \lambda_1) E_m \big) }{2 \sqrt{g_+} \K \sqrt{-\lambda_1 \lambda_2} (\lambda_1 - \lambda_2)(\lambda_2 - \lambda_3)  }\,,
\\
\partialderiv S{\lambda_3} - \frac{S}{k_1} \partialderiv{k_1}{\lambda_3} = 
\dfrac{\sqrt{-\lambda_1 \lambda_2}  (E_m-\K)}{2 \sqrt{g_+} \K \sqrt{\lambda_3}  (\lambda_3 - \lambda_2) }\,.
\end{gather}
\ese
Since now $\lambda_1\le0\le\lambda_2\le\lambda_3$, their expression in terms of the Riemann invariants must differ.
Explicitly, instead of~\eqref{e:lambda_riemanninvariants_DSIIdefocusing},
similarly to the case of the focusing NLS equation, we take \cite{Kamchatnov}
\vspace*{-1ex}
\bse
\label{e:lambda_riemanninvariants_DSIIfocusing}
\begin{gather}
U_1 = \half\,(r_1 + r_2 + r_3 + r_4), \qquad 
\lambda_1 = -\tfrac14 g_+\,(r_1-r_2+r_3-r_4)^2\,,\\
\lambda_2 = -\tfrac14 g_+\,(r_1-r_2-r_3+r_4)^2\,,\qquad
\lambda_3 = -\tfrac14 g_+\,(r_1+r_2-r_3-r_4)^2\,.
\end{gather}
\ese
Recall that for the focusing NLS equation, the Riemann invariants form complex conjugate pairs.
Here we take $r_3 = r_1^*$ and $r_4 = r_2^*$, with $\Im r_1>0$ and $\Im r_2>0$, 
which is consistent with the constraint $\lambda_3 \ge \lambda_2\ge0$. 
Then, \eqref{e:U1s1} and~\eqref{e:kJs1s2s3} remain unchanged, but \eqref{e:rhobarri} is replaced by
\be
\rhobar = -g_+ \,\bigg( \tfrac14 (r_1 - r_2 + r_3 - r_4)^2 + (r_1-r_4)(r_2-r_3) \frac{E_m}{\K} \bigg)\,.
\ee
After carrying out the above substitutions, we can still perform a change of variables analogous to the one used in the defocusing DSII case. 
The end result is that equations~\eqref{e:rWhitham}, \eqref{e:DqDt}, \eqref{e:DpDt} and~\eqref{e:qpQ0P0} also comprise the Whitham modulation equations for the focusing DSII system in Riemann-type variables.
On the other hand, the dependence of $\_A$, $\_B$ and $\_C$ on $\@R$ are now different.
Specifically, for the $\_A$, we still have the structure given by equation~\eqref{e:defocusingDSIIAmatrix}, but the form of the parameters $a_j$ changes due to the modification of the relations between $\lambda_1,\dots,\lambda_3$ to $a_j$ for $j=1,\dots,4$. 
Explicitly, the expressions of $a_j$ for the focusing DSII system are as follows:
\bse
\begin{gather}
a_1=\frac{\left(r_1-r_2\right) \K+ \left(r_2-r_3\right) E_m }{2 \left(r_1-r_2\right) \left(r_1-r_3\right) \K}\,,\qquad
a_2=\frac{\left(r_1-r_2\right) \K+ \left(r_4-r_1\right) E_m }{2 \left(r_1-r_2\right) \left(r_2-r_4\right)
\K}\,,\\
a_3=\frac{\left(r_4-r_3\right) \K+ \left(r_1-r_4\right) E_m }{2 \left(r_1-r_3\right) \left(r_3-r_4\right) \K}\,,\qquad
a_4=\frac{\left(r_3-r_4\right) \K+ \left(r_2-r_3\right) E_m }{2 \left(r_2-r_4\right) \left(r_4-r_3\right) \K}\,.
\end{gather}
\ese

\paragraph{Whitham equations for the DSI system.}
Recall that the character of the periodic solutions of both DSI systems changes depending on whether $q^2$ is greater or less than one.
Here we discuss the case of the the defocusing DSI system (i.e., $\sigma=-1$ and $\nu=1$) with $q^2<1$ at first. 
In this case, $\kappa$ has the same sign as in the focusing DSII system, which implies that we need to use the same change of dependent variables from $\lambda_1,\dots,\lambda_3$ and $U_1$ to $\@r = (r_1,\dots,r_4)^\t$ in~\eqref{e:lambda_riemanninvariants_DSIIfocusing}. 
Consequently, the Whitham modulation equations take the same form as in the focusing DSII case, although the resulting system is not identical due to the different parameter values of $\sigma$ and $\nu$.
By virtue of the symmetry~\eqref{e:DSIsymmetry}, the case of the focusing DSI system ($\sigma=\nu=-1$) with $q^2>1$ 
is equivalent to that of the defocusing DSI system with $q^2<1$, which shares the same sign of $\kappa$ as in the defocusing DSII system as long as the value of $\sigma$ is taken into account.
Similarly, the Whitham modulation equations of the defocusing DSI case take the same form as in the defocusing DSII case.

\section{Reductions of the Whitham modulation equations}
\label{s:reductions}

The DSWS~\eqref{e:DSWS} admits several self-consistent reductions, as we discuss next.

\paragraph{Reduction to the the Whitham modulation equations for the NLS equation.}
One can verify that, when $q=p=Q_o=P_o=0$ and derivatives with respect to~$y$ vanish, the Whitham modulation equations for the defocusing DSII system reduce to the four-component genus-one Whitham modulation equations for the defocusing NLS equation.
Similarly, the Whitham modulation equations for the focusing DSI system reduce to those for the focusing NLS equation.

\paragraph{Harmonic limit of the Whitham equations.}
In Appendix~\ref{a:modulationalstability} we show that the defocusing DSII equation $(\sigma=\nu=1$)
is the only variant for which the uniform background solution is stable, whereas all other cases suffer from modulational instability.
For this reason, here we limit ourselves to only discussing the harmonic limit of the Whitham equations for the defocusing DSII equation.

In the harmonic limit $r_2 \to r_1^+$ (i.e., $\lambda_2 \to \lambda_1^+$), 
we have $\rhobar = \lambda_1$, so 
\eqref{e:whithamphysical_drhodt} becomes an evolution equation for~$\lambda_1$.
Moreover, combining \eqref{e:kappadef} and~\eqref{e:AaB}, 
we obtain an expression for $\lambda_3$ in terms of $k_1$ and $k_2$ and $\lambda_1$:
\be
\lambda_3 = \lambda_1 + \pi^2 (k_1^2 + k_2^2)\,.
\ee
Collectively,~\eqref{e:whithamphysical_dkdt},~\eqref{e:whithamphysical_dvdt},~\eqref{e:whithamphysical_drhodt} and~\eqref{e:whithamphysical_dPQdt} then provide seven modulation equations for the seven dependent variables $k_1$ and $k_2$, $v_1$ and $v_2$, $\rhobar$, $\overline{P}$ and $\overline{Q}$. 
Note here we eliminate the dependent variable $J$ by using the relation $J =\phi_{o,1}\rhobar$ in the harmonic limit. We can write the resulting equations as
\bse
\label{e:DSWS_HarmonicLimit}
\begin{gather}
\rhobar_t + \slowgradient\cdot \big(\rhobar\, \Sigma_3\,\@v\big) = 0 \,,
\label{e:DSWS_HarmonicLimit1}
\\
\@v_t + \slowgradient \big( \half \mathbf{v}\cdot(\Sigma_3\mathbf{v}) - \rhobar - \overline{P}   \big) = 0\,,
\label{e:DSWS_HarmonicLimit3}
\\
\overline{P}_x - \overline{Q}_y + 2\rhobar_x = 0,\qquad
    \overline{Q}_x + \overline{P}_y = 0\,,
\label{e:DSWS_HarmonicLimit4}
\\
\@k_t + \slowgradient \omega = 0\,,
\label{e:DSWS_HarmonicLimit2}
\end{gather}
\ese
where $\@D = (\partial_x,\partial_y)^\t$ as before
and $\omega = \@k\cdot(\Sigma_3\@v) - \sqrt{\lambda_3}\, \@k\cdot(\Sigma_3\@k) /\|\@k\|$.
Equations~\eqref{e:DSWS_HarmonicLimit1}, \eqref{e:DSWS_HarmonicLimit3} and~\eqref{e:DSWS_HarmonicLimit4} above are equivalent to the dispersionless Davey-Stewartson system~\cite{GeYi2020},
which governs the evolution of a slowly varying mean field,
whereas~\eqref{e:DSWS_HarmonicLimit2} governs the propagation of a linear wave packet on top of this mean field.

\paragraph{Soliton limit of the Whitham equations.} 
Similarly to the harmonic limit, we only compute the soliton limit of the modulation equations for the defocusing DSII equation, since in section~\ref{s:stability} we will show that the soliton solutions of all other variants of the DS system are unstable.

In the soliton limit of the defocusing DSII case, \eqref{e:whithamphysical_drhodt} and~\eqref{e:whithamphysical_dPQdt} remain the same as in the harmonic limit.
In this limit,~\eqref{e:PQperiodic} gives $P_o = P + 2 \cos^2 \alpha \rho$. 
Combining this result with~\eqref{e:omegamudefousingDSII} and~\eqref{e:whithamphysical_dvdt} we obtain the evolution equation for $\@v$.
Moreover, dividing Eq.~\eqref{e:whithamphysical_dkdt} by $\kappa$ and using~\eqref{e:omegamudefousingDSII}, we obtain the modulation equation for $\@k_o$. 
Since $\rhobar = A^2$, we then have seven modulation equations for the seven dependent variables, $A$, $\theta$, $\alpha$, $v_1$ and $v_2$, $\overline{P}$ and $\overline{Q}$.
\bse
\label{e:DSWS_SolitonLimit}
\begin{gather}
\rhobar_t + \slowgradient\cdot \big(\rhobar\,\Sigma_3\,\@v\big) = 0 \,,
\label{e:DSWS_SolitonLimit1}
\\
\@v_t + \slowgradient \big(\half \mathbf{v}\cdot (\Sigma_3\mathbf{v}) - 2 \cos^2 \alpha \rhobar  - \overline{P}  + A^2 \cos(2\alpha) \big) = 0\,,
\label{e:DSWS_SolitonLimit2}
\\
\overline{P}_x - \overline{Q}_y + 2\rhobar_x = 0,\qquad
    \overline{Q}_x + \overline{P}_y = 0\,,
\label{e:DSWS_SolitonLimit4}
\\
k_{o,t} + \slowgradient \omega_o  = 0\,.
\label{e:DSWS_SolitonLimit3}
\end{gather}
\ese
Similarly to the harmonic limit, \eqref{e:DSWS_SolitonLimit1}, \eqref{e:DSWS_SolitonLimit2} and~\eqref{e:DSWS_SolitonLimit4} are equivalent to dispersionless soliton system, whereas \eqref{e:DSWS_SolitonLimit3} governs the direction of the soliton field.
Note however that an evolution equation for $\theta$ is currently missing, since the time derivatives of $\theta$ cancel out from Eq.~\eqref{e:averagedmomentum2}.
The reason for this degeneracy is that, as $m\to1$, each period contains a single soliton whose width is $O(1)$ in the variable $z=2\K Z$, and therefore $O(1/\K)$ in the variable~$Z$, 
so the soliton only affects the period averages at $O(1/\K)$.
Indeed, \eqref{e:rhoDSIIdefocusing} and~\eqref{a:cnperiodaverages} yield exactly $\rhobar = \lambda_3 - (\lambda_3-\lambda_1)E_m/\K$, i.e.,
\vspace*{-1ex}
\be
\rhobar = A^2 - 2E_m\,A\sin\theta\,\|\@k\|\,,
\label{e:rhobarsolitoncorrection}
\ee
where, for all $0<m<1$, we have extended the definitions $A=\sqrt{\lambda_3}$ and $\cos\theta=\sqrt{\lambda_1/\lambda_3}$ introduced in section~\ref{sec:soliton-limit} [not to be confused with the parameter $A=\lambda_2$ in~\eqref{e:AaB}], so that~\eqref{e:rhobarsolitoncorrection} is exact. Moreover, since $E_m = 1 + O(\K\,\e^{-2\K})$ as $m\to1$, replacing $E_m$ by 1 in~\eqref{e:rhobarsolitoncorrection} only introduces exponentially small corrections.
In the limit $m\to1$, the quantity $\|\@k\| = \sqrt{\kappa} = A\sin\theta/(2\K)\to0$ plays the role of a soliton density
and $2A\sin\theta = \int\nolimits_\Real (A^2-\rho)\,\d\xi$, with $\xi = \@k_o\cdot\@x/\epsilon$, is the mass deficit of the dark soliton.
All the information about $\theta$ is therefore contained in the $O(\|\@k\|)$ corrections to the averaged conservation laws, which are lost when the limit is taken directly in~\eqref{Whitham-vector}.

A natural way to recover the missing equation is to compute the soliton limit using the Riemann-type variables introduced in section~\ref{s:whitham} [cf.~\eqref{e:rWhitham}], where the limit is not degenerate.
For the defocusing DSII system, $m\to1$ corresponds to $r_2,r_3\to r_\star$, and 
\eqref{e:lambda_riemanninvariants_DSIIdefocusing}, together with $q=\tan\alpha$ (so that $g_+ = \sec^2\alpha$), yields
(taking $|\alpha|<\pi/2$ for definiteness)
\vspace*{-0.6ex}
\be
r_1 = \half v_1 - A\cos\alpha\,,\qquad 
r_\star = \half v_1 - A\cos\theta\cos\alpha\,,\qquad 
r_4 = \half v_1 + A\cos\alpha\,,
\label{e:rsolitonlimit}
\ee
so that $r_1$ and $r_4$ describe the background (and reduce to the Riemann invariants of the dispersionless NLS equation when $\alpha=0$), while $r_\star\in[r_1,r_4]$ encodes the soliton parameter through $\cos\theta = (r_1+r_4-2r_\star)/(r_4-r_1)$.
Correspondingly, $U_1 = v_1 - A\cos\theta\cos\alpha$, and one can verify that the component normal to the soliton line of the velocity field $\big(U_1,-(qU_1+p)\big)$ that defines $D_t$ in~\eqref{e:Ddef} is exactly~$\omega_o$ in~\eqref{e:omegamudefousingDSII}.
In other words, in the soliton limit $D_t$ differentiates along the direction corresponding to the motion of the soliton.
We therefore conjecture that the missing modulation equation can be obtained from the second and third components of~\eqref{e:rWhitham}, which yields
\vspace*{-0.6ex}
\be
D_t r_\star + h_\star = 0\,,\qquad h_\star = \lim_{r_2,r_3\to r_\star} h_2 = \lim_{r_2,r_3\to r_\star} h_3\,,
\label{e:rstarequation}
\ee
which, combined with~\eqref{e:DSWS_SolitonLimit} and~\eqref{e:rsolitonlimit}, provides an evolution equation for~$\theta$.
In the one-dimensional reduction ($q=p=0$ and no $y$-dependence), this is precisely how the soliton limit of the Whitham equations for the defocusing NLS equation is obtained.
The characteristic velocities associated with $r_2$ and $r_3$ coalesce into the dark soliton velocity $U_1 = \half(r_1+2r_\star+r_4)$, so that $h_\star=0$ and~\eqref{e:rstarequation} reduces to $r_{\star,t} + U_1 r_{\star,x} = 0$ \cite{Kamchatnov,ElHoefer2016}.
An analogous procedure was used to derive the soliton--mean flow modulation equations for the KP equation from the soliton limit of the KP-Whitham system in Riemann-type variables~\cite{NLTY2021}.
In the present case, however, the limit in~\eqref{e:rstarequation} is delicate, since, 
as $r_2,r_3\to r_\star$, the prefactor $\Delta_{ilm}/(|\Delta|\,\partial_{r_j}\!\log k_1)$ in $h_2$ and $h_3$ grows like $\K$, 
so a finite limit requires the corresponding combination of $W_1,W_2,W_3'$ and $W_4'$ to vanish like $1/\K$,
and it is precisely these $O(1/\K)$ terms that carry the $\theta$-dependence identified in~\eqref{e:rhobarsolitoncorrection}.
Moreover, the terms involving $D_yk_1/k_1$ in~\eqref{e:Ws} diverge logarithmically as $m\to1$, so the necessary cancellations must be verified explicitly.
Carrying out this computation and obtaining the explicit form of $h_\star$ is therefore left as a problem for future work.

\section{Stability analysis of the periodic solutions of the DS system}
\label{s:stability}

In this section we employ the Whitham modulation equations that were derived in section~\ref{s:whitham} to study the stability properties of the periodic traveling wave solutions of all variants of the DS system which were derived in section~\ref{s:periodic}.
We also compare the predictions from Whitham theory to those coming from a direct linearization of the DS system and to direct numerical simulations.
We should point out that these approaches can be applied to the general class of periodic solutions derived in section~\ref{s:periodic}. 
For simplicity, however, here we will restrict ourselves to the flat-phase solutions, in order to keep the presentation concise.

\subsection{Stability analysis via linearization of the Whitham modulation equations}
\label{s:whithamstabilityanalysis}

One of the most direct applications of Whitham modulation theory is a study of the stability of the exact periodic traveling wave solutions of the nonlinear PDE in question.
(Indeed, this was Whitham's original motivation to derive the modulation equations.)
In this section we show how the Whitham modulation system~\eqref{e:DSWS} can be used to study the stability of the periodic solutions of the DS system~\eqref{e:DS}.
In sections~\ref{s:linearizationstabilityanalysis} and~\ref{s:DNS}, we will then compare the resulting predictions with those arising from two alternative approaches:
(i)~direct linearization of the Davey-Stewartson system, and 
(ii)~direct numerical simulations.

We should reiterate that~\eqref{e:Ppde} only determines $P(x,y,t)$
up to an arbitrary solution of the corresponding homogeneous equation
(i.e., Laplace's equation for the DSII equation and the wave equation for the DSI equation).
In all the above-mentioned approaches, we will take all such homogeneous solutions to be a constant.


\paragraph{General setting.} 
Recall that the DSWS can be written compactly in matrix form as \eqref{e:DSWS} for the vector-valued
dependent variable $\@R=(r_1,r_2,r_3,r_4,q,p,P_o,Q_o)^\t$. 
Recall also that when these eight quantities are independent of $x,y,t$,
the cnoidal wave solutions derived in section~\ref{s:periodic} 
are exact periodic traveling wave solutions of the DS system~\eqref{e:DS2}.
To study the stability of such solutions, 
we therefore look for solutions of the DSWS~\eqref{e:DSWS} in the form of a constant solution $\@R\O0$ plus a small perturbation 
$\@R\O1$, where
$\@R\O{j}=(q\O{j},p\O{j},r_1\O{j},r_2\O{j},r_3\O{j},r_4\O{j},P\O{j},Q\O{j})$ for $j=0,1$. 
Namely, we look for $\@R = \@R\O0 + \delta\,\@R\O1$, with 
with $0<\delta\ll1$.
Or, componentwise,
\bse
\begin{gather}
r_j(x,y,t) = r_j\O0 + \delta r_j\O1(x,y,t),\quad j=1,2,3,4\,,
\\
q(x,y,t) = q\O0 + \delta q\O1(x,y,t)\,, \quad 
p(x,y,t) = p\O0 + \delta p\O1(x,y,t)\,, 
\\
P_o(x,y,t) = P_o\O0 + \delta P_o\O1(x,y,t)\,, \quad 
Q_o(x,y,t) = Q_o\O0 + \delta Q_o\O1(x,y,t)\,.
\end{gather}
\ese
Substituting this ansatz into \eqref{e:DSWS} and neglecting terms $O(\delta^2)$ and higher,
we then obtain the linearized DSWS system
\be
\_A\O0 \@R\O1_t + \_B\O0 \@R\O1_x + \_C\O0 \@R\O1_y = \@0\,,
\label{e:LDSWS}
\ee
where $\_A\O0 = \_A(\@R\O0)$, $\_B\O0 =\_B(\@R\O0)$ and $\_C\O0= \_C(\@R\O0)$. 
Since \eqref{e:LDSWS} is a linear system of PDEs with constant coefficients,
it is sufficient to study plane wave solutions of it, which correspond to periodic perturbations of the underlying cnoidal wave solution.
We therefore look for solutions of~\eqref{e:LDSWS} in the form
\be
\@R\O1(x,y,t) = \@C\,\e^{i(\xi x + \eta y - \zeta t)},
\label{e:planewaveansatz}
\ee
where $\@C$ is a constant vector, and $\xi$, $\eta$ and $\zeta$ are respectively the perturbation's wavenumbers in $x$ and $y$ and the perturbation's angular frequency.
(Note that the perturbations in~\eqref{e:planewaveansatz} are not required to be coperiodic with the underlying cnoidal wave solutions of the DS system at this point.)
Substituting the expression~\eqref{e:planewaveansatz} into \eqref{e:LDSWS}
then yields the homogeneous linear system of equations 
$(\xi\,\_B\O0 + \eta\,\_C\O0 - \zeta\,\_A\O0 )\,\@C = \@0$, which is
equivalent to the generalized eigenvalue problem
\be
(\xi\,\_B\O0 + \eta\,\_C\O0)\,\@C = \zeta\,\_A\O0\@C\,.
\label{e:matrixeigenvalueproblem}
\ee
The eigenvalues (corresponding to nontrivial solutions~$\@C$) 
are the roots of the characteristic polynomial $p(\xi,\eta,\zeta) = \det( \xi\,B\O0 + \eta\,C\O0 - \zeta\,A\O0 )$. 
In turn, the condition $p(\xi,\eta,\zeta)=0$ determines the linearized dispersion relation
$\zeta = \zeta(\xi,\eta)$.
If $\zeta\in\Real$, the DSWS system~\eqref{e:DSWS} is hyperbolic, 
and the periodic solutions of the DS system associated with the constant solution $\@R=\@R\O0$ of the DSWS are linearly stable.
Conversely if $\zeta\notin\Real$, the DSWS is elliptic, and the corresponding periodic solutions are unstable.

\begin{figure}[t!]
	\centerline{%
		\includegraphics[width=0.4875\textwidth]{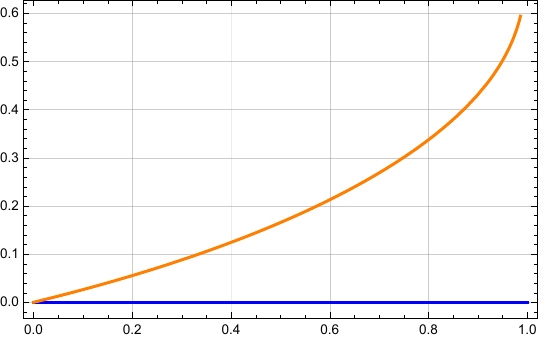}\quad
		\includegraphics[width=0.4875\textwidth]{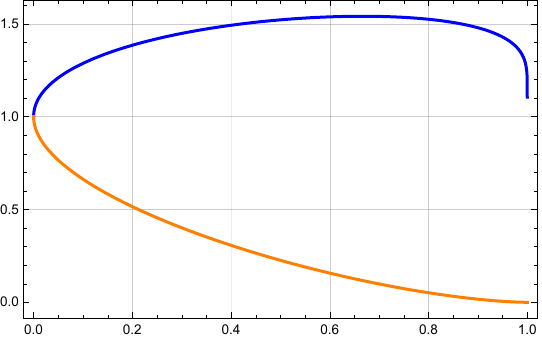}}
    \bigskip
    \centerline{%
		\includegraphics[width=0.4875\textwidth]{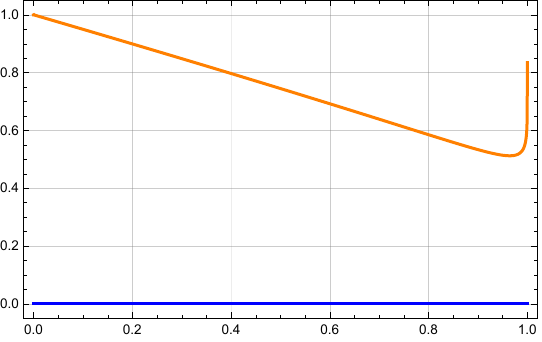}\quad
		\includegraphics[width=0.4875\textwidth]{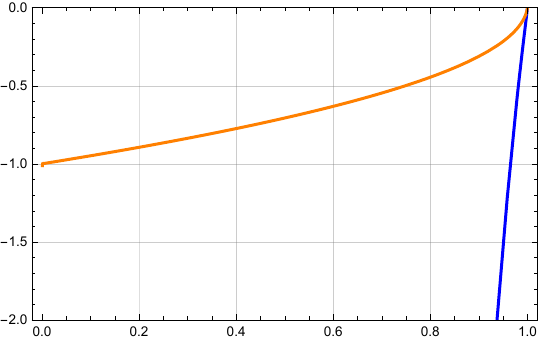}}
    \bigskip
    \centerline{%
        \includegraphics[width=0.4875\textwidth]{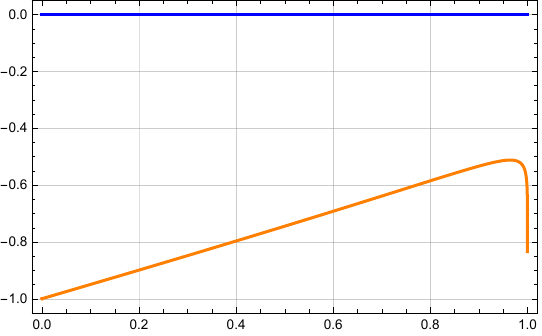}\quad
		\includegraphics[width=0.4875\textwidth]{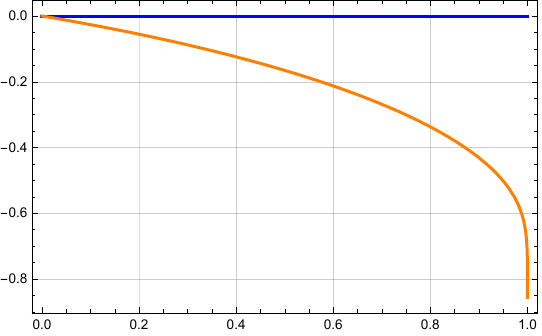}}
        \medskip
        \caption{
        Top left panel:         
        $f_0(m)=0$ (blue) and $f_1(m)$ (orange) as a function of $m$ 
        (horizontal axis) for the defocusing DSII system in the case $\xi=0$ (purely transversal perturbations).
        Top right panel:  
        $f_{1}(m)$ (blue) and $f_2(m)$ (orange) as functions of $m$ (horizontal axis) for the defocusing DSII system in the case $\eta=0$ (purely longitudinal perturbations).
        Middle left panel:
        Same as the top left panel, but for purely transversal perturbations of the focusing DSII system.
        Middle right panel: Same as the top right panel, but for purely longitudinal perturbations of the focusing DSII system.
        Bottom left panel: Same as the top left panel, but for the defocusing DSI system with $q^2<1$.
        Bottom right panel: Same as the top left panel, but for the focusing DSI system with $q^2<1$.
	}
	\label{f:f23}
	\kern-\medskipamount
\end{figure}

\paragraph{Defocusing DSII.}
First we look at the stability of the periodic solutions of the defocusing DSII system, i.e., we take $\sigma=\nu=1$.
Setting $p=q=P_o=Q_o=s_1=U_1=0$ for simplicity, and 
choosing $\lambda_1=0$, $\lambda_2=m$ and $\lambda_3=1$, 
from~\eqref{e:lambda_riemanninvariants_DSIIdefocusing} we get
\be
\label{e:rmstability}
r_1 = - \tfrac12 (1+\sqrt{m})\,,\quad
r_2 = - \tfrac12 (1-\sqrt{m}), \quad 
r_3 = \tfrac12 (1-\sqrt{m}), \quad 
r_4 = \tfrac12 (1+\sqrt{m})\,.
\ee
For generic values of $\xi$ and $\eta$, finding the linearized dispersion $\zeta(\xi,\eta)$ involves computing the roots of a complicated quartic polynomial.
On the other hand, a particularly simple scenario is obtained when  $\xi=0$ (and $\eta\ne 0$), which corresponds to purely transversal perturbations
(That is, the perturbations have no $x$ dependence in this case.)
In this case we obtain 
\be
p(\xi,\eta,\zeta) = \tfrac13 \eta^2 \zeta^2\,\big( \zeta^2 - f_1(m) \eta^2 \big)^2\,, 
\ee
which yields the linearized dispersion relation as 
\be
(\zeta/\eta)^2 = f(m),
\ee
with two possible branches of solutions:
\vspace*{-1ex}
\begin{gather}
f_0(m) = 0\,,\qquad
f_1(m) = \frac{(K_m-E_m)^2 (2E_m + (m-1)K_m)}{K_m ( m K_m^2 - (K_m - E_m)^2)}\,,
\end{gather}
with $f_1(m)$ corresponding to a double branch.
Note that $f_1(0) = 0$, and $f_1(m)>0$ for all $m\in(0,1]$. 
We can therefore conclude that the periodic solutions of defocusing DSII are linearly stable
with respect to purely transversal perturbations. 
These predictions are in agreement with the results of direct numerical simulations as well as those obtained from the direct linearization of the defocusing DSII system (see section~\eqref{s:linearizationstabilityanalysis}).
In particular, taking the limit $m\to1$, we obtain the linearized dispersion relations for the black solitons of the 
defocusing DSII equation: $f_0(1) = 0$ and $f_1(1) = 1$.
Figure~\ref{f:f23} (top left) shows a plot of $f_0(m)$ and $f_1(m)$ as functions of~$m$ for $m\in[0,1]$.

Next we look at the case $\eta=0$ (and $\xi\ne0$) in~\eqref{e:planewaveansatz}, corresponding to purely longitudinal perturbations.
Similarly to before, we obtain the characteristic polynomial as
\be
p(\xi,\eta,\zeta) = \alpha(m)\,\xi^2 \zeta^2\,\big( \zeta^2 - f_1(m) \xi^2 \big)\,\big( \zeta^2 - f_2(m) \xi ^2 \big)
\ee
(where the overall factor $\alpha(m)$ is inessential for our purposes and is therefore omitted for brevity),
which also yields three branches of the linearized dispersion relation as $(\zeta/\xi)^2 = f(m)$:
\be
f_0(m) = 0\,,\quad
f_{1,2}(m)= \dfrac{m \K^2 ( ((1-m)\K - E_m)^2 + mE_m^2 ) \pm 2 mK_m^2 E_m \sqrt{ m ((1-m)\K - E_m)^2}}{ ( (\K-E_m)^2 - mK_m^2  )^2 }\,.
\ee
Note that $f_1(0) = f_2(0) = 1$, and $f_j(m)>0$ for all $m\in(0,1)$ and $j=1,2$.
We therefore conclude that the periodic solutions of defocusing NLS are linearly stable to longitudinal perturbations in the long wave limit, in agreement with the results of \cite{DeconinckBottman}.
Figure~\ref{f:f23} (top right) shows a plot of $f_1(m)$ and $f_2(m)$ as functions of~$m$ for $m\in[0,1]$.

\paragraph{Focusing DSII.}
Next we look at the stability of the periodic solutions of the focusing DSII system, i.e., we take $\sigma=1\,,\nu=-1$.
Similarly, setting $p=q=P_o=Q_o=s_1=U_1=0$ for simplicity as before, and 
choosing $\lambda_1=0$, $\lambda_2=1-m$ and $\lambda_3=1$, 
from~\eqref{e:lambda_riemanninvariants_DSIIfocusing} we get
\be
r_1 = \tfrac i2 (1+\sqrt{1-m})\,,\quad
r_2 = \tfrac i2 (1-\sqrt{1-m}), \quad 
r_3 = -\tfrac i2 (1+\sqrt{1-m}), \quad 
r_4 = -\tfrac i2 (1-\sqrt{1-m})\,.
\ee
Looking again at the case $\xi=0$ (i.e., purely transversal perturbations) 
first, we again find three possible branches of solutions, 
namely $(\zeta/\eta)^2 = f(m)$, with $f_0(m)\equiv 0$ as before, but now with a double branch identified by 
\begin{gather}
f_1(m) = \frac{E_m^2 ((2-m) \K - 2E_m) }{ \K (E_m^2 - (1-m)\K^2 )}\,
\end{gather}
Figure~\ref{f:f23} (middle left) shows a plot of $f_0(m)$ and $f_1(m)$ as a function of~$m$ for $m\in[0,1]$.
Note that, as before, $f_1(m)>0$ for all $m\in[0,1]$.
We therefore conclude that all periodic solutions of focusing DSII are also stable to purely transversal perturbations.
These predictions are in agreement with the results obtained from the direct linearization of the focusing DSII system (see section~\eqref{s:linearizationstabilityanalysis}).
These findings should be compared with 
the results of \cite{BrownPerry}, where it was shown that the lump solutions of the focusing DSII system are spectrally unstable.  
\unskip\footnote{To avoid confusion, we point out that the system displayed in \cite[top of p.2]{BrownPerry} carries a sign that would make it the defocusing variant.  On the other hand, the system actually analyzed there, and the one to which the lump solutions belong, is the focusing DSII equation, in agreement with our naming convention.}
The present result applies instead to the periodic traveling wave solutions of the focusing DSII system. 

Conversely, in the case $\eta=0$ (i.e., purely longitudinal perturbations), 
we obtain the following three branches of linearized dispersion relation as $(\zeta/\xi)^2 = f(m)$, with
\bse
\begin{gather}
    f_0(m) = 0\,,\qquad
    f_{1,2}(m) = - \dfrac{  (1-m)\K^2  }{( (1-m) \K^2 - E_m^2 )^2} F_{1,2}(m)\,,\\
    F_{1,2}(m) = (2-m) ( (1-m) \K^2 + E_m^2 )- 4(1-m) \K E_m \pm 2 (\K-E_m) \sqrt{ (1-m) ( (1-m) \K - E_m )^2}\,.
\end{gather}
\ese
Note that $f_1(1) = f_2(1) = 0$, but $f_2(m)< f_1(m) < 0$ for all $m\in(0,1]$,
which allows to conclude that all these solutions are linearly unstable to long-wavelength longitudinal perturbations, in agreement with the results of~\cite{DeconinckSegal}.
Note also that, while $\lim_{m\to0} f_1(m) = -1$ and $\lim_{m\to0} f_2(m) = -\infty$.
This is because $\sqrt{|f_2(m)|}$ is proportional to the ratio $|\zeta/\eta|$, and in this case 
the growth rate remains finite in the limit $\eta\to0$, as we will show in the following sections. 
Figure~\ref{f:f23} (middle right) shows a plot of $f_1(m)$ and $f_2(m)$ as functions of~$m$ for $m\in[0,1]$. 

\paragraph{Defocusing DSI.}
Now we look at the stability of the periodic solutions of the defocusing DSI system ($\sigma=-1$ and $\nu=1$).
Owing to the $x\leftrightarrow y$ symmetry~\eqref{e:DSIsymmetry} of the DSI system which sends the focusing DSI into the defocusing DSI and viceversa, it is sufficient to study both of these variants in the case $q^2<1$.
Taking $\xi=0$, $q^2<1$ and the same setting as the focusing DSII case, we then obtain three possible branches of solutions:
\vspace*{-1ex}
\bse
\begin{gather}
f_0(m) = 0\,,\qquad
f_1(m) = -\frac{E_m^2 ((2-m) \K - 2E_m) }{ \K (E_m^2 - (1-m)\K^2 )}\,,
\end{gather}
\ese
again with $f_1(m)$ corresponding to a double branch. 
Figure~\ref{f:f23}(bottom left) shows a plot of $f_0(m)$ and  $f_1(m)$ as a function of~$m$ for $m\in[0,1]$.
Note that $f_1(m) < 0$ for all $m\in[0,1]$. 
We can therefore conclude that all periodic solutions of the defocusing DSI equation are unstable to transverse perturbations,
with relative growth rate equal to $g(m) = \sqrt{|f_{1,2}(m)|}$ .
In particular, taking the limit $m\to1$, we obtain the instability growth rate for the sech solitons of the focusing DSII equation: $\lim_{m\to1}g(m) = 1$.
Conversely, in the case $\eta=0$, the results for the defocusing DSI system coincide with those for the focusing DSII system (because both the unperturbed solution and the perturbations are independent of $y$).
Therefore, we will not repeat them here.

\paragraph{Focusing DSI with $q^2<1$.}
Finally, we look at the stability of the periodic solutions of the focusing DSI system ($\sigma=-1$ and $\nu=-1$) with $q^2<1$ and $\xi=0$, by taking the same setting as the defocusing DSII case, we obtain three possible branches of solutions:
\vspace*{-1ex}
\bse
\begin{gather}
f_0(m) = 0\,,\qquad
f_1(m) = -\frac{(\K-E_m)^2 (2E_m + (m-1)\K)}{\K ( m \K^2 - (\K - E_m)^2)}\,,
\end{gather}
\ese
again with $f_1(m)$ corresponding to a double branch. 
Note that $f_1(m) < 0$ for all $m\in(0,1]$ ,
which means that all non-constant periodic solutions of the focusing DSI equation are also unstable to transverse perturbations,
with relative growth rate equal to $g(m) = \sqrt{|f_1(m)|}$.
In particular, taking the limit $m\to1$, we obtain the instability growth rate for the black solitons of the focusing DSI equation: $\lim_{m\to1}g(m) = 1$.
Figure~\ref{f:f23}(bottom right) shows a plot of $f_0(m)$ and  $f_1(m)$ as functions of~$m$ for $m\in[0,1]$.
Note that $f_0(0) = f_1(0)=0$, indicating stability of the constant background with respect to the perturbations with $q^2<1$.  
In appendix~\ref{a:modulationalstability}, however, we show that the background solution of the focusing DSI equation is unstable to perturbations with $q^2>1$.
Conversely, in the case $\eta=0$, the results for the focusing DSI system coincide with those for the defocusing DSII system, similarly to above.
Therefore, like before, we will not repeat them.

\subsection{Stability analysis via direct linearization of the DS system}
\label{s:linearizationstabilityanalysis}

We now study the stability properties of the periodic solutions of the DS system by directly linearizing the DS system itself.  

\paragraph{General setting.}
Even though the perturbation approach can be applied to the general class of periodic solutions derived in section~\ref{s:periodic}, 
here we restrict ourselves to the flat-phase solutions in order to keep the presentation as simple as possible.  
We study the stability of the periodic solutions with respect to co-periodic perturbations.
Specifically, we look for a perturbed solution of the DS system in the form (cf.\ \cite{CarterSegur})
\bse
\label{e:linearizationexpansion}
\begin{gather}
\psi(\@x,t) = \e^{i S(\@x,t)} \big(\,R(Z) + \delta w_1(Z,\@x,t) + i\delta w_{-1}(Z,\@x,t)\,\big), 
\\
P(\@x,t) = -2\nu R^2(Z) + \delta w_0(Z,\@x,t), 
\end{gather}
\ese
where $R(Z)$ is a real-valued (but not necessarily positive) function expressing the exact periodic traveling wave solutions,
$S(\@x,t) = v_1 x + v_2 y- \mu t$ as before
and where for simplicity we set $\epsilon=1$ throughout this section.
Here and below,
$w_1$, $w_{-1}$ and $w_0$ are real-valued functions quantifying the corrections to the leading-order solution, 
and $0<\delta\ll1$ is the smallness parameter of the perturbations.

For simplicity, we also take $v_1=v_2 = \omega =0$ and we also focus on the case when $k_2=0$, i.e., the case of wave crests parallel to the $y$ axis.
Plugging~\eqref{e:linearizationexpansion} into the DS system~\eqref{e:DS1}, 
the $O(1)$ terms in $\delta$ recover the exact traveling wave solutions for $R(Z)$ and $\phi(Z)$.
Then, neglecting terms at $O(\delta^2)$ and higher,
we obtain the linearized DS system
\vspace*{-0.6ex}
\bse
\label{e:directlinearizationPDEs}
\begin{gather}
\kern-14em
i (w_{1,t} + i w_{-1,t})
  + \half\big[ w_{1,xx} - \sigma w_{1,yy} + i ( w_{-1,xx} - \sigma w_{-1,yy} ) \big]
\nonumber\\\kern14em
+ \mu (w_1 + i w_{-1} ) + \sigma\nu R^2 (w_1 - i w_{-1} ) + \sigma R w_0 = 0\,,
\label{e:directlinearizationPDE1}
\\
	w_{0,xx} + \sigma w_{0,yy} + 4\nu (Rw_1)_{xx} = 0\,,
\label{e:directlinearizationPDE2}
\end{gather}
\ese
where 
$(Rw_1)_{xx} = k_1^2 R''(Z) w_1 + 2 k_1 R'(Z)w_{1,x} + R(Z)w_{1,xx}$.
Next, we look for plane-wave transverse perturbations.  
Since equations~\eqref{e:directlinearizationPDEs} are linear, we can do so using the following ansatz: 
\vspace*{-1ex}
\be
\label{e:linearizationansatz}
	w_j(Z,\@x,t) = W_j(2K_mk_1x)\,e^{i(\eta y - \zeta t)},\qquad 
    j = \pm 1,0\,, 
\ee
so that $\eta\in\mathbb{R}$ is the wavenumber of the transverse perturbations, and 
$\zeta$ is the linearized dispersion relation (when real-valued) or the growth/decay rate (when imaginary).
We then have
\be
\label{e:wderiv}
	w_{j,t} = - i\zeta W_j\,\e^{i(\eta y - \zeta t)}, \quad 
	w_{j,xx} = W_j''\,\e^{i(\eta y - \zeta t)}, \quad  
	w_{j,yy} = - \eta^2W_j\,\e^{i(\eta y - \zeta t)}\,,\qquad j = \pm 1,0\,.
\ee
Finally, we restrict ourselves to the case $\lambda_1 = 0$ and $\lambda_3 = 1$, 
in which case the leading-order periodic solutions simplify considerably.
Indeed, recalling \eqref{e:kappadef}, \eqref{e:AaB} and~\eqref{e:AaB:variant} 
in this case one has $k_1=1/(2K_m)$ for all variants of the DS system, which implies $2K_mk_1 = 1$ in the above setting.
Then, \eqref{e:directlinearizationPDE2} and the sum and difference of~\eqref{e:directlinearizationPDE1} yield the system
\bse
\begin{gather}
	\half \big(W_1'' + \sigma \eta^2 W_1 \big) + (\mu + \sigma\nu R^2) W_1 + \sigma R W_0 + i\zeta W_{-1} = 0, \\
	\half \big(W_{-1}'' + \sigma \eta^2 W_{-1} \big) + (\mu - \sigma\nu R^2) W_{-1} - i\zeta W_1 = 0, \\
	W_0'' - \sigma \eta^2 W_0 + 4\nu (R'' W_1 + 2R'W_1' + RW_1'') = 0,
\end{gather}   
\ese
where primes denote differentiation with respect to~$x$.
The above system can be recast as the following generalized eigenvalue problem for a matrix differential operator:
\be
\left[\begin{array}{ccc}
	   0 & \frac12(\partial'' + \sigma \eta^2) + \mu - \sigma\nu R^2 & 0 \\
	   - \frac12 (\partial'' + \sigma \eta^2) - \mu - \sigma\nu R^2 & 0 & -\sigma R \\
	   4\nu(R\partial'' + 2R'\partial' + R'') & 0 & \partial'' - \sigma \eta^2
    \end{array}\right]\,\@W
    = i\zeta I_{2,3}\@W\,,
\label{e:linearizedeigensystem}
\ee
where $\@W(x,t) = (W_1,W_{-1},W_0)^\t$
and $I_{n,m}$ is as in~\eqref{e:Inmdef}.

The eigenvalue problem~\eqref{e:linearizedeigensystem}
could be numerically solved in the Fourier domain. 
Explicitly, letting 
$\^W_j(\xi) = \F_\xi[W_j(x)] = \int\nolimits_\Real \e^{-i\xi x}W_j(x)\,\d x$
denote the Fourier transform of $W_j(x)$, 
the system~\eqref{e:linearizedeigensystem} becomes a generalized matrix eigenvalue problem
for  $\^{\@W}(\xi) = (\^W_1,\^W_{-1},\^W_0)^\t$, involving the convolutions 
$\R_n \^W_j = \F_\xi[ R^n W_j ] = 
 \int\nolimits_\Real \F_{\~\xi-\xi}[R^n(x)] \^W_j(\xi) \d\~\xi$
for $n=1,2$.
In practice, after discretizing the Fourier domain by setting up a spatial grid $x_n = n\Delta x$ 
corresponding to a Fourier grid $k_n = n\Delta k$ for $n = - N/2,\dots,N/2-1$,
with $\Delta k = 2\pi/(N\Delta x)$,
the convolutions become matrix multiplications with circulant Toeplitz matrices \cite{Fornberg,Trefethen}.
Alternatively, however, and more directly, the eigenproblem~\eqref{e:linearizedeigensystem} can be also discretized in the physical domain using spectral differentiation matrices \cite{Fornberg,Trefethen}.
Indeed, in all the calculations discussed below we chose the latter route,
using a computational period equal to at least 16 times the fundamental period of the unperturbed solution and at least $2^9$ spatial grid points.

\begin{figure}[t!]
    \centerline{\kern-1em
		\includegraphics[trim = 0 0 0 0,clip,scale=0.575]{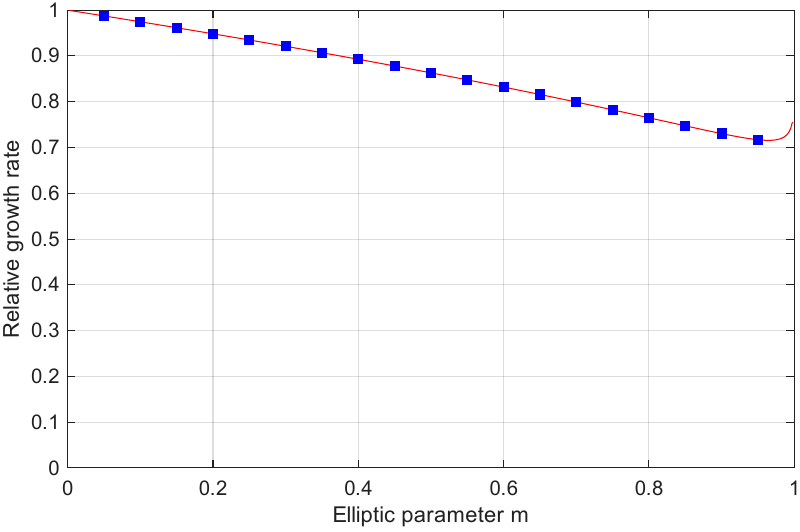}\quad
		\includegraphics[trim = 0 0 0 0,clip,scale=0.575]{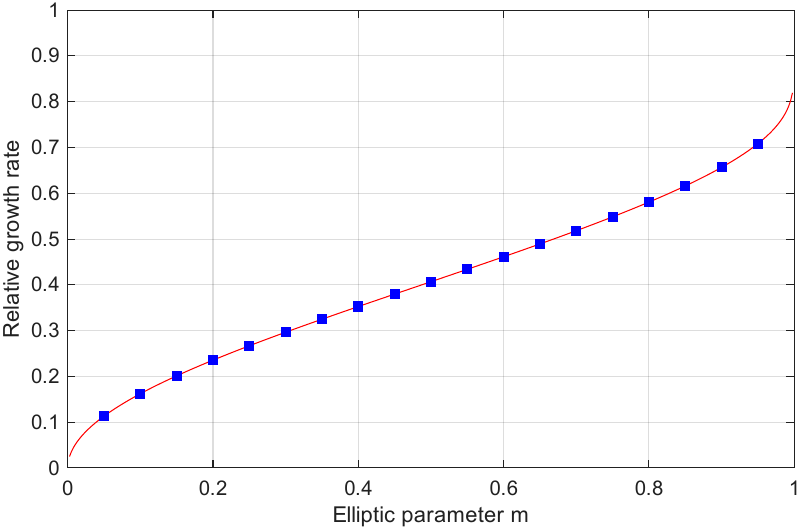}\kern-1em}
    \bigskip
	\centerline{\kern-1em
		\includegraphics[trim = 0 0 0 0,clip,scale=0.575]{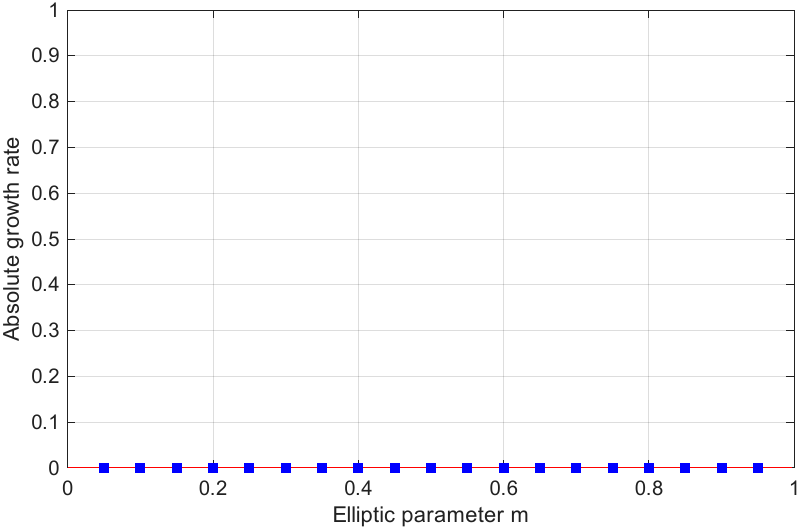}\quad
		\includegraphics[trim = 0 0 0 0,clip,scale=0.575]{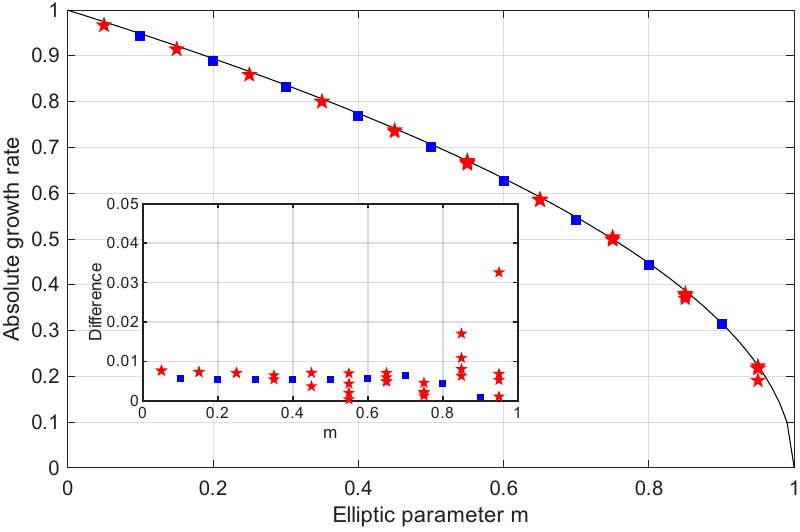}\kern-1em}
    \medskip
        \caption{%
        Top row: Relative perturbation growth rate as a function of $m$ for purely transversal perturbations of the DSI system, 
        computed via the numerical solution of the linearized system~\eqref{e:linearizedeigensystem} (blue squares)  
        with $\eta = 0.01$,
        compared to the predictions from Whitham modulation theory (red curves). 
        Left column: defocusing case.  Right column: focusing case. 
        Bottom row: Absolute perturbation growth rate as a function of $m$ for purely longitudinal perturbations of the DSII system, computed
        from the modified linearized system obtained from~\eqref{e:modifiedlinearizationansatz} (blue squares) and from direct numerical simulations (red stars).
        Left column: defocusing case.
        Right column: focusing case.
        The inset in the bottom right panel shows the difference between the numerical values and the function $\sqrt{1-m}$ (black curve in the main plot). 
    }
    \label{f:linearization}
    \kern-\medskipamount
\end{figure}

\paragraph*{Direct linearization of defocusing DSI.} ~
Taking $\lambda_2 = m$, from~\eqref{e:DSIdef_dn} we have
$R(Z) = \dn(x,m)$ and $\mu = -1+\half m$.
The numerical solution of the eigenproblem~\eqref{e:linearizedeigensystem} then returns complex values of $\zeta$ for all $m\in(0,1)$, confirming the instability of the periodic solutions~\eqref{e:DSIdef_dn}.
Moreover, the numerical values for the relative growth rate (shown in in the top left panel of  Fig.~\ref{f:linearization}) 
agree very well with those coming from the linearization of the Whitham modulation equations in section~\ref{s:whithamstabilityanalysis}, thereby providing a quantitative validation of the modulation equations derived in section~\ref{s:whitham}. 

\paragraph*{Direct linearization of focusing DSI.} ~
In this case, taking $\lambda_2 = m$, from~\eqref{e:DSIfoc_sn} we have
$R(Z) = \sqrt{m}\,\sn(x,m)$ and $\mu = \half (m+1)$.
The numerical solution of the eigenproblem~\eqref{e:linearizedeigensystem} again returns complex values of $\zeta$ for all values of $m\in(0,1)$ tested, confirming the instability of the periodic solutions~\eqref{e:DSIfoc_sn}.
Moreover, the numerically obtained values for the relative growth rate as a function of $m$ (shown in the top right panel of Fig.~\ref{f:linearization}) again agree very well with the predictions from the linearization of the Whitham modulation equations in section~\ref{s:whithamstabilityanalysis} (blue curve in Fig.~\ref{f:linearization}a), again providing an important, quantitative validation of the modulation equations derived in section~\ref{s:whitham}. 

\paragraph*{Direct linearization of focusing and defocusing DSII: transversal perturbations.}
For the defocusing DSII system,  
taking $\lambda_2 = m$, from \eqref{e:DSIIdef_sn} we have
$R(Z) = \sqrt{m}\,\sn(x,m)$ and $\mu = \half (m+1)$.
For the focusing DSII system, taking $\lambda_2 = 1-m$, from \eqref{e:DSIIfoc_dn} we have
$R(Z) = \dn(x,m)$ and $\mu = - 1 + \half m$.
In both cases, the numerical solution of the eigenproblem~\eqref{e:linearizedeigensystem} 
returns purely real values of $\zeta$ for all $m\in(0,1)$
up to numerical error, confirming the stability of the corresponding periodic solutions with respect to transversal perturbations 
as well as the predictions coming from the linearization of the Whitham modulation equations in section~\ref{s:whithamstabilityanalysis}, once more providing a validation of the modulation equations derived in section~\ref{s:whitham}. 

\paragraph*{Direct linearization of focusing and defocusing DSII: longitudinal perturbations.}
The above results concern stability with respect to purely transversal perturbations.
On the other hand,
the same setting can be used to test the stability of the periodic solutions with respect to longitudinal perturbations,
by simply replacing~\eqref{e:linearizationansatz} with
\be
\label{e:modifiedlinearizationansatz}
	w_j(Z,\@x,t) = W_j(2K_mk_1x)\,e^{i(\xi x - \zeta t)},\qquad 
    j = \pm 1,0\,.
\ee
The resulting linearized system is omitted for brevity.
In this case, however, the results for the focusing and defocusing cases are markedly different.
Indeed, while in the defocusing case the resulting values of $\zeta$ are still real,
in the focusing case one obtains nonzero growth rates.
The bottom right panel of Fig.~\ref{f:linearization} shows the absolute growth rate $|\Im\zeta|$ for the focusing DSII~system.
As can be seen from the figure, the numerically computed values for $|\Im\zeta|$ agree very well with the curve $\sqrt{1-m}$, and also agree with the results of direct numerical simulations of the focusing DSII system, which are described in section~\ref{s:DNS}.
We emphasize that this last result does not invalidate the Whitham modulation theory.
Recall that the Whitham modulation equations provide a prediction for the linearized dispersion relation in the long wavelength limit, i.e., in the limit $\eta\to0$,
and in all cases one obtains that $\zeta$ is proportional to $\eta$.
What the numerical results mean then is simply that for the focusing DSII system there are faster growing perturbations, and those are the ones that are picked up by the linearization and direct numerics.

\subsection{Direct numerical stability analysis for the DSII system}
\label{s:DNS}

The stability of the periodic and soliton solutions of the DS system can be independently investigated by numerical computations in which the DS system is integrated starting from an initial condition consisting of the exact solution plus a small random broad-spectrum perturbation, any ensuing exponential divergence from the unperturbed solution can be observed, and the growth rate (quantified as the coefficient $c$ in any exponential term $e^{c t}$) measured, using the numerical integration technique and power method described in appendix~\ref{a:numericalmethods}. 
Next we discuss in detail the computations and the resulting outcomes for both variants of the DSII system.
(Note that, as discussed in appendix~\ref{a:numericalmethods}, it was not possible to run direct numerical simulations for the DSI system.)

\begin{figure}[t!]
\centering
\includegraphics[width=1.0\textwidth]{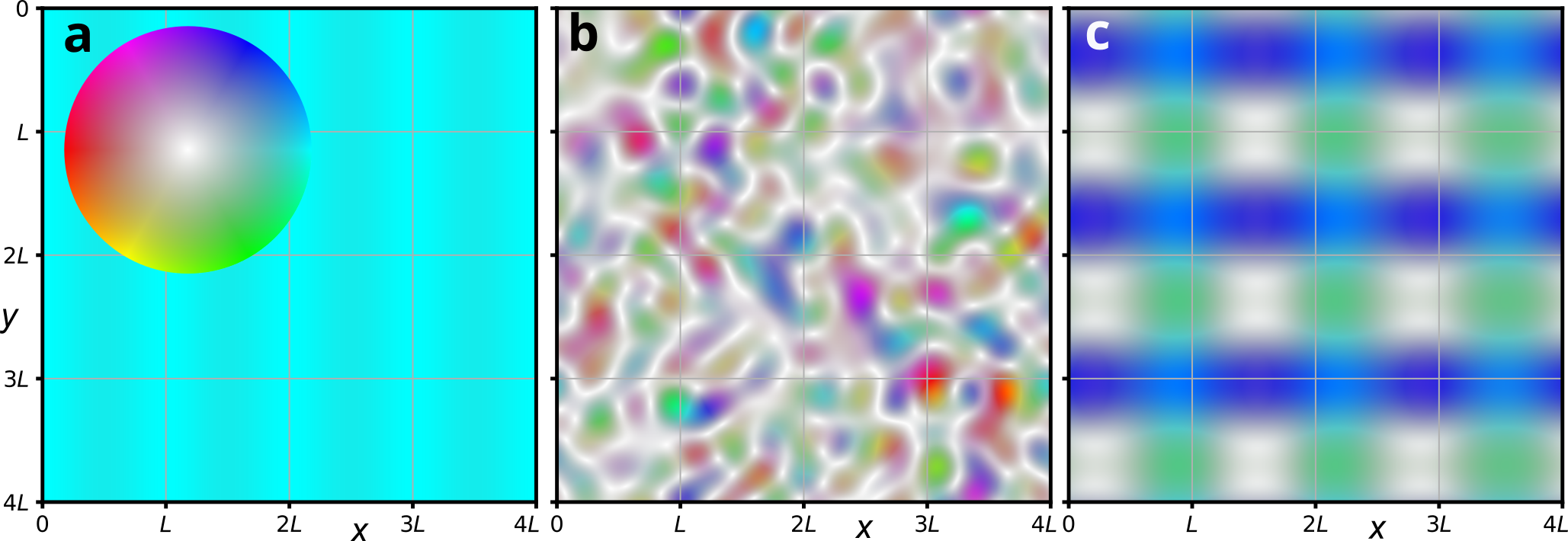}
\caption{Numerically observed instability of periodic state in focusing DSII: power method on a $4\wavelength \times 4\wavelength$ domain with $m=0.15$. 
(a) the field $\psi(\@x,t)$ of the periodic state at $t=0$ (which has small amplitude since $m$ is small). 
The inset displays the color map used for the unit disk in the complex $z$-plane in all the images 
of this figure and Fig.~\ref{f:DNS_defocusingDSII}, in which hue identifies the phase and the saturation is proportional to the modulus, with white corresponding to $z=0$. 
(b) A small random initial perturbation applied to the periodic state; 
(c) The deviation of $\psi(\@x,t)$ after many iterations of the power method.
All complex fields in this figure are normalized to maximum modulus 1 for the purposes of visualization.}
\label{f:DNS_focusingDSII}
\end{figure}

\paragraph{Instability of periodic solutions of focusing DSII.}
We computed the dominant modes of growth for the deviations from the longitudinal periodic state
of the focusing DSII system, namely, 
\be
\psi(x,y,t) = \dn(x,m)\,\e^{ i\mu^{\prime} t},\qquad 
P(x,y,t) = \dn^2(x,m) - \dn^2(0,m),
\label{e:unperturbed}
\ee
where  
$\mu^{\prime}=\mu-2\operatorname{dn}^2(0,m)$, 
$\mu=1-m/2$, and
$m$ is the elliptic parameter in the Jacobi elliptic function $\dn$, as before,
and where the invariance of the solution with respect to constant shifts in $P$ was used
(cf.\ appendix~\ref{a:identities}).
We considered periodic rectangular domains whose dimensions we specify in terms of the wavelength (or period $\wavelength$ of the periodic state, which is $\wavelength = 2 K(m)$.

Because of the unavoidable use of a finite computational domain, which in our case used periodic boundary conditions as discussed in appendix~\ref{a:numericalmethods}, each numerical stability experiment supports only perturbations whose Fourier components have wave vectors belonging to a discrete set, and which in particular have a maximum wavelength in the longitudinal and transverse directions equal to the corresponding domain dimension (transverse meaning the $y$-direction, along which the unperturbed periodic solution~\eqref{e:unperturbed} is constant). We therefore performed computations with a variety of domain sizes and aspect ratios. To support the periodic state, the longitudinal dimension must be an integer multiple of $\wavelength$.
Conversely, the transverse dimension may be varied continuously. On each domain we considered a variety of $m$ values between 0.05 and 0.95. 
A plot of a typical unperturbed solution, the initial perturbation and the final observed state is 
illustrated in Fig.~\ref{f:DNS_focusingDSII}.

Modes with positive growth rate were observed in most cases, 
confirming the instability predicted in sections~\ref{s:whithamstabilityanalysis} and~\ref{s:linearizationstabilityanalysis}. 
Importantly, however, we point out that instabilities were present only when the domain size was
greater than~$\wavelength$ in at least one spatial dimension. 
In other words, the periodic solution~\eqref{e:unperturbed} appears to be stable with respect to perturbations that are both longitudinally and transversely co-periodic with the unperturbed periodic solution. 

\begin{figure}[t!]
\centering
\includegraphics[width=1.0\textwidth]{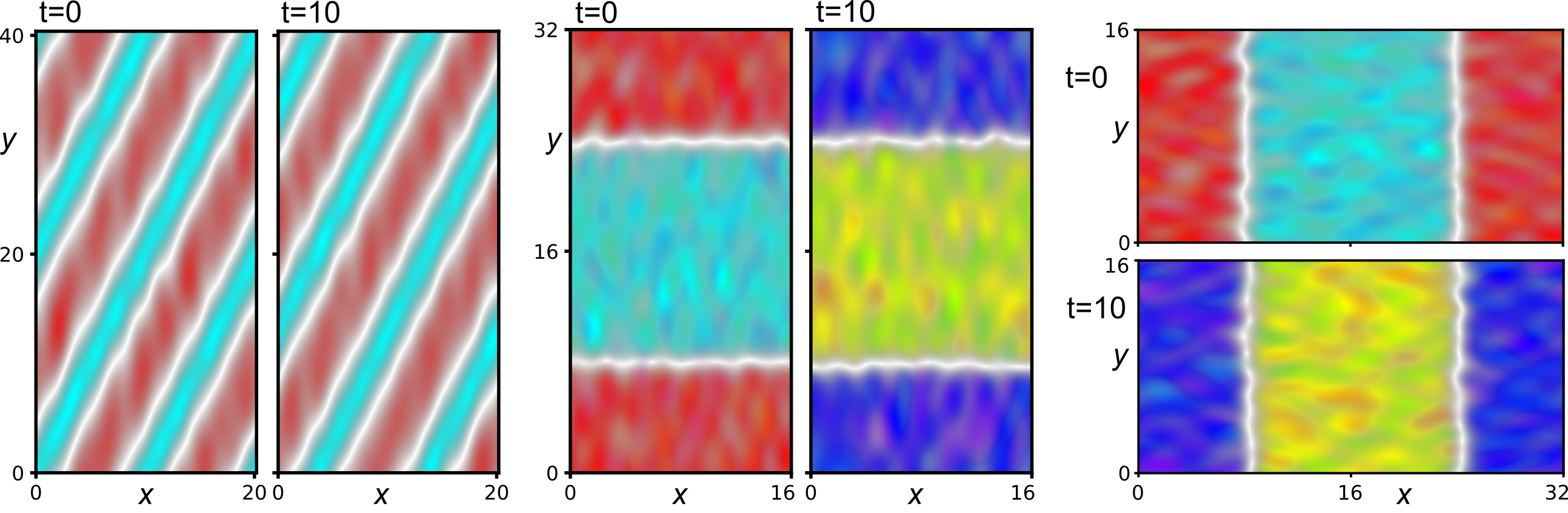}
\caption{
Stability in defocusing DSII. On the left, a perturbed oblique periodic state: snapshots of $P$ at $t=0$ and $t=10$. The unperturbed solution was obtained by taking $\lambda_1=0.1$, $\lambda_2=0.3$, $\lambda_3=0.35$, the wave vector is along $[1,2]$ and the axes span 2 periods along both axes ($\approx20.189$ and $\approx40.378$ for $x$ and $y$ axes respectively). The function $P$ is pictured because the relative amplitude of $\psi$ is small. 
In the center panel, a perturbed pair of back-to-back horizontal dark solitons $\psi(x,y,t) = e^{it} \tanh(y)$, $P(x,y,t)=0$: snapshots of $\psi$ at $t=0$ and $t=10$. The right panel shows the same as the middle panel, but for vertical solitons. 
The color map is the same as in Figure 3: white denotes 0.}
\label{f:DNS_defocusingDSII}
\end{figure}

For all the domain geometries that were tested, the measured growth rate of the dominant instability was found to approach~1 as $m\rightarrow 0$ and to be a monotonic decreasing function of~$m$ that lies very close to $\sqrt{1-m}$.
The numerically computed growth rate of the instabilies of the periodic state in the focusing DSII equation as a function of $m$ for a $4\wavelength \times 4\wavelength$ spatial domain were shown in the bottom right panel of Fig.~\ref{f:linearization}. 
For each $m$ value the computation was conducted four times, with different initial perturbations, in order to obtain a measure of uncertainty in the growth rate estimates. 
The convergence of the power method is slower for values of $m$ closer to~1. 
We emphasize that the numerically computed values are in excellent quantitative agreement with the predictions arising from the linearized DS system in section~\ref{s:linearizationstabilityanalysis}.
This excellent agreement between the numerical values and the curve $\sqrt{1-m}$ is at present only empirical.
At the same time, it is interesting to note that 
$\sqrt{1-m} = \dn(K_m,m) = \min_{\@x\in\Real^2}|\psi(\@x,t)|$.
Thus, the observed growth rate appears to equal the minimum amplitude of the periodic state, rather than any average of it. 
Recall that, for the modulational instability, the growth rate is proportional to the constant amplitude of the background. 
Hence, it is as if the minimum value of the periodic state acts as a background value for the perturbations.

For all values of $m$, the dominant mode of instability was observed to have a predominant Fourier component of wavelength around $1.4\wavelength$. For example, in Fig.~\ref{f:DNS_focusingDSII}(c), where the domain size is $4\wavelength \times 4 \wavelength$, 
one can clearly see 3~cycles of the perturbation in both directions (with a corresponding wavelength $\frac43\wavelength\approx 1.33\wavelength$). 
We also note that the observed dominant mode of instability has a predominantly transverse variation when $m$ is small (near zero), when the periodic state has small amplitude, whereas it has a predominantly longitudinal variation when $m$ is large (near 1), when the periodic state has large amplitude.

\paragraph{Stability of periodic and soliton solutions of defocusing DSII.}

Very different numerical results were observed for the defocusing DSII system.
We generated trajectories starting from small random perturbations of periodic solutions (horizontal, vertical, and oblique) and of dark solitons (horizontal and vertical). 
In no case was exponential growth of the perturbation observed. 
Examples of a perturbed oblique periodic solution and of horizontal and vertical states resembling two dark solitons are shown in Figure~\ref{f:DNS_defocusingDSII}.
The absence of any instabilities is consistent with both the Whitham stability analysis of section~\ref{s:whithamstabilityanalysis} and the predictions from the linearized DS system in~\ref{s:linearizationstabilityanalysis}.
At the same time, it is important to realize that the theoretical predictions only discussed 
purely longitudinal or purely transverse perturbations.
In contrast, the direct numerical results of the present section imply that the periodic solutions of the defocusing~DSII system are stable with respect to arbitrary perturbations, not only purely transversal or longitudinal ones, thereby providing a much stronger stability claim.

\subsection{Oblique stability analysis for defocusing DSII}
\label{s:oblique}

In the previous sections we have seen that only the periodic solutions of the defocusing DSII system are stable to both purely longitudinal and purely transversal perturbations,
whereas the solutions of all other variants of the DS system are unstable.
On the other hand, a natural question is whether the periodic solutions of the defocusing DSII system are also stable with respect to more general kinds of perturbations. 
In this section we address this question.
We do so first using the Whitham modulation equations; 
then we confirm the results using the direct linearization of the defocusing DSII system.
Note that, since the solutions of all other variants of the DS system have already been shown to be unstable, we can limit our analysis to the defocusing DSII system.

To investigate oblique perturbations, we extend the analysis of section~\ref{s:whithamstabilityanalysis}
and consider the linearized system~\eqref{e:LDSWS} but replacing the normal-mode ansatz~\eqref{e:planewaveansatz} by
\be
\mathbf{R}^{(1)}(x,y,t) = \@C\, \e^{\i\xi (x\cos\alpha + y\sin\alpha - \gamma t)},
\label{e:obliquenormalmodeansatz}
\ee
where $\xi>0$ is the perturbation wavenumber, $\alpha$ is the angle of the perturbation wavevector relative to the $x$-axis, and $\gamma$ is the (possibly complex) phase velocity. 
The values $\alpha=0$ and $\alpha=\pi/2$ correspond respectively to longitudinal and transverse perturbations, which allows us to recover the results of section~\ref{s:whithamstabilityanalysis}
as special cases.

Substituting the ansatz~\eqref{e:obliquenormalmodeansatz} into the linearized system~\eqref{e:LDSWS} and canceling the common factor $\mathrm{i}\xi$ yields the generalized eigenvalue problem
\be
\left( \cos\alpha\, B^{(0)} + \sin\alpha\, C^{(0)} \right)\mathbf{C} = \gamma A^{(0)}\mathbf{C}.
\ee
The corresponding characteristic equation is 
\be
p(\alpha,\gamma;m) = \det\left[ \cos\alpha\,B^{(0)}(m) + \sin\alpha\,C^{(0)}(m) - \gamma A^{(0)}(m) \right]=0,
\ee
where the matrices are evaluated at the constant background specified by ~\eqref{e:rmstability}, with $q=p=P_o=Q_o=0$. 
For each pair of values $(\alpha,m)$, we numerically solved the $8\times8$ generalized eigenvalue problem. The results show that all finite characteristic velocities remain real within numerical accuracy throughout the sampled range of $\alpha\in (0,\pi/2)$ and $m \in (0,1)$. 
Thus, the Whitham modulation analysis detects no oblique modulational instability.
These results suggest modulational stability of the defocusing DSII periodic waves with respect to oblique perturbations, extending the longitudinal and transverse stability results discussed in sections~\ref{s:whithamstabilityanalysis}.

We can also investigate oblique perturbations by directly linearizing
the DS system about the periodic solution, extending the analysis of section~\ref{s:linearizationstabilityanalysis}.
To this end, we replace the ansatz in
\eqref{e:linearizationansatz} by
\be
w_j(Z,\@x,t) = W_j(2K_mk_1x)\,e^{i (\xi_1 x+ \xi_2 y - \zeta t)},\qquad 
j = \pm 1,0\,, 
\ee
where $\xi_1=\xi\cos\alpha$ and $\xi_2=\xi\sin\alpha$.
The derivative formulas in \eqref{e:wderiv} are then replaced by 
\bse
\begin{gather}
    w_{j,t} = - i\zeta W_j\,\e^{i(\xi_1 x+ \xi_2 y - \zeta t)}\,, \qquad 
	w_{j,yy} = - \xi_2^2\alpha W_j\,\e^{i(\xi_1 x+ \xi_2 y - \zeta t)}\,, \\
	w_{j,xx} =( W_j'' + 2i \xi_1^2 W_j' - \xi_1^2 W_j  ) \,\e^{i(\xi_1 x+ \xi_2 y - \zeta t)}\,,\qquad j = \pm 1,0\,.
\end{gather}
\ese
Introducing the shifted differential operator
\be
\mathcal D_\alpha=\partial'+ i\xi_1\,, \qquad
\mathcal D_\alpha^2 = \partial'' + 2i\xi_1\partial' - \xi_1^2\,,
\ee
we obtain the generalized eigenvalue problem
\be
\begin{pmatrix}
0 & \frac12(\mathcal D_\alpha^2 + \sigma\xi_2^2) + \mu - \sigma\nu R^2 & 0
\\
-\frac12(\mathcal D_\alpha^2 + \sigma\xi_2^2) - \mu - \sigma\nu R^2 & 0 & -\sigma R
\\
4\nu\left( R\mathcal D_\alpha^2 + 2R'\mathcal D_\alpha + R'' \right) & 0 &
\mathcal D_\alpha^2-\sigma\xi_2^2
\end{pmatrix}
\@W = i\zeta I_{2,3}\@W,
\label{e:obliqueeigensystem}
\ee
where $\@W(x,t)$ and $I_{n,m}$ is again as in~\eqref{e:Inmdef}.
We can solve \eqref{e:obliqueeigensystem} numerically using the same discretization approach described in section~\ref{s:linearizationstabilityanalysis}.
For the defocusing DSII waves, no eigenvalues with positive imaginary part were detected within numerical accuracy over the parameter ranges examined. 
These results agree with the oblique modulational stability predicted by the Whitham system.
At the same time, the above calculations provide a stronger result, since the direct linearization of the DS system can also capture 
finite-wavenumber instabilities that would not be detected by the Whitham stability analysis.
We also note that the absence of any instabilities is further confirmed by the direct numerical simulations reported in the previous section.

\section{Discussion and concluding remarks}
\label{s:discussion}

The final outcome of this work is a complete and validated modulation theory for all four variants of the Davey-Stewartson (DS) system, together with a sharp and somewhat surprising stability picture: of the four variants of the DS system, only the defocusing DSII system possesses linearly stable periodic wave trains and line solitons. 
For the other three variants of the DS system, we obtained the instability growth rate explicitly as a function of the elliptic parameter.
These predictions have been confirmed independently by a direct linearization of the DS system and, for the DSII variants, by direct numerical simulation.
Along the way we also performed a detailed study of the parametric dependence of the periodic solutions of all variants of the DS system, including their harmonic and soliton limits.

The fact that the coefficient matrices $\_A$, $\_B$ and~$\_C$ of the DSWS~\eqref{e:DSWS} are independent of $P_o$ and $Q_o$
implies that their value does not affect the stability properties of the periodic solutions
of any of the variants of the DS system.
At the same time, however, it is crucial to realize that $Q_o$ and $P_o$ cannot be neglected as dependent variables, otherwise one would obtain an incorrect modulation system.
The same is not true for the variables $q$ and $p$, which do appear explicitly in $\_B$ and $\_C$ and whose values do enter the stability properties of the periodic solutions, as we have seen in section~\ref{s:stability}.

It is worth to briefly comment on the physical significance of the results of section~\ref{s:stability}.
It is well known that the dark line solitons of the two-dimensional defocusing NLS equation are unstable with respect to long-wavelength transverse perturbations~\cite{KuznetsovTuritsyn1988} (the so-called snake instability, see the review~\cite{KivsharPelinovsky2000}), and the same is true of their periodic counterparts.
Since the defocusing DSII system reduces to the defocusing NLS equation for solutions independent of~$y$, the transverse dynamics is where the two models depart from each other, and they do so in two distinct respects: the DSII system has hyperbolic rather than elliptic transverse dispersion, and it is coupled to a mean field.
Disentangling the relative contribution of these two mechanisms to the transverse stabilization of dark wave trains, and more generally a modulation-theoretic diagnosis of when mean-field coupling stabilizes and when it destabilizes a two-dimensional wave train, seems to us one of the more interesting questions opened up by this work.

Beyond the DS system itself, we would like to highlight three features of the present calculations that we believe are not specific to it.
The first is the role of the mean-field constants $P_o$ and $Q_o$ mentioned above. 
The same phenomenon occurs for the KP equation~\cite{RSPA2017}, and we expect that it will arise it whenever a first-order reduction of a non-evolutionary constraint introduces an auxiliary potential.
The second is the diagnostic value of comparing the modulation predictions against a direct linearization of the original PDE: 
As the case of the focusing DSII system illustrates, the two need not agree regarding the fastest growing mode, and the discrepancy is not a failure of modulation theory, but a signal that the dominant instability may lie outside the long-wavelength regime that modulation theory describes.
A practitioner comparing Whitham predictions with a numerically computed spectrum must therefore track the correct branch of perturbations (the one whose growth rate vanishes as the transverse wavenumber tends to zero) rather than the largest growth rate in the spectrum.
The third is the observation that the qualitative form of the periodic solutions can depend on the propagation direction, as it does for the DSI systems across $|k_1|=|k_2|$; this is a purely two-dimensional phenomenon, and it should be anticipated to arise in any system whose linear part is not rotationally invariant.
 
The results of this work open up a number of other interesting avenues for further study.
From a practical point of view, one of the most interesting possibilities will be the application of the DS-Whitham system (or, more likely, suitable reductions thereof) to study the dynamics of various kinds of initial value problems 
for those variants of the DS system for which the periodic and soliton solutions are stable, similarly to what was recently done in \cite{JFM2021,NLTY2021,RSPA2022,JFM2026} for the KP equation.

We also emphasize that, even though in this work we limited ourselves to studying the integrable variants of DS for brevity, the present methodology applies to more general versions of DS that are not integrable (e.g., such as hyperbolic-hyperbolic variants, elliptic-elliptic variants as well as variants in which the values of the coefficients differ from those that give rise to the integrable variants).
On the other hand, discussing four additional variants DS system (elliptic-elliptic versus hyperbolic-hyperbolic and focusing versus defocusing) would produce a significant increase in the total length of this work, which could make it more cumbersome and confusing.
For these reasons, the treatment of these other variants is left for a future publication.

From a theoretical point of view, an interesting problem will be a rigorous study of the stability properties of the periodic solutions.
In this respect, we mention that a framework exists for making the predictions of Whitham theory rigorous \cite{ClarkeMarangellPerkins,JohnsonOregero,JohnsonOregeroPerkins,JohnsonPerkins}.
We believe that the availability of the Whitham modulation equations obtained in this work will make it possible to obtain analogous rigorous results for the Davey-Stewartson systems.

Also from a theoretical point of view, recall that the DS system is completely integrable.
On one hand, we emphasize that the methodology and the results of this work do not in any way rely on the integrability of the DS system.
On the other hand, even though integrability did not play any role in the analysis, 
it is widely believed that exact asymptotic reductions of integrable systems are themselves integrable.
One may wonder whether the DS-Whitham system is itself integrable.
At the moment, this is an interesting open question.
Again, it may be useful here to consider what is known for KP equation, where the integrability of the corresponding full Whitham modulation system is still an open question, but at the same time it was shown that its harmonic limit, soliton limit, and dimensional reductions are integrable \cite{NLTY2020,NLTY2024}.

We also believe that yet a further interesting direction for further investigation will be the study of well-posedness of DS with periodic boundary conditions.
We believe this problem is particularly interesting in the case of DSI, since in that case~\eqref{e:Ppde} is a forced linear wave equation, which is well-known to be ill-posed as a boundary value problem.  

Finally, recall that the second equation of the DS system [i.e., \eqref{e:Ppde}] only determines $P(x,y,t)$ up to an arbitrary solution of the corresponding homogeneous equation
(i.e., Laplace's equation for the DSII equation and the wave equation for the DSI equation).
Here, we took all such homogeneous solutions to be a constant.
Obviously one could also study the stability of solutions with respect to more general class of perturbations. 
This is a nontrivial task, however, which is therefore left as an interesting problem for further study.

We hope that the results of this work and the present discussion will stimulate further work on the above-mentioned issues as well as on the formulation and applications of Whitham modulation theory to other multi-dimensional nonlinear evolution equations.

\bigskip
\paragraph*{Acknowledgments.}  
We thank Asela Abeya, Mark Ablowitz, Gennady El, Mark Hoefer, Dmitri Kireyev, Dmitry Pelinovsky, Barbara Prinari and Paolo Santini for many insightful conversations on topics related to the present work.
G.B.\ acknowledges support from the Simons foundation under grant number SFI-MPS-TSM-00013369.

\bigskip
\paragraph*{AI disclosure.}
After completing the work and writing the first draft of this manuscript, we asked Anthropic's Opus~5 and OpenAI's Astra~6 to review it, which revealed a few minor issues that were later resolved.

\medskip
\addcontentsline{toc}{section}{Appendix}
\section*{Appendix}
\setcounter{section}1
\setcounter{subsection}0
\setcounter{equation}0
\def\theequation{\Alph{section}.\arabic{equation}}
\def\thesection{\Alph{section}}
\let\z=\zeta

\subsection{Identities}
\label{a:identities}

Here we list various identities used in several places in the main portion of the work.

\begingroup
\def\listitem{\noindent\stepcounter{enumi}\arabic{enumi}.~}

\listitem
The DS system~\eqref{e:DS} possesses a number of invariances that are used at various points in the the main text.  Specifically:

\textit{Phase rotations.}
If $\psi(\@x,t)$ and $\Psi(\@x,t)$ solve \eqref{e:DS},
so do the new dependent variables
\vspace*{-0.6ex}
\begin{gather}
\label{a:phaseinvariance}
\~\psi(\@x,t) = \e^{i\Phi_o/\epsilon}\,\psi(\@x,t)\,,
\qquad
\~\Psi(\@x,t) = \Psi(\@x,t)
\end{gather}
for all $\Phi_o\in\Real$. The same is true for equations~\eqref{e:DS1} and~\eqref{e:DS2}  
with $\~P = P$ and $\~Q = Q$.
In terms of the Madelung variables~\eqref{Madelung}, 
the transformation~\eqref{a:phaseinvariance} simply amounts to the shift $\Phi \mapsto \Phi + \Phi_o/\epsilon$ of the fluid velocity potential, with the density~$\rho$ left unchanged.
Accordingly, any additive constant in the phase of the solutions can be set to zero without loss of generality.
 
\textit{Constant shifts of the mean field~$P$.}
If $\psi(\@x,t)$ and $P(\@x,t)$ solve \eqref{e:DS1},
so do the new dependent variables
\vspace*{-1ex}
\begin{gather}
\label{a:Pshiftinvariance}
\~\psi(\@x,t) = \e^{i\sigma p_o t/\epsilon}\,\psi(\@x,t)\,,
\qquad
\~P(\@x,t) = P(\@x,t) + p_o
\end{gather}
for all $p_o\in\Real$ (and with $\~Q = Q$ for~\eqref{e:DS2}). 
The same is true for the DS system in~\eqref{e:DS} with $\~\Psi = \Psi + p_o$.
Thus, the additive constant left undetermined in~$P$ by \eqref{e:Ppde} 
is inessential, since it can always be removed by a phase rotation of~$\psi$.
(Similarly, the additive constant left undetermined in~$Q$ by \eqref{e:DS-p} is obviously inessential,
since \eqref{e:DS2} only involves $Q$ through its derivatives.)
 
\textit{Galilean transformations.}
If $\psi(\@x,t)$ and $P(\@x,t)$ solve \eqref{e:DS},
so do the new dependent variables
\vspace*{-1ex}
\bse
\label{a:Galileaninvariance}
\begin{gather}
\~\psi(x,y,t) = \e^{i[c_1x+c_2y -\frac12(c_1^2-\sigma c_2^2)t]/\epsilon}\psi(x-c_1t,y+\sigma c_2t,t)\,,
\\
\~P(x,y,t) = P(x-c_1t,y+\sigma c_2t,t)
\end{gather}
\ese
for all $\@c = (c_1,c_2)^\t\in\Real^2$.

\listitem
For any twice differentiable function  $\rho(Z)$, one has
\vspace*{-1ex}
\bse
\begin{gather}
(\log\rho)'' + \frac{\rho''}{\rho} = 
\left(\frac{1}{\rho}\rho'\right)' + \frac{\rho''}{\rho}
	= \frac{2\rho''\rho-(\rho')^2}{\rho^2}\,,
\label{a:lnrho''}
\\
\left( \frac{(\rho')^2}{\rho^2} \right)' = 
\frac{2\rho'\rho''\rho^2 - (\rho')^2 2\rho\rho'}{\rho^4} = 
	\frac{2\rho'\rho''}{\rho^2} - \frac{2(\rho')^3}{\rho^3}.
\label{a:rho'sqroverrhosqr'}
\end{gather}
\ese

\listitem
The Jacobi elliptic amplitude $\am(\phi,m)$ and Jacobian elliptic function $\cn(z,m)$, $\sn(z,m)$ and $\dn(z,m)$ satisfy the identities%
\vspace*{-1ex}
\bse
\begin{gather}
\am(nK_m,m) = n\pi/2,\qquad n\in\Integer\,,
\label{a:Jacobiamplitude}
\\
\cn^2(z,m) + \sn^2(z,m) = 1\,,
\label{e:cnsqr+snsqr=1}
\\
\dn^2(z,m) = 1-m \sn^2(z,m)\,.
\label{a:dnidentity}
\end{gather}
\ese

\listitem
The complete elliptic integral of the first kind $K_m = K(m)$ has the following asymptotic behaviors:%
\bse
\begin{gather}
\label{a:K0}
K_m = \half\pi + \txtfrac18 m + O(m)^2\,,\quad m\to0^+\,,\\
\label{a:Kinfty}
K_m = - \txtfrac12[(1-m)\log(1-m) - 4\log2 ] - \txtfrac18(1-m)[\log(1-m) - 4\log2 + 2] + O(1-m)^2\,,\quad m\to1^-\,. 
\end{gather}
\ese

\listitem
For all $c\in\Real$, the incomplete elliptic integral of the third kind $\Pi(z;\phi,m)$ 
has the following asymptotic behavior:
\vspace*{-1ex}
\begin{gather}
\Pi(1-c/\lambda;\phi,m) = \frac{\pi\sqrt\lambda}{2\sqrt c} + o(1)\,,\qquad \lambda\to0^+\,.
\label{a:ellipticpiidentity}
\end{gather}
Also, as $m\to 1^-$, it has the following asymptotic series:
\be
\Pi(X^2,m) = \frac{K(m)}{1-X^2} - \frac{X}{2(1-X^2)}\log\left(\frac{1+X}{1-X}\right) + o(1-m), \quad m\to 1^-
\label{a:Pim1asymptotic}
\ee

\listitem
For all $c\in\Real$, the Jacobian elliptic functions $y = \sn(cz,m)$ and $y = \cn(cz,m)$ satisfy respectively 
the following two nonlinear first-order ODEs \cite{NIST}:
\bse
\begin{gather}
(y'/c)^2 = (1-y^2)(1-my^2)\,,  
\label{e:snODE}
\\
(y'/c)^2 = (1-y^2)(1-m+my^2)\,. 
\label{e:cnODE}
\end{gather}
\ese

\listitem
The complete elliptic integrals of the first, second and third kind satisfy respectively the following differential identities \cite{ByrdFriedman,NIST}:
\bse
\label{a:KEPiderivs}
\begin{gather}
\frac{dK_m}{dm} = \frac{E_m - (1-m)\K}{2m(1-m)}\,, \qquad 
\frac{dE_m}{dm} = \frac{E_m - \K}{2m}\,,
\\
\frac{\partial\Pi(\gamma,m)}{\partial m} = 
\frac{E_m - (1-m)\Pi}{2(1-m)(m-\gamma)}\,, 
\\
\frac{\partial\Pi(\gamma,m)}{\partial \gamma} = \frac{(m-\gamma^2)\Pi - \gamma E_m - (m-\gamma)\K}{2\gamma(1-\gamma)(m-\gamma)}\,.
\end{gather}
\ese

\listitem
The period averages of the powers of $\rho$, which are needed to write the Whitham modulation equations, are computed noting that
\vspace*{-1ex}
\bse
\label{a:cnperiodaverages}
\begin{gather}
C_2(m) = \int_0^1 \cn^2(2K_m Z,m)\,\d Z = \frac{E_m - (1-m)\K}{mK_m}\,,
\\
C_4(m) = \int_0^1 \cn^4(2K_m Z,m)\,\d Z = \frac{2(2m-1)E_m - (1-m)(3m-2)\K}{3 m^2 \K}\,.
\end{gather}
Accordingly, the period average of $\sn^2(2K_mZ)$ is 
\vspace*{-1ex}
\be
\int_0^1 \sn^2(2K_m Z,m)\,\d Z = 1 - C_2(m) = \frac{\K - E_m}{m \K}\,.
\ee
\ese

\listitem 
For all $c_1, c_2\in\Real^+$, one has 
\vspace*{-1ex}
\begin{equation}
\label{e:sechinvintegral}
\int_0^z \frac1{c_1 + c_2\tanh^2s}\,\d s = \frac 1{c_1+c_2}\left( z + 
    \sqrt{\frac{c_2}{c_1}} \arctan\bigg(\sqrt{\frac{c_2}{c_1}}\tanh z\bigg) \right)\,.
\end{equation}

\listitem
To simplify the inverse of the Vandermonde matrix, it is convenient to use the identities
\bse
\begin{align}
r_i r_l r_m &= -r_j^3 + s_1 r_j^2 - \tfrac12{(s_1^2-s_2)} r_j + \tfrac16{(2s_3-3s_1 s_2 + s_1^3)}, \\
r_i r_l + r_l r_m + r_m r_i &= r_j^2 - s_1 r_j + \tfrac12{(s_1^2-s_2)} = r_j^2 - 2Ur_j + 2 U^2 - \tfrac12 s_2, \\
r_i + r_l + r_m &= s_1 - r_j = 2U_1 - r_j, \\
(r_i - r_j)(r_l - r_j)(r_m-r_j) &= - 4r_j^3 + 3s_1 r_j^2 - (s_1^2-s_2)r_j + \tfrac16{(2s_3-s_1 s_2 + s_1^3)}\,,
\end{align}
\ese
where the symmetric polynomials $s_1,s_2,s_3$ were defined in \eqref{e:s_n(r)}.

\endgroup

\subsection{Numerical methods for the DS system}
\label{a:numericalmethods}

As mentioned in the main text, in order to verify the predictions of Whitham modulation theory
we performed direct numerical simulations of the DS system on rectangular domains with periodic boundary conditions and initial conditions that are small random perturbations of the exact solutions whose stability is of interest. 
In this appendix we briefly describe the numerical methods used for this purpose.

\paragraph{Time integration.}
For all four variants of the DS system, \eqref{e:psipde} is integrated for $\psi(x,y,t)$ using an integrating factor Runge-Kutta~4 (IFRK4) method (e.g., similarly to \cite{milewskitabak,driscollfornberg}). 
Namely, one introduces the change of dependent variable
\be
\^w(\xi_1,\xi_2,t) = \e^{\frac i2(\xi_1^2 - \sigma \xi_2^2)t}\,\^\psi(\xi_1,\xi_2,t)\,, 
\label{e:psi2wmap}
\ee
where $\^\psi(\xi_1,\xi_2,t) = \F_{\xi_1,\xi_2}[\psi(x,y,t)]$,
with $\F_{\xi_1,\xi_2}$ and $\F^{-1}_{x,y}$ denoting the direct and inverse two-dimensional periodic Fourier transforms
and $\xi_1$ and $\xi_2$ are the Fourier variables associated respectively with $x$ and~$y$.
It is straightforward to see that the transformation~\eqref{e:psi2wmap} maps \eqref{e:psipde} into
\be
\^w_t = i\sigma \e^{\frac i2(\xi_1^2 - \sigma \xi_2^2)t} \F_{\xi_1,\xi_2}[(P + \nu|\psi|^2)\psi]\,.
\label{e:whatODE}
\ee
That is, \eqref{e:psi2wmap} integrates the stiff linear part of the PDE~\eqref{e:psipde} exactly.
One can then now integrate \eqref{e:whatODE} numerically using an explicit method, 
e.g., such as RK4, where at each stage of each RK4 step, the right-hand side of~\eqref{e:whatODE} 
is evaluated by recovering $\psi(x,y,t)$ from $w(x,y,t)$ by inverting~\eqref{e:psi2wmap}.

\paragraph{Computation of the auxiliary field.}
At each stage of each RK4 step, one of course also needs to integrate~\eqref{e:Ppde} to update the value of $P(x,y,t)$.
On the other hand, the integration of~\eqref{e:Ppde} differs between the DSII and DSI variants of the system.
For the DSII equation ($\sigma=1$), we can simply take
\begin{equation}
P(x,y,t) = - 2\nu\F^{-1}_{x,y}\bigg[ \frac{\xi_1^2}{\xi_1^2 + \sigma \xi_2^2} \F_{\xi_1,\xi_2}[|\psi(x,y,t)|^2]\bigg] + P_o\,,
\label{e:PFTsol}
\end{equation}
where we chose an arbitrary harmonic function corresponding to the solution of the homogeneous version of \eqref{e:Ppde} to be simply a constant.
Note that the denominator in the right-hand side of~\eqref{e:PFTsol} is zero at $\xi_1 = \xi_2 = 0$.
This however only contributes a solution of the homogeneous version of~\eqref{e:Ppde} (i.e., Laplace's equation), and therefore simply amounts to a redefinition of~$P_o$.

The above approach for does not work for the DSI equation ($\sigma = -1$), however, because the denominator in~\eqref{e:PFTsol} changes into $\xi_1^2 - \xi_2^2$, which is singular along $\xi_1 = \pm\xi_2$.
Efficient methods have been developed to numerically integrate the DSI equation on the line \cite{KleinStoilovII}.
However, those methods are not applicable to the case when $\psi(x,y,t)$ and $P(x,y,t)$ are periodic functions of $x$ and $y$,
and it is not clear at present what an effective approach would be.
For this reason, in this work we refrained from numerically integrating the DSI equation.

\paragraph{Power method for perturbations.}
To obtain the dominant mode of growth of disturbances from the periodic or soliton state
and that mode's growth rate, a form of the power method is used. Denoting the 
base trajectory whose instability is being studied by $\hat{w}_0(\xi_1,\xi_2,t)$, and a nearby trajectory by $\hat{w}(\xi_1,\xi_2,t)$, 
we choose $\hat{w}(\xi_1,\xi_2,0) = \hat{w}_0(\xi_1,\xi_2,0) + \rho(\xi_1,\xi_2)$,
where $\rho$ is a random function of small norm with a high-wavenumber cut-off so that all components present initially are well-resolved on the spatial grid. 
The trajectory $\hat{w}(\xi_1,\xi_2,t)$ is repeatedly advanced a small distance in time (0.1 time units, 100 RK timesteps). 
After each advance, $\hat{w}(\xi_1,\xi_2,t)$ is altered by shrinking its separation from the known $\hat{w}_0(\xi_1,\xi_2,t)$ such that the norm of that separation is re-set to $10^{-6}$, an arbitrarily chosen small number. 
In this way, $\hat{w}(\xi_1,\xi_2,t)$ is kept close to $\hat{w}_0(\xi_1,\xi_2,t)$ as $t$ increases, and the separation $\hat{w}(\xi_1,\xi_2,t)-\hat{w}_0(\xi_1,\xi_2,t)$ asymptotically aligns with the fastest growing mode of the deviation from $\hat{w}_0(\xi_1,\xi_2,t)$ if such an exponentially growing mode exists. 
The growth rate can be approximated from the asymptotic change in the norm of $\hat{w}(\xi_1,\xi_2,t)-\hat{w}_0(\xi_1,\xi_2,t)$ over a single advance.

\subsection{Modulational stability analysis for the background solution of the DS system}
\label{a:modulationalstability}

In this section we study the modulational stability of the constant background solutions of all four variants of the DS system.  
For simplicity we set $\epsilon=1$ for all the calculations in this section.

Similarly to section~\ref{s:linearizationstabilityanalysis},
we look for a perturbation of the uniform solution of the DS system~\eqref{e:DS1} in the form
\be
\label{e:perturbedbackgroundDSII}
\psi(\@x,t) = A\,e^{\sigma\nu i A^2 t} \big( 1 + \delta (w_1 + i w_{-1})\,\big)\,,\qquad
P(\@x,t) = \delta w_0\,,
\ee
where $w_{\pm1}$ and $w_0$ are real-valued functions of $\@x$ and $t$,
and $0<\delta\ll1$ is a parameter expressing the smallness of the perturbations.
(The difference from section~\eqref{s:linearizationstabilityanalysis} is that now the unperturbed spatial profile is constant.)
Substituting~\eqref{e:perturbedbackgroundDSII} into~\eqref{e:DS1}, the leading-order terms cancel automatically. Next, neglecting higher order terms in $\delta$, we obtain
\bse
\begin{gather}
    2 \sigma \nu A^2  u + i (u_t + iv_t) + \half\big[ (u_{xx}- \sigma u_{yy}) + i (v_{xx}- \sigma v_{yy})\big] + \sigma A w = 0\,,\\
    2 \sigma \nu A^2  u - i (u_t - iv_t) + \half\big[ (u_{xx}- \sigma u_{yy}) - i (v_{xx}- \sigma v_{yy})\big] + \sigma A w = 0\,,\\
    4 \nu A^2  u_{xx} + w_{xx} + \sigma w_{yy}  = 0\,.
\end{gather}
\ese
By taking the sum and difference of the first two equations, we obtain three real PDEs for the three perturbed fields:
\be
\label{e:perturbedPDEsDSII}
    u_t + \half(v_{xx} - \sigma v_{yy}) = 0\,, \quad
    v_t + \half(u_{xx} - \sigma u_{yy}) -\sigma A( 2 \nu  A  u + w ) = 0\,, \quad
    4 \nu A^2  u_{xx} + w_{xx} + \sigma w_{yy}  = 0\,.
\ee
Since~\eqref{e:perturbedPDEsDSII} are linear, we look for plane wave solution of the form
\be
\label{e:planewave}
u = W_1 e^{i( kx + ly - \omega t)} \,,\qquad 
v = W_{-1} e^{i( kx + ly - \omega t)} \,,\qquad 
w = W_0 e^{i( kx + ly - \omega t)} \,,
\ee
with $W_{\pm1}$ and $W_0$ arbitrary constants.
Substituting~\eqref{e:planewave} into~\eqref{e:perturbedPDEsDSII} we obtain the homogeneous linear system
\be 
\begin{bmatrix}
    -i\omega & - \half( k^2 - \sigma l^2 ) & 0\\
    -2\sigma\nu A^2 + \half ( k^2 - \sigma l^2 ) & -i\omega & - \sigma A \\
    -4 \nu A^2 k^2 & 0 & -k^2 - \sigma l^2
\end{bmatrix} \,\@W = \@0\,,
\ee
where $\@W = (W_1,W_{-1},W_0)^\t$,
which is the analogue of~\eqref{e:linearizedeigensystem}.
Looking for nontrivial solutions, we then obtain the linearized dispersion relation:
\be
\omega = \pm \frac{(k^2 - \sigma l^2) \sqrt{4\sigma\nu A^2 + k^2 + \sigma l^2 }}{ 2 \sqrt{k^2 + \sigma l^2}}\,.
\label{e:omegaMI}
\ee
We now look at the values that \eqref{e:omegaMI} can take for the four variants.

For the defocusing DSII system ($\sigma = \nu = 1$), $\omega$ is always real, and therefore no modulational instability occurs.
In contrast, for the focusing DSII system ($\sigma = 1$ and $\nu=-1$), modulational instability occurs for all $(k,l)$ satisfying 
$k^2 + l^2 < 4 A^2$, unless $k^2 = l^2$.
Moreover, the relative growth rate of long-wavelength perturbations of the constant background of the focusing DSII system is $1$, as obtained in the limit $k^2 + l^2 \to 0$ with $A=1$.

For the defocusing DSI system ($\sigma = -1$ and $\nu=1$), modulational instability occurs for all $(k,l)$ satisfying 
$0 < k^2 - l^2 < 4 A^2$.
Conversely, for the focusing DSI system ($\sigma = -1$ and $\nu=-1$), modulational instability occurs for all $(k,l)$ satisfying 
$l^2 - k^2 > 4 A^2$.

The above results agree with the predictions from Whitham modulation theory for $m=0$
(see section~\ref{s:stability}).

\subsection{Further details on the determination of the periodic solution for the fluid density}
\label{a:periodicdensity}

In this appendix we provide some further details on the determination of the periodic solutions for the fluid density $\rho(Z)$ described in section~\ref{s:leadingorderdensity}.

Using the identity~\eqref{a:lnrho''},  we can rewrite~\cref{J2rho} as
\begin{equation}
\rho''' + \frac{(\rho')^3}{\rho^2} - \frac{2\rho'\rho''}{\rho} 
	- \frac{4\rho'\rho}{\kappa} 
	+ \frac{4J^2\rho'}{\rho^2} = 0\,,
\label{e:rho'''}
\end{equation}
which, integrating, yields
\begin{equation}
\rho'' - \frac{(\rho')^2}{\rho} - \frac{2\rho^2}{\kappa} 
	- \frac{4J^2}{\rho} + 2c_1 = 0,
\label{e:rho''}
\end{equation}
where $2c_1$ is a constant of integration.
Multiplying by $2\rho'/\rho^2$, we get
\begin{equation}
\frac{2\rho'\rho''}{\rho^2} - \frac{2(\rho')^3}{\rho^3} 
	- \frac{4\rho'}{\kappa} - \frac{8J^2\rho'}{\rho^3} 
	+ \frac{4c_1 \rho'}{\rho^2} = 0.
\label{e:rho''2}
\end{equation}
Next, using~\eqref{a:rho'sqroverrhosqr'}, and 
integrating~\cref{e:rho''2}, we then obtain~\eqref{e:rho'sqr}.

Next, substituting~\eqref{e:rho2ymap} into~\cref{e:rho'sqr} and~\eqref{rhopSq} yields,
respectively
\bse
\begin{gather}
\left(y'\right)^2 = 
\frac{1}{B^2}\left[ 
\frac{B^3}{\kappa} y^4 + \left(\frac{3A}{\kappa}B^2 - c_2 B^2\right)y^2 
	+ \left( \frac{3A^2B}{\kappa} + c_1 B - 2c_2AB \right)
+ \left(\frac{A^3}{\kappa} + c_1 A - c_2 A^2 - J^2\right)\frac1{y^2} \right],
\\
\left(y'\right)^2 = \frac{1}{\kappa B^2}\bigg[ 
    B^3 y^4 + B^2(3A - s_1 ) y^2 + B(3A^2 - 2As_1 + s_2 )  
        + (A^3 - A^2s_1 + As_2 - s_3 ) \frac1{y^2} \bigg]\,,
\label{yprime}
\end{gather}
\ese
and comparing these expressions one finds \eqref{ccJJ}.

\subsection{Simplification of the modulation equations using convective derivatives}
\label{a:convectivederivatives}

In this appendix we show how the modulation equations
\eqref{e:dqdt}, \eqref{e:dpdt}, \eqref{e:dk1dt}, \eqref{e:dv1dt}, 
\eqref{e:drhobardt}, \eqref{e:drhobarudt} and the two equations~\eqref{e:dPodxdQodx}
can be rewritten as the system~\eqref{e:physicalDSWS} using the convective derivative $D_t$ defined in~\eqref{e:Ddef}.
We then show how to obtain~\eqref{RHSw3} and~\ref{RHSw4}.

1. To rewrite equation~\eqref{e:dqdt} and~\eqref{e:dpdt}, one simply replaces the temporal derivative of $q$ and $p$ with the convective derivative $D_t$.
Next, we look at the modulation equation for $k_1$, namely~\eqref{e:dk1dt}.
We have
\vspace*{-1ex}
\begin{multline}
k_{1,t} + (k_1(g_- U_1 - \sigma p q))_x 
\\
= D_tk_1 - U_1k_{1,x} + \sigma (qU_1 +p) k_{1,y} + k_{1,x} (g_- U_1 - \sigma p q) + k_1 (g_- U_1 - \sigma p q)_x
\\
= D_tk_1 + \sigma (qU_1 +p) D_y k_1  + k_1 (g_- U_1 - \sigma p q)_x \,.
\end{multline}
Using this expression, one obtains the modulation equation~\eqref{e:DkDt} for $k_1$.
Next, introducing the quantity $S = k_1 J \overline{\rho^{-1}}$ as in~\eqref{e:Sdef}, we have $v_1 = U_1 + S$, so we can rewrite equation~\eqref{e:dv1dt} as
\be
v_{1,t} + w_2 + \left(v_1 (g_-U_1 - \sigma p q)\right)_x = 0\,,
\label{e:dv1dt_tmp}
\ee
where 
$w_2=- \half \left(g_- U_1^2 + \sigma p^2 + 2\sigma P_o-k_1^2 g_- c_2\right)_x$. 
Moreover, we write the left-hand side of~\eqref{e:dv1dt_tmp} as
\vspace*{-1ex}
\be
D_t v_1 - U_1 v_{1,x} + \sigma (q U1 + p )v_{1,y} + w_2 + \left(v_1 (g_-U_1 - \sigma p q)\right)_x\,,
\ee
or equivalently, substituting $v_{1,y}=v_{2,x} = (q v_1)_x + p_x$, as
\be
D_t v_1 - U_1 v_{1,x} + \sigma (q U_1 + p )(p_x+ v_1q_x + qv_{1,x}) + w_2 + v_{1,x} (g_-U_1- \sigma p q) + v_1 (g_-U_1- \sigma p q)_x\,.
\ee
Then, collecting the coefficients of $v_{1,x}$, 
we note that they all cancel each other out,
which allows us to rewrite \eqref{e:dv1dt_tmp} as
\be
D_t v_1 + \sigma (q U_1 + p )p_x + v_1 (\sigma (q U_1 + p ) q_x + (g_-U_1- \sigma p q)_x) + w_2 = 0\,.
\ee
Next, by substituting $D_tv_1 = D_tU_1 + D_tS$, we have
\be
D_tU_1 + D_tS + W_2' = 0\,,\qquad
W_2' = w_2 + \sigma (q U_1 + p )p_x + (U_1+S) (\sigma (q U_1 + p ) q_x + (g_-U_1- \sigma p q)_x) \,.
\label{e:DU1Dt1}
\ee
Finally, multiplying equation~\eqref{e:DkDt} by $U_1-S$ and then adding it to equation~\eqref{e:DU1Dt1},  
one obtains~\eqref{e:DU1}.

2. Next we rewrite the modulation equation~\eqref{e:drhobardt} for $\rhobar$ respectively as
\bse
\begin{align*}
\rhobar_t + D_x (k_1 J+U_1\rhobar) - \sigma (k_1 J+U_1 \rhobar)q_y - \sigma (p\rhobar)_y = 0\,,\\
\noalign{\noindent and}
D_t \rhobar +\rhobar D_xU_1 + D_x (k_1 J) - \sigma((k_1 J+U_1 \rhobar)q_y +\rhobar p_y) = 0\,.
\end{align*}
\ese
This yields a modulation equation for $\rhobar$ as
\be
D_t \rhobar + w_3 = 0\,,\qquad
w_3 =\rhobar D_xU_1 + D_x (k_1 J) - \sigma((k_1 J+U_1 \rhobar)q_y +\rhobar p_y) \,.
\label{e:DrhobarDt1}
\ee
We then multiply equation~\eqref{e:DkDt} by $-\rhobar$ and add it to equation~\eqref{e:DrhobarDt1}.  
Doing so, one finally obtains~\eqref{e:Drhobar}.

3. For equation~\eqref{e:drhobarudt} for $\rhobar u$, we rewrite the first four terms in the equation as
\bse
\begin{multline*}
(k_1J + U_1\rhobar)_t  + D_x\big( v_1 k_1 J + U_1(k_1 J +U_1 \rhobar ) \big) - \sigma \big( v_1 k_1 J + U_1(k_1 J +U_1 \rhobar ) \big) q_y - \sigma \big( p(k_1J+U_1\rhobar) \big)_y 
\\
= D_t(k_1 J) + \rhobar D_t(U_1) + U_1 D_t{\rhobar}  +  (k_1 J+U_1\rhobar) D_x(U_1)   - \sigma \big( v_1 k_1 J + U_1(k_1 J +U_1 \rhobar ) \big) q_y
\\
+ D_x ( v_1 k_1 J) - \sigma (k_1J+U_1\rhobar)p_y\,. 
\end{multline*}
\ese
Now we substitute~\eqref{e:DrhobarDt1} in the above expression, obtaining
\vspace*{-1ex}
\bse
\begin{multline*}
D_t(k_1 J) + \rhobar D_t(U_1) + U_1 D_t\rhobar + (k_1 J+U_1\rhobar)D_x(U_1)
\\
  - \sigma\big(v_1 k_1 J+U_1(k_1 J+U_1\rhobar)\big)q_y + D_x(v_1 k_1 J)- \sigma(k_1J+U_1\rhobar)p_y 
\\
= D_t(k_1 J)+\rhobar D_t(U_1) + k_1 J D_x(U_1)-\sigma v_1 k_1 Jq_y + D_x(v_1 k_1 J)-U_1D_x(k_1 J)-\sigma k_1Jp_y.
\end{multline*}
\ese
In this way we obtain a new modulation equation for $k_1 J$ as
\vspace*{-1ex}
\be
\label{e:Dk1JDt1}
D_t(k_1 J) + \rhobar D_t(U_1) + w_4 = 0\,,
\ee
where
\vspace*{-1ex}
\bse
\begin{multline}
w_4 
  = 2k_1 J D_x(U_1) - \sigma U_1 k_1 J q_y - \sigma k_1 J p_y
  + \Bigg( \frac{k_1^2 (2+g_-)}{2\kappa}\overline{\rho^2} \Bigg)_x
  - \sigma (P_o \rhobar)_x + \Bigg( \dfrac{\overline{Q^2}-\sigma\overline{P^2}}{4\nu} \Bigg)_x
  + \Bigg( \dfrac{\overline{PQ}}{2\nu} \Bigg)_y
  \\
  +\bigg(k_1^2 \bigg(\frac{1}{\kappa}\overline{\rho^2} + c_1 - c_2\rhobar\bigg)\bigg)_x 
  -\sigma\bigg(k_1^2 q\bigg(\frac{1}{\kappa}\overline{\rho^2} + c_1 - c_2\rhobar\bigg) \bigg)_y\,. \nonumber    
\end{multline}
\ese
Multiplying equation~\eqref{e:DkDt} by $-2k_1 J$ and adding it to equation~\eqref{e:Dk1JDt1},
one then finally obtains~\eqref{e:Dk1JDt}.

4. Finally, we can rewrite the equation~\eqref{e:dPodxdQodx} using the fact that
\bse
\label{e:DPoDxDQoDx}
\begin{multline*}
\bigg(-\frac{2k_1^2}{\kappa\sigma}\rhobar + P_o + 2\frac{k_1^2+\sigma k_2^2}{\kappa\sigma} \rhobar\bigg)_x - \sigma \bigg(\frac{2k_1^2 q}{\kappa\sigma}\rhobar + Q_o\bigg)_y 
 = \bigg(\frac{2 \sigma k_1^2 q^2}{\kappa\sigma}\rhobar + P_o \bigg)_x
    - \sigma \bigg(\frac{2k_1^2 q}{\kappa\sigma}\rhobar + Q_o\bigg)_y  \\
 = (P_o)_x - \sigma (Q_o)_y+ (2k_1^2q^2\rhobar/\kappa)_x - 2(k_1^2 q\rhobar/\kappa)_y\,,
\end{multline*}
and
\begin{equation*}
\bigg(\frac{2k_1^2 q}{\kappa\sigma}\rhobar + Q_o\bigg)_x 
  + \bigg(-\frac{2k_1^2}{\kappa\sigma}\rhobar + P_o\bigg)_y 
  = \sigma (Q_o)_x + \sigma (P_o)_y - 2D_y( k_1^2 \rhobar/\kappa) + 2 (k_1 ^2 \rhobar/\kappa) q_x \,.
\end{equation*}
\ese

5. For \eqref{e:Drhobar}, one can first verify by direct calculations that
\begin{multline}
\frac{3\sigma \nu}{2g_+} \left(D_t\overline{\rho} - \frac{\rhobar}{k_1}D_tk_1+W_3\right)
+ \frac{3}{4} s_1 \times \eqref{RHSw2} 
+ \frac{3}{4}(s_2 - \frac12 s_1^2) \times \eqref{RHSw1} 
\\ 
= \frac32\sigma \nu\left( D_t (\overline{\rho}/g_+) - \frac{\rhobar}{g_+ k_1}D_t k_1 + \frac{W_3}{g_+} \right) 
+ \frac{3q \rhobar}{ g_+^2}  D_t q
+ \frac{3}{4}\left( s_1 \times \eqref{RHSw2} + \left(s_2-\frac12 s_1^2\right) \times \eqref{RHSw1} \right) 
\\ 
= \frac{3}{k_1}\sum_{i=1}^4 r_i^2\frac{\partial k_1}{\partial r_i}D_t r_i 
- \frac{3q \rhobar}{ g_+^2}W_q
+ \frac{3}{2g_+}W_3
+ \frac{3}{4}\left( s_1 W_2 + \left(s_2-\frac12 s_1^2\right)W_1 \right)\,.
\end{multline}
In turn, 
using the chain rule in the second step, 
this yields~\eqref{RHSw3},
and where $W_3'$ is given by~\eqref{e:W_3'}.

6. For equation\eqref{e:Dk1JDt}, one can also verify that
\begin{multline}
\frac{\sigma\nu}{g_+}\left(D_t(k_1J) + \overline{\rho} D_tU_1 - 2k_1J\frac{D_tk_1}{k_1} + W_4\right)
+ s_1 \eqref{RHSw3}
+ \frac12 (s_2 - s_1^2)\eqref{RHSw2} 
+ \left(s_3 - s_1s_2 + \frac{1}{4}s_1^3\right)\eqref{RHSw1}
\\ 
= \sigma \nu D_t(k_1J/g_+) + \sigma \nu(\overline{\rho}/g_+)D_tU_1 - 2\sigma \nu (k_1J/g_+)\frac{D_tk_1}{k_1} + (\sigma \nu W_4/g_+) 
+ 2\sigma q (k_1J/g_+^2)D_tq 
\\ 
+ s_1\eqref{RHSw3} + \frac12 (s_2-s_1^2)\eqref{RHSw2}
+ \left(s_3 - s_1s_2 + \frac{1}{4}s_1^3\right)\eqref{RHSw1} 
\\
= \frac{4}{k_1}\sum_{i=1}^4 r_i^3 \frac{\partial k_1}{\partial r_i} D_tr_i 
- 2 \sigma q (k_1J/g_+^2)W_q
+ \frac{1}{g_+} W_4
+ s_1 W_3' + \frac12 (s_2-s_1^2)W_2 
+ \left(s_3 - s_1s_2 + \frac{1}{4}s_1^3\right)W_1\,,
\end{multline}
which finally yields~\eqref{RHSw4}, and where $W_4'$ is given by~\eqref{e:W_4'}.

\addcontentsline{toc}{section}{References}
\def\reftitle#1{``#1''}
\def\booktitle#1{\textit{#1}}


\begin{thebibliography}{O}
\makeatletter
\def\@biblabel#1{#1.}
\def\v#1{{\bf#1}}
\makeatother
\def\journal#1#2{\textit{#1} {\bf#2}}
\advance\itemsep-3pt
\small

\bibitem{ABH_JPA2023}
A. Abeya, G. Biondini and M. A. Hoefer,
\reftitle{Whitham modulation theory for the nonlinear Schr\"odinger equation in two and three spatial dimensions},
\journal{J. Phys. A}{56}, 025701 (2023)

\bibitem{PRE63p046605}
M. J. Ablowitz, G. Biondini and S. Blair,
\reftitle{Nonlinear Schr\"odinger equations with mean terms in non-resonant multi-dimensional quadratic materials},
\journal{Phys. Rev. E}{63}, 046605 (2001)

\bibitem{ABR2018}
M. J. Ablowitz, G. Biondini and I. Rumanov,
\reftitle{Whitham modulation theory for (2+1)-dimensional equations of Kadomtsev-Petviashvili type},
\journal{J. Phys. A}{51}, 215501 (2018)

\bibitem{RSPA2017}
M. J. Ablowitz, G. Biondini and Q. Wang,
\reftitle{Whitham modulation theory for the Kadomtsev-Petviashvili equation},
\journal{Proc. Roy. Soc. A}{473}, 20160695 (2017)

\bibitem{ABW_2DBO}
M. J. Ablowitz, G. Biondini and Q. Wang,
\reftitle{Whitham modulation theory for the two-dimensional Benjamin-Ono equation},
\journal{Phys. Rev. E}{96}, 032225 (2017)

\bibitem{AC1991}
M. J. Ablowitz and P. A. Clarkson,
\booktitle{Solitons, nonlinear evolution equations and inverse scattering}
(Cambridge University Press, 1991)

\bibitem{ACR_2DNLS}
M. J. Ablowitz, J. Cole and I. Rumanov,
\reftitle{On the Whitham system for the radial nonlinear Schr\"odinger equation},
\journal{Stud. Appl. Math.}{142}, 269--313 (2019)

\bibitem{AS1979}
M. J. Ablowitz and H. Segur,
\reftitle{On the evolution of packets of water waves},
\journal{J. Fluid Mech.}{92}, 691 (1979)

\bibitem{AS1981}
M. J. Ablowitz and H. Segur,
\booktitle{Solitons and the inverse scattering transform}
(SIAM, 1981)

\bibitem{Assainova}
O. Assainova, C. Klein, K. D. T.-R. McLaughlin and P. D. Miller,
\reftitle{A study of the direct spectral transform for the defocusing Davey-Stewartson II equation in the semiclassical limit},
\journal{Commun. Pure Appl. Math.}{72}, 1474 (2019)

\bibitem{BenneyNewell1967}
D. J. Benney and A. C. Newell,
\reftitle{The propagation of nonlinear wave envelopes},
\journal{J. Math. Phys. (Stud. Appl. Math.)}{46}, 133 (1967)

\bibitem{BenneyRoskes}
D. J. Benney and G. J. Roskes,
\reftitle{Wave instabilities},
\journal{Stud. Appl. Math.}{48}, 377 (1969)

\bibitem{NLTY2024}
G. Biondini, A. J. Bivolcic, M. A. Hoefer and A. Moro,
\reftitle{Two-dimensional reductions of the Whitham modulation system for the Kadomtsev-Petviashvili equation},
\journal{Nonlinearity}{37}, 025012 (2024)

\bibitem{BC2024}
G. Biondini and A. Chernyavsky,
\reftitle{Whitham modulation theory for the Zakharov-Kuznetsov equation and transverse stability of its periodic traveling wave solutions},
\journal{Stud. Appl. Math.}{152}, 596 (2024)

\bibitem{supplement}
Please see the supplementary material.

\bibitem{NLTY2020}
G. Biondini, M. A. Hoefer and A. Moro,
\reftitle{Integrability, exact reductions and special solutions of the KP-Whitham equations},
\journal{Nonlinearity}{33}, 4114--4132 (2020)

\bibitem{EAJAM2022}
G. Biondini and D. Kireyev,
\reftitle{On the periodic solutions of the Davey-Stewartson systems},
\journal{East Asian J. Appl. Math.}{12}, 1 (2022)

\bibitem{JPA55p305701}
G. Biondini, D. Kireyev and K.-i. Maruno,
\reftitle{Soliton resonance and web structure in the Davey-Stewartson system},
\journal{J. Phys. A}{55}, 305701 (2022)

\bibitem{BLMP:PLA1988}
M. Boiti, J. J. Leon, L. Martina and F. Pempinelli,
\reftitle{Scattering of localized solitons in the plane},
\journal{Phys. Lett. A}{132}, 432 (1988)

\bibitem{DeconinckBottman}
N. Bottman, B. Deconinck and M. Nivala,
\reftitle{Elliptic solutions of the defocusing NLS equation are stable},
\journal{J. Phys. A}{44}, 285201 (2011)

\bibitem{BrownPerry}
R. M. Brown and P. A. Perry,
\reftitle{Soliton solutions and their (in)stability for the focusing Davey-Stewartson II equation},
\journal{Nonlinearity}{31}, 4290 (2018)

\bibitem{ByrdFriedman}
P. Byrd and M. Friedman,
\booktitle{Handbook of elliptic integrals for engineers and scientists}
(Springer-Verlag, Berlin, 1971)

\bibitem{CalogeroEckhaus1987}
F. Calogero and W. Eckhaus,
\reftitle{Nonlinear evolution-equations, rescalings, model PDEs and their integrability},
\journal{Inv. Probl.}{3}, 229 (1987)

\bibitem{CarterSegur}
J. Carter and H. Segur,
\reftitle{Instabilities in the two-dimensional cubic nonlinear Schr\"odinger equation},
\journal{Phys. Rev. E}{68}, 045601R (2003)

\bibitem{ClarkeMarangellPerkins}
W. A. Clarke, R. Marangell and W. R. Perkins,
\reftitle{Rigorous justification of the Whitham modulation equations for equations of Whitham type},
\journal{Stud. Appl. Math.}{149}, 297 (2022)

\bibitem{DaveyStewartson}
A. Davey and K. Stewartson,
\reftitle{On three-dimensional packets of surface waves},
\journal{Proc. Roy. Soc. A}{338}, 101--110 (1974)

\bibitem{DeconinckSegal}
B. Deconinck and B. L. Segal,
\reftitle{The stability spectrum for elliptic solutions to the focusing NLS equation},
\journal{Phys. D}{346}, 1--19 (2017)

\bibitem{driscollfornberg}
T. A. Driscoll and B. Fornberg,
\reftitle{A fast spectral algorithm for nonlinear wave equations with linear dispersion},
\journal{J. Comput. Phys.}{155}, 456--467 (1999)

\bibitem{ElHoefer2016}
G. A. El and M. A. Hoefer,
\reftitle{Dispersive shock waves and modulation theory},
\journal{Phys. D}{333}, 11--65 (2016)

\bibitem{Fokas2003}
A. S. Fokas,
\reftitle{The Davey-Stewartson I equation on the quarter plane with homogeneous Dirichlet boundary conditions},
\journal{J. Math. Phys.}{44}, 3226--3244 (2003)

\bibitem{FokasAblowitz1983}
A. S. Fokas and M. J. Ablowitz,
\reftitle{Method of solution for a class of multidimensional nonlinear evolution equations},
\journal{Phys. Rev. Lett.}{51}, 7--10 (1983)

\bibitem{FokasAblowitz1984}
A. S. Fokas and M. J. Ablowitz,
\reftitle{On the inverse scattering transform of multidimensional nonlinear equations related to first-order systems in the plane},
\journal{J. Math. Phys.}{25}, 2494--2505 (1984)

\bibitem{FokasPelinovskySulem}
A. S. Fokas, D. E. Pelinovsky and C. Sulem,
\reftitle{Interaction of lumps with a line soliton for the DSII equation},
\journal{Phys. D}{152--153}, 189--198 (2001)

\bibitem{FS1990}
A. S. Fokas and P. M. Santini,
\reftitle{Dromions and a boundary value problem for the Davey-Stewartson 1 equation},
\journal{Phys. D}{44}, 99--130 (1990)

\bibitem{Fornberg}
B. Fornberg,
\booktitle{A practical guide to pseudospectral methods}
(Cambridge University Press, 1998)

\bibitem{GuoMichalache2022}
L. Guo, L. Chen, D. Mihalache and J. He,
\reftitle{Dynamics of soliton interaction solutions of the Davey-Stewartson I equation},
\journal{Phys. Rev. E}{105}, 014218 (2022)

\bibitem{GungorOzemir}
F. Gungor and C. Ozemir,
\reftitle{Dispersionless Davey-Stewartson system: Lie symmetry algebra, symmetry group and exact solutions},
\journal{Eur. Phys. J. Plus}{136}, 715 (2021)

\bibitem{HasegawaKodama}
A. Hasegawa and Y. Kodama,
\booktitle{Solitons in optical communications}
(Oxford University Press, Oxford, 1995)

\bibitem{HuLuoWang2024}
L. Hu, X.-D. Luo and Z. Wang,
\reftitle{Diffraction and interaction of interfacial solitons in a two-layer fluid of great depth},
\journal{SIAM J. Appl. Math.}{84}, 1385 (2024)

\bibitem{IR2000}
E. Infeld and G. Rowlands,
\booktitle{Nonlinear waves, solitons and chaos}
(Cambridge University Press, 2000)

\bibitem{JohnsonOregero}
M. A. Johnson and J. A. Oregero,
\reftitle{Modulational stability of wave trains in the Camassa-Holm equation},
\journal{J. Diff. Eq.}{446}, 113627 (2025)

\bibitem{JohnsonOregeroPerkins}
M. A. Johnson, J. A. Oregero and W. R. Perkins,
\reftitle{On the modulation of wave trains in the Ostrovsky equation},
arXiv:2505.21466

\bibitem{JohnsonPerkins}
M. A. Johnson and W. R. Perkins,
\reftitle{Modulational instability of viscous fluid conduit periodic waves},
\journal{SIAM J. Math. Anal.}{52}, 277 (2020)

\bibitem{Kamchatnov}
A. M. Kamchatnov,
\booktitle{Nonlinear periodic waves and their modulations}
(World Scientific, 2000)

\bibitem{KivsharPelinovsky2000}
Yu. S. Kivshar and D. E. Pelinovsky,
\reftitle{Self-focusing and transverse instabilities of solitary waves},
\journal{Phys. Rep.}{331}, 117--195 (2000)

\bibitem{KleinMcLaughlinStoilovPhysD2019}
C. Klein, K. McLaughlin and N. Stoilov,
\reftitle{Spectral approach to the scattering map for the semi-classical defocusing Davey-Stewartson II equation},
\journal{Phys. D}{400}, 132126 (2019)

\bibitem{KleinMcLaughlinStoilovPRSA2019}
C. Klein, K. McLaughlin and N. Stoilov,
\reftitle{High precision numerical approach for Davey-Stewartson II type equations for Schwartz class initial data},
\journal{Proc. Roy. Soc. A}{476}, 20190864 (2019)

\bibitem{KleinRoidot2014}
C. Klein and K. Roidot,
\reftitle{Numerical study of the semiclassical limit of the Davey-Stewartson II equations},
\journal{Nonlinearity}{27}, 2177 (2014)

\bibitem{KleinSaut}
C. Klein and J.-C. Saut,
\booktitle{Nonlinear dispersive equations: Inverse scattering and PDE methods}
(Springer, 2021)

\bibitem{KleinStoilovI}
C. Klein and N. Stoilov,
\reftitle{Numerical study of blow-up mechanisms for Davey-Stewartson II systems},
\journal{Stud. Appl. Math.}{141}, 89 (2018)

\bibitem{KleinStoilovII}
C. Klein and N. Stoilov,
\reftitle{Numerical study of Davey-Stewartson I systems},
\journal{Stud. Appl. Math.}{149}, 76 (2022)

\bibitem{KuznetsovTuritsyn1988}
E. A. Kuznetsov and S. K. Turitsyn,
\reftitle{Instability and collapse of solitons in media with a defocusing nonlinearity},
\journal{Zh. Eksp. Teor. Fiz.}{94}, 119--129 (1988)
[\journal{Sov. Phys. JETP}{67}, 1583--1588 (1988)]

\bibitem{Leblond}
H. Leblond,
\reftitle{Electromagnetic waves in ferromagnets: a Davey-Stewartson-type model},
\journal{J. Phys. A}{32}, 7907 (1999)

\bibitem{JFM2026}
H. Lin, A. Granquist, G. Biondini, G. A. El and M. A. Hoefer,
\reftitle{Exact theory of two-dimensional dispersive shallow water flow over corners},
in preparation

\bibitem{ManasSantini}
M. Manas and P. M. Santini,
\reftitle{Solutions of the Davey-Stewartson II equation with arbitrary rational localization and nontrivial interaction},
\journal{Phys. Lett. A}{227}, 325--334 (1997)

\bibitem{milewskitabak}
P. A. Milewski and E. G. Tabak,
\reftitle{A pseudospectral procedure for the solution of nonlinear wave equations with examples from free surface flows},
\journal{SIAM J. Sci. Comput.}{21}, 1102 (1999)

\bibitem{Newell1985}
A. C. Newell,
\booktitle{Solitons in mathematics and physics}
(SIAM, 1985)

\bibitem{NMPZ1984}
S. P. Novikov, S. V. Manakov, L. P. Pitaevskii and V. E. Zakharov,
\booktitle{Theory of solitons. The inverse scattering method}
(Plenum, New York, 1984)

\bibitem{NIST}
F. W. J. Olver, D. W. Lozier, R. Boisvert and C. W. Clark,
\booktitle{NIST handbook of mathematical functions}
(Cambridge University Press, 2010)

\bibitem{Pelinovsky1994}
D. Pelinovsky,
\reftitle{Instability and decay of solitary waves in the Davey-Stewartson I system},
\journal{Phys. Lett. A}{196}, 181--186 (1994)

\bibitem{Perry2016}
P. A. Perry,
\reftitle{Global well-posedness and long-time asymptotics for the defocussing Davey-Stewartson II equation in $H^{1,1}(\Complex)$},
\journal{J. Spectral Theory}{6}, 429--481 (2016)

\bibitem{RaoHeCheng2022}
J. Rao, J. He and Y. Cheng,
\reftitle{The Davey-Stewartson I equation: doubly localized two-dimensional rogue lumps on the background of homoclinic orbits or constant},
\journal{Lett. Math. Phys.}{112}, 75 (2022)

\bibitem{NLTY2021}
S. Ryskamp, M. A. Hoefer and G. Biondini,
\reftitle{Oblique interactions between solitons and mean flows in the Kadomtsev-Petviashvili equation},
\journal{Nonlinearity}{34}, 3583--3617 (2021)

\bibitem{RSPA2022}
S. Ryskamp, M. A. Hoefer and G. Biondini,
\reftitle{Modulation theory for soliton resonance and Mach reflection},
\journal{Proc. Roy. Soc. A}{478}, 20210823 (2022)

\bibitem{JFM2021}
S. Ryskamp, M. D. Maiden, G. Biondini and M. A. Hoefer,
\reftitle{Evolution of truncated and bent gravity wave solitons: the Mach expansion problem},
\journal{J. Fluid Mech.}{909}, A24 (2021)

\bibitem{SenChowdhury}
S. Sen and A. Roy Chowdhury,
\reftitle{On the stability of Davey-Stewartson system: a dressing operator approach},
\journal{Ann. Phys.}{7}, 99 (1989)

\bibitem{SulemSulem}
C. Sulem and P. L. Sulem,
\booktitle{The nonlinear Schr\"odinger equation: Self-focusing and wave collapse}
(Springer, 1999)

\bibitem{TajiriArai}
M. Tajiri and T. Arai,
\reftitle{Periodic soliton solutions of the Davey-Stewartson equation},
\journal{Proc. Inst. Math. Nat. Acad. Sci. Ukraine}{30}, 210--217 (2000)

\bibitem{TajiriTakeuchiArai}
M. Tajiri, K. Takeuchi and T. Arai,
\reftitle{Soliton stability to the Davey-Stewartson I equation by the Hirota method},
\journal{J. Phys. Soc. Japan}{70}, 1505--1511 (2001)

\bibitem{Trefethen}
L. N. Trefethen,
\booktitle{Spectral methods in Matlab}
(SIAM, Philadelphia, 2000)

\bibitem{WangWang2019}
D. Wang and H. Wang,
\reftitle{An asymptotic preserving scheme for the defocusing Davey-Stewartson II equation in the semiclassical limit},
\textit{IMA J. Numer. Anal.}, draf019 (2025)

\bibitem{Whitham1974}
G. B. Whitham,
\booktitle{Linear and nonlinear waves}
(Wiley, 1974)

\bibitem{Xu2017}
T. Xu, M. Li, Y. Huang, Y. Chen and C. Yu,
\reftitle{Nonsingular localized wave solutions for the nonlocal Davey-Stewartson I equation with zero background},
\journal{Mod. Phys. Lett. B}{35}, 1750338 (2017)

\bibitem{GeYi2020}
G. Yi,
\reftitle{On the dispersionless Davey-Stewartson system: Hamiltonian vector field Lax pair and relevant nonlinear Riemann-Hilbert problem for dDS-II system},
\journal{Lett. Math. Phys.}{110}, 445--463 (2020)

\bibitem{GeYi2022}
G. Yi,
\reftitle{On the dispersionless Davey-Stewartson hierarchy: Zakharov-Shabat equations, twistor structure and Lax-Sato formalism},
\journal{Z. Angew. Math. Mech.}{102}, e202100435 (2022)

\bibitem{YiLiaoTianWang2022}
G. Yi, X. Liao, K. Tian and Z. Wang,
\reftitle{The semi-classical limit of Davey-Stewartson hierarchy},
\journal{Appl. Math. Lett.}{131}, 108012 (2022)

\bibitem{Zakharov1972}
V. E. Zakharov,
\reftitle{Collapse of Langmuir waves},
\journal{Sov. Phys. JETP}{35}, 908 (1972)

\end{thebibliography}
\end{document}